\documentclass[aps,prd,reprint,superscriptaddress,nofootinbib]{revtex4-1}
\usepackage[all]{xy}
\usepackage{amsmath,amsthm,amssymb}
\usepackage{graphicx}
\usepackage{color}
\usepackage{comment}
\usepackage{array}
\usepackage{bm}
\allowdisplaybreaks[1]
\usepackage{hyperref}

\definecolor{CiteColor}{rgb}{0,0.6,0}
\hypersetup{citecolor=CiteColor}

\newcommand{\ie}{{\it i.e.}}
\newcommand{\eg}{{\it e.g.}}

\definecolor{red  }{rgb}{1,0,0}
\definecolor{blue }{rgb}{0,0,1}
\definecolor{green}{rgb}{0,1,0}

\date{\today}
\begin{document}

\thispagestyle{empty}

%%%-----------------------------------------------------------------%%%
%%%-----------------------------------------------------------------%%%
\title{Self-force effects on the marginally bound zoom-whirl orbit 
in Schwarzschild spacetime }

\def\addOpen{The Open University of Japan, Chiba 261-8586, Japan}
\def\addIIP{International Institute of Physics, 
Universidade Federal do Rio Grande do Norte, 59070-405, Natal, Brazil}
\def\addSoton{School of Mathematics, University of Southampton, 
Southampton SO17 1BJ, United Kingdom}
\def\addIHES{Institut des Hautes Etudes Scientifiques, 
91440 Bures-sur-Yvette, France}
\def\addKyushu{Faculty of Arts and Science, Kyushu University, 
Fukuoka 819-0395, Japan}
\def\addUIB{Departament de Física, Universitat de les Illes Balears, IAC3--IEEC, Crta.\ Valldemossa km 7.5, E-07122 Palma, Spain}

\author{Leor Barack}
\affiliation{\addSoton}

\author{Marta Colleoni}
%\email{marta.colleoni_at_soton.ac.uk}
\affiliation{\addUIB}

\author{Thibault Damour}
\affiliation{\addIHES}

\author{Soichiro Isoyama}
%%\affiliation{\addGuelph}
\affiliation{\addSoton}
\affiliation{\addOpen}
\affiliation{\addIIP}

\author{Norichika Sago}
\affiliation{\addKyushu}

\date{\today}

%%%%%%%%%%%%%%%%%%%%%%%%%%%%%%%%%%%%%%%%%%%%%%%%%%
\begin{abstract}
%%%%%%%%%%%%%%%%%%%%%%%%%%%%%%%%%%%%%%%%%%%%%%%%%%
For a Schwarzchild black hole of mass $M$, we consider a test particle falling from rest at infinity and becoming trapped, at late time, on the unstable circular orbit of radius $r=4GM/c^2$. %(the IBCO: innermost energetically-bound circular orbit). 
When the particle is endowed with a small mass, $\mu\ll M$, it experiences an effective gravitational self-force, whose conservative piece shifts the critical value of the angular momentum and the frequency of the asymptotic circular orbit away from their geodesic values. By directly integrating the self-force along the orbit (ignoring radiative dissipation), we numerically calculate these shifts to $O(\mu/M)$.  Our numerical values are found to be in agreement with estimates first made within the Effective One Body formalism, and with predictions of the first law of black-hole-binary mechanics (as applied to the asymptotic circular orbit). Our calculation is based on a time-domain integration of the Lorenz-gauge perturbation equations, and it is a first such calculation for an unbound orbit. We tackle several technical difficulties specific to unbound orbits, illustrating how these may be handled in more general cases of unbound motion. Our method paves the way to calculations of the self-force along hyperbolic-type scattering orbits. Such orbits can probe the two-body potential down to the ``light ring'', and could thus supply strong-field calibration data for eccentricity-dependent terms in the Effective One Body model of merging binaries.
%%%%%%%%%%%%%%%%%%%%%%%%%%%%%%%%%%%%%%%%%%%%%%%%%%
\end{abstract}
%%%%%%%%%%%%%%%%%%%%%%%%%%%%%%%%%%%%%%%%%%%%%%%%%%

%%%%%%%%%%%%%%%%%%%%%%%%%%%%%%%%%%%%%%%%%%%%%%%
%%%%%%%%%%%%%%%%%%%%%%%%%%%%%%%%%%%%%%%%%%%%%%%%%%%
\maketitle
  
%%%%%%%%%%%%%%%%%%%%%%%%%%%%%%%%%%%%%%%
\section{Introduction}
\label{sec:Intro}
%%%%%%%%%%%%%%%%%%%%%%%%%%%%%%%%%%%%%%%%

The extreme mass-ratio regime of the gravitational two-body problem in general relativity is amenable to a perturbative treatment based on a systematic expansion of Einstein's field equations in the small mass ratio $\eta$.
At leading order one recovers the geodesic approximation: the smaller object (assumed sufficiently compact) reduces to a pointlike test particle, and it traces a geodesic orbit in the spacetime associated with the larger object (say, a Kerr black hole).  At subsequent orders, the expansion accounts for the particle's interaction with its own gravitational perturbation (``self-force''), as well as for any effects of its internal structure. In this effective picture, the motion of the small object is described in terms of an accelerated worldline in the background geometry of the larger object. The equation of motion for this worldline is now known through $O(\eta^2)$ in the local effective acceleration \cite{Pound:2012nt,Gralla:2012db,Pound:2017psq}\footnote{These derivations assume the small object is nonspinning. There is a nonperturbative formulation \cite{Harte:2014wya} that accounts for spin and higher structure multipoles but does not apply when the small object is a black hole.},
and a program for computing the self-force and its effects in astrophysically relevant binaries has been ongoing for over two decades.  Recent achievements include numerical calculations of the first-order self-force [$O(\eta)$ self-acceleration] for generic bound orbits in Kerr geometry \cite{vandeMeent:2017bcc}, and a first direct calculation of a second-order effect of the self-force [$O(\eta^2)$ self-acceleration] \cite{Pound:2019lzj}. Ref.\ \cite{Barack:2018yvs} is a recent review of self-force theory and its application to the astrophysical problem of compact-object inspiral into massive black holes.

A central goal of the self-force program is to obtain an accurate model of the gravitational waves from extreme-mass-ratio inspiral sources, which are prime targets for the Laser Interferometer Space Antenna (LISA). But many of the program's intermediate results have proven valuable on their own. In particular, a fruitful synergy emerged with other approaches to the binary inspiral problem. Calculations of self-force contributions to physical quantities like orbital and spin precession, Detweiler's redshift \cite{DetwCO}, or the small object's tidal fields, provide useful benchmarks against which other methods can be tested. Thus, self-force results have informed studies of the performance of the post-Newtonian (PN) expansion in the strong field regime \cite{Favata:2010yd}, played a role 
(notably Refs.\ \cite{Bini:2013zaa,vandeMeent:2016hel})
in the recent derivation of the fourth-PN equation of motion \cite{Damour:2014jta,Damour:2015isa,Bernard:2015njp,Damour:2016abl,Bernard:2016wrg,Foffa:2019rdf,Foffa:2019yfl}, helped test the validity of the conjectured ``first law of black hole binary mechanics'' \cite{alt1} in the strong-field regime, and were even successfully compared with results from fully nonlinear simulations in numerical relativity \cite{LeTiec:2011bk,Damour:2011fu,Tiec:2013twa}. Self-force calculations also play an important role in the ongoing program to refine the Effective One Body (EOB) approach \cite{Buonanno:1998gg,Buonanno:2000ef,Damour:2000we}  to binary dynamics, by providing ``calibration'' data for the EOB potentials (see, e.g., \cite{Bini:2018aps} and references therein). This synergistic program is an area of intensive current activity; we refer readers to \cite{Tiec:2014lba} or Sec.\ 8 of \cite{Barack:2018yvs} for reviews. 

All direct self-force calculations so far have been restricted to adiabatic bound-orbit configurations,\footnote{Perhaps a sole exception is the early work in \cite{Barack:2002ku}, which considered a radial infall trajectory into a Schwarzschild black hole as a first test case, concentrating on method development.} relevant to the astrophysical inspiral problem. In fact, self-force computation methods tend to assume---and rely on---approximate periodicity of the orbit. This is strongly the case for methods based on a frequency-domain treatment of the field equations, but even time-domain methods often rely on periodicity, for reasons explained further below. Of course, in the absence of self-force results, synergistic studies have also been restricted to bound orbits so far.  

There is now a strong drive to extend self-force calculations to unbound, scattering-type problems, and in this paper we report a first step in that direction. We can list at least four motivating factors. First, scattering orbits (especially high-energy ones) can probe the black hole geometry deep inside the gravitational well, below the innermost stable orbit. As such, they can provide valuable calibration data for EOB theory, in a strong-field domain that is inaccessible to bound orbits. This potential was identified by one of us (TD) already in 2010 \cite{damour} (a work that set off the synergy programme between self-force and EOB), and the prospects for its realization are becoming ever more promising with the ongoing work to 
translate the physics of (classical and quantum) post-Minkowkian scattering into a
Hamiltonian description (notably within the EOB formalism)  \cite{Damour:2016gwp,Damour:2017zjx,Bini:2017xzy,Vines:2017hyw,Bini:2018ywr,Vines:2018gqi,Cheung:2018wkq,Bjerrum-Bohr:2018xdl,Bern:2019nnu,Antonelli:2019ytb}. 
The latter works, as well as other gravitational scattering 
computations \cite{Bini:2017wfr,Bini:2017pee,Guevara:2018wpp}, bring with them new opportunities for interfacing with self-force theory.
 As a third motivation, we mention that unbound orbits have a special role in studies of black-hole ``overspinning'' scenarios \cite{CB,Colleoni:2015ena}, on account of their being {\it a priori} most serious candidates for challenging the censorship conjecture; self-force calculations along such orbits are necessary within such analyses.   

Our final reason for studying self-force on unbound orbits is a more fundamental one. There is a sense in which unbound orbits offer a better access to unambiguous information about the conservative sector of the two-body dynamics than bound orbits do. A bound-orbit configuration in black-hole perturbation theory does not admit an obvious (local) notion of conserved energy, as it lacks a local time-translation symmetry (except in the geodesic limit).\footnote{See, however, our discussion below of the first law of binary black-hole mechanics, where a time-averaged notion of such energy is introduced, neglecting dissipation.} An {\it unbound} orbit, on the other hand, has a vanishing interaction potential at $t\to -\infty$ (and also at $t\to +\infty$, if the orbit scatters back to infinity), and therefore a readily identifiable (Bondi-type) invariant mass and binding energy. 
%(Specifically, we are referring here to a Fokker-Wheeler-Feynman-type energy \cite{Fokker1929,Wheeler:1949hn}; see our discussion below.)} 
This direct handle on the energetics of the scattering process is invaluable in establishing a common language between self-force and other approaches (e.g., PN or EOB), which must be based on a catalogue of physically unambiguous, gauge-invariant calculable quantities.  

With these motivations in mind, we set out in this paper to calculate the self-force and its effects in a first example of an unbound orbit. We work in Schwarzschild geometry, and consider the special geodesic orbit that starts at rest at infinity (``zero binding energy'') and has just the right amount of angular momentum to eventually get trapped---dissipation neglected---in eternal motion on an unstable circular orbit. We refer to this unique orbit as the ``Zero (binding) Energy Zoom-whirl Orbit'' (ZEZO). We let $M$ denote the Schwarzschild background mass, and $\mu$ denote the particle's mass, with $\mu/M=\eta\ll 1$.\footnote{Beware that the notation more commonly found in EOB or PN literatures is  $\{m_1,m_2\}$, instead of $\{\mu,M\}$.} In the geodesic approximation (i.e., in the limit $\eta\to 0$, with the self-force fully neglected), the required fine-tuned value of angular momentum is $L=4M\mu$, and the radius of the asymptotic circular orbit is $r=4M$, with associated frequency $\Omega:=d\varphi/dt=(8M)^{-1}$. [Here, and throughout this paper, we use units in which $G=c=1$, and $(t,r,\theta,\varphi)$ are standard Schwarzschild coordinates.] We ask how these values change under the effect of the conservative piece of the first-order self-force (dissipation ignored), insisting that the orbit still starts at rest at infinity and that at late time it asymptotically approaches some circular orbit. Our numerical computation in this paper gives 
%\bl{[Sis: Update 2019/08/21]}
%\red{[MUST Fix notation problem $\hat L\leftrightarrow {\cal L}$]}
\begin{align}
\label{Omega_final}
{\hat \Omega} &= (8M)^{-1}\left[1+ 0.5536(2)\eta\right], \\
\label{L_final}
{\hat L} &= 4M\mu\left[1- 0.304(3)\eta\right].
\end{align}
Here overhats indicate values as corrected by the self-force, and parenthetical figures show the estimated magnitude of the error bar on the last displayed decimal(s). 

We note that the above definition of our ``self-force-perturbed'' ZEZO is unambiguous, since it alludes only to invariant (asymptotic) symmetries of spacetime : flat-space symmetries at $t\to -\infty$ and helical symmetry at $t\to +\infty$ (and, as we describe in Sec.\ \ref{sec:MBMSsf}, the ``conservative piece'' of the first-order self-force is also defined unambiguously). Thus, our computed $O(\eta)$ corrections to $\Omega$ and $L$ serve as unambiguous, ``gauge invariant'' (in a sense to be made more precise later) diagnostics of the post-geodesic conservative dynamics. Indeed, these quantities were among the useful invariants proposed by one of us already in \cite{damour} for establishing links between self-force theory and EOB (and PN).

The computation leading to Eqs.\ (\ref{Omega_final}) and (\ref{L_final}) requires one to integrate certain components of the self-force along the entire geodesic ZEZO coming from infinity (and also, for reasons explained in Sec.\ \ref{sec:MBMSsf}, along the time-reversed ZEZO going out to infinity). As we have mentioned, such a calculation of the self-force, along an unbound orbit, has not been attempted before and involves having to deal with several new technical difficulties. Most advanced self-force codes (such as van de Meent's \cite{vandeMeent:2017bcc}) rely on a discrete Fourier-harmonic decomposition of the perturbation equations, suitable for discrete-spectrum problems. Such codes cannot be easily adapted for handling a source on an unbound orbit, whose perturbation has a continuous spectrum. Recent initial work by Hopper \cite{Hopper:2017qus,Hopper:2017iyq} has demonstrated how the asymptotic flux of radiation from unbound orbits can be computed in a frequency-domain framework, but the extension and application of his method to a calculation of the local self-force is nontrivial and yet to be achieved. In this paper we choose to base our calculation on a time-domain method, whose utility and efficacy are essentially agnostic to whether the perturbation's frequency spectrum is discrete or continuous. 

Our time-domain method is based on a direct integration of the linearized Einstein's equations in the Lorenz gauge, and represents an extension of the method and code developed in Refs.\ \cite{baracklousto,bsago1,bsago2} where it was applied for bound (circular or eccentric) orbits. We list here a few of the technical issues that arise in extending the method to unbound orbits. First, and most obvious, our integration domain for the self-force becomes infinite (and subtle at $r\simeq 4M$), demanding the introduction and control of suitable integration cutoffs, and/or the use of suitable extrapolations. Second, ``junk radiation'' from imperfect initial conditions is potentially much more of a problem for an unbound orbit than it is for a bound one, both because such radiation takes longer to dissipate away and because its effect on the physical self-force data is harder to isolate and remove (in the case of a bound orbit, one can simply discard the perturbation produced by the first few orbital cycles, dominated by the junk radiation). 

A third technical hurdle turned out to be the hardest to deal with. The monopole and dipole modes of the Lorenz-gauge perturbation, which can have an important contribution to the self-force, cannot be obtained via the time-domain integration method of \cite{baracklousto,bsago1,bsago2}, due to the occurrence of certain spurious linear-in-time gauge modes that appear to grow during the numerical evolution of the field equations. This problem has been analysed in detail in Ref.\ \cite{Dolan}. A complete satisfactory resolution for it is not yet known despite recent progress \cite{JT_capra21}. In the case of bound orbits, the problem has been circumvented (at least in the Schwarzschild case) by constructing suitable monopole and dipole solutions analytically. This, however, is not easily done for unbound orbits, due to the nontrivial time dependence of their perturbations. In this work we propose and implement a method for dealing with this problem in the specific case of the ZEZO.

The $O(\eta)$ terms of the ZEZO's $\hat\Omega$ and $\hat L$ were first estimated within the
EOB formalism in Ref.\ \cite{damour}. Specifically, these self-force terms were shown to be 
precisely proportional to $a(\frac14)$ and  $a'(\frac14)$, respectively, where $a(u)$ denotes the self-force piece of the basic radial potential $A(u; \nu)=1-2 u+ \nu a(u)+ O(\nu^2)$ of EOB dynamics; see Eqs. \eqref{EOB1} below. [Here, $u:= (M+\mu)/r_{\rm EOB}$, while $\nu:= \mu M/(M+\mu)^2 =\eta/(1+\eta)^2$ denotes the symmetric mass ratio.]
At the time of Ref.\ \cite{damour}, the numerical values of 
$a(\frac14)$ and  $a'(\frac14)$ could only be coarsely estimated by using (third-order) PN theory, 
together with initial results from self-force theory \cite{Barack:2009ey}, 
and some early numerical-relativity calibration of EOB theory \cite{Damour:2009kr}. 
Expressed in the notation of the present paper, Ref.\ \cite{damour} predicted 
$
\hat L = 4M\mu\left[1- 0.288(80) \eta\right]\, ,
$
 and that the correction to $\Omega$ should be positive. No concrete value was ventured for $\Omega$, for which the estimate was less certain, but based on information given in \cite{damour}, one gets $\hat\Omega\simeq (8M)^{-1}(1+ 0.32\eta)$. The proximity to our ``exact'' self-force results (\ref{Omega_final}) and (\ref{L_final}), especially for $\hat L$, is notable.

An independent way of calculating $\hat\Omega$ and $\hat L$ is provided by the so-called {\it first law of binary black hole mechanics}
\cite{LeTiec:2011ab,Tiec:2015cxa,Blanchet:2017rcn}: a variational formula that links local quantities constructed from the metric perturbation evaluated on the orbit (specifically, Detweiler's redshift $z$) to certain global energy and angular momentum of the binary system.\footnote{We shall discuss in detail below the relation between the notions of energy and angular momentum in the first law (and in EOB dynamics), and the usual  Arnowitt--Deser--Misner (ADM), or Bondi, ones. The first law of binaries was originally formulated in a PN context. Later work \cite{Tiec:2013kua,isoyama14,Fujita:2016igj} established Hamiltonian formulations of the first law directly in the context of self-force theory.}
When applied to a circular orbit \cite{LeTiec:2011dp}, the formula gives the $O(\eta^2)$ contributions to the binding energy and angular momentum in terms of the $O(\eta)$ pieces of $z(\Omega)$ and $dz(\Omega)/d\Omega$, two functions that are known numerically with a very high precision \cite{Akcay}. The first-law formula does not apply directly to the ZEZO, but it does apply to the asymptotic circular orbit at $r\simeq 4M$. And since the ZEZO and the circular orbit to which it asymptotes necessarily possess the same energy and angular momentum, it follows that the first-law formula can be used to compute these for the ZEZO as well, in terms of the known values of $z(\Omega)$ and $dz(\Omega)/d\Omega$ on the circular geodesic orbit at $r=4M$. A simple manipulation, detailed in Sec.\ \ref{sec:1st}, also gives the asymptotic frequency of the ZEZO. Thus, the first law independently predicts the $O(\eta)$ terms of the ZEZO's $\hat\Omega$ and $\hat L$. We find these first-law predictions to be in agreement with our direct self-force results (\ref{Omega_final}) and (\ref{L_final}), to within our error bars. This serves to corroborate the evidence supporting the validity of the first law even in the strong-field regime.

The first law can also be used to provide a simple link (first obtained in \cite{Barausse:2011dq}) between the self-force piece of the redshift $z$ and the EOB potential $a(u)$. This link has been used in the past to compute accurate numerical values and analytical representations of $a(u)$ from numerical self-force computations \cite{Akcay}. The latter allow one to accurately compute $a(\frac14)$ and  $a'(\frac14)$ and thereby refine the EOB predictions for the self-force corrections to $\hat L$ and $\hat\Omega$. In Sec.\ \ref{sec:EOB} we shall go through this calculation and show how these EOB predictions are in full agreement with the direct self-force results  (\ref{Omega_final}) and (\ref{L_final}).

%*Alternatively (and, as we shall check, equivalently), the first law was shown \cite{Barausse:2011dq} to provide a simple link between the self-force correction to the function $z(x, \eta)$ (where $ x\equiv \left[(M+\mu)\Omega \right]^{\frac23}$) and the self-force correction $a(u)$ to the EOB radial potential. This link has been used in the past to compute accurate values
%and representations of $a(u)$ from numerical self-force computations \cite{Akcay}. The latter results allow
%one to accurately compute the two quantities entering the EOB predictions for the self-force corrections
%to  $\hat L$ and $\hat\Omega$, namely, as mentioned above,  $a(\frac14)$ and  $a'(\frac14)$. To complement our discussion using the self-force corrections to $z(x, \eta)$ (and its derivative), we shall
%show how the use of the accurate fits for the function $a(x)$ obtained in Ref. \cite{Akcay}
%directly leads to full agreement with the values, Eqs.\ (\ref{Omega_final}) and (\ref{L_final}),
%calculated in the present paper.*

The plan of this paper is as follows. We begin, in Sec.\ \ref{sec:MBMSgeo}, with a description of the ZEZO and its properties in the geodesic limit. In Sec.\ \ref{sec:MBMSsf} we define the conservative piece of the self-force, add it to the equation of motion, and describe the resulting effects on the ZEZO. In Secs.\ \ref{sub2sec:IBCO-shift} and \ref{sub2sec:IBCO-AMshift} we derive formulas for the self-force corrections to $\Omega$ and $L$ (respectively), written explicitly in terms of the self-force components (and certain worldline integrals thereof). In Sec.\ \ref{sec:Numeric-MBMS} we review our numerical method, describe the details of its implementation, and obtain the raw self-force data needed for our analysis. In Sec.\ \ref{results} we then calculate $\hat\Omega$ and $\hat L$ and arrive at our main results (\ref{Omega_final}) and (\ref{L_final}). Sections \ref{sec:EOB} and \ref{sec:1st} contain our comparisons with the theoretical
predictions made, respectively, from EOB theory, and directly from the first-law.
We conclude in Sec.\ \ref{sec:conclusion} with a discussion of foreseeable future applications.

%%%%%%%%%%%%%%%%%%%%%%%%%%%%%%%%%%%%%%%%%%%%%%%%%%%%%%%%%%%%%%%%%%%%%%%%%
\section{Zero binding energy zoom-whirl orbit in the geodesic approximation}
\label{sec:MBMSgeo}
%%%%%%%%%%%%%%%%%%%%%%%%%%%%%%%%%%%%%%%%%%%%%%%%%%%%%%%%%%%%%%%%%%%%

Consider a test particle of mass $\mu$ moving on a timelike geodesic orbit in the exterior of a Schwarzschild black hole of mass $M$. Denote the (Schwarzschild-)coordinate position along the orbit by $x^{\alpha}_{\mathrm p}(\tau)$, with tangent four-velocity $u^{\alpha} := {\dot x}^{\alpha}_{\mathrm p}$, where $\tau$ is proper time and an overdot denotes $d/d\tau$. Without loss of generality, we place the orbit in the equatorial plane, i.e.\ take $\theta_{\rm p}=\pi/2$ and $u^{\theta}=0$. The particle's energy,  $E := -\mu u_t$, and azimuthal angular momentum, $L := \mu u_\varphi$, are conserved along the geodesic, i.e.\ $\dot{E} = \dot{L} = 0$; here $u_{\alpha}=g_{\alpha\beta}u^{\beta}$, with $g_{\alpha\beta}$ being the background Schwarzschild metric.
The geodesic equations of motion can then be written in a first-integral form,
\begin{align}
\label{tEq}
\mu\dot{t}_{\mathrm p} &= \frac{E}{f(r_{\mathrm p})}\,,\\
\label{phiEq}
\mu\dot{\varphi}_{\mathrm p} &= \frac{L}{r_{\mathrm p}^2}\,, \\
\label{radEq}
\mu\dot{r}_{\mathrm p} &= \pm\left[E^2 - V(r_{\mathrm p};L)\right]^{1/2} ,
\end{align}  
where we have introduced $f(r) := (1 - 2M/r)$ 
and the radial effective potential 
$V(r; L) := f(r) (\mu^2 + {L^2}/{r^2})$. 
From Eq.\ (\ref{radEq}), the effective radial force acting on particle is given by
\begin{eqnarray}\label{geodacc}
\mu\ddot{r}_{\mathrm p} &=&
-\frac{1}{2\mu}
\frac{\partial V(r_{\mathrm p};L)}{\partial r_{\mathrm p}}
=\frac{L^2(r_{\rm p}-3M)-M\mu^2r_{\rm p}^2}{\mu r_{\rm p}^4}
\nonumber\\
&=:& {\cal F}^r(r_{\rm p};L) .
\end{eqnarray}

The ZEZO is defined by the requirements that (i) $\dot r_{\mathrm p}\to 0$ for $r_{\mathrm p}\to \infty$, and (ii) the trajectory asymptotes to a circular orbit in either the infinite future or the infinite past, i.e.\ $\dot r_{\mathrm p}=0=\ddot r_{\mathrm p}$ for some $r_{\rm p}=R$. The first requirement determines $E$ using Eq.\ (\ref{radEq}), and subsequently the second requirement determines $L$ and $R$ using (\ref{radEq}) and (\ref{geodacc}). One obtains
\begin{equation}
\label{MBMS-test}
E = \mu \,, \quad L = 4\mu M \,, \quad R=4 M \, .
\end{equation} 
Here we have adopted the convention that $L$ is positive, i.e.\ $\dot{\varphi}_{\rm p}>0$.
The parameters in (\ref{MBMS-test}) describe both (disjoint) branches of the ZEZO geodesic: the one going out to infinity [$+$ sign in Eq.\ (\ref{radEq})], as well as the one coming in from infinity [$-$ sign in Eq.\ (\ref{radEq})]; we shall refer to the former as the ``outbound ZEZO'' (oZEZO) and to the latter as the ``inbound ZEZO'' (iZEZO).

\begin{figure}[htb]
	\begin{center}
        \includegraphics[width=\columnwidth]{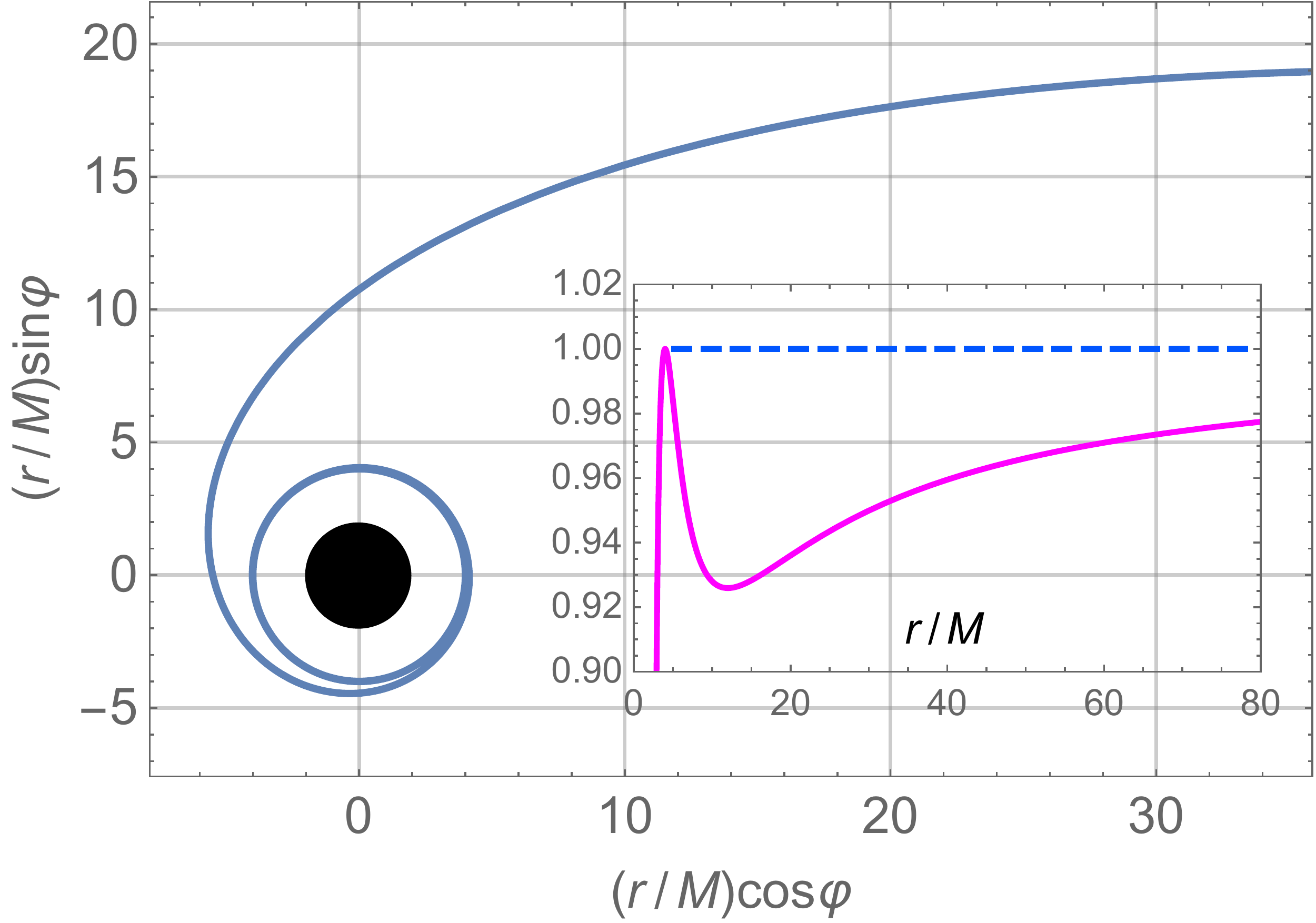}
\caption{The Zero-(binding-)energy zoom-whirl geodesic orbit (ZEZO), depicted here in the orbital plane. The inset shows the radial effective potential $V(r;4\mu M)$ [cf.\ Eq.\ (\ref{radEq})], with the dashed line representing the radial range of the ZEZO orbit. The orbit asymptotes to the innermost bound circular orbit (IBCO) at $r=4M$, corresponding to the maximum of the effective potential. }  
        \label{fig:ZEZO}
	\end{center}
\end{figure}    

The ZEZO geodesic is depicted in Fig.\ \ref{fig:ZEZO}.  The asymptotic circular geodesic of radius $r=4M$ corresponds to a maximum of the effective potential $V(r)$ (see the inset), and it is unstable: a small perturbation would send the particle flying either to infinity or into the black hole. This circular orbit is marginally bound in the following sense: For a timelike circular geodesic (stable or unstable) at any constant $r>4M$, an arbitrarily small perturbation cannot send the orbiting particle to infinity, while it can do so for any timelike circular geodesic with $r<4M$. Thus we refer to the circular geodesic orbit at $r=4M$ as the Innermost Bound Circular Orbit (IBCO). The IBCO's frequency is given by 
\begin{equation}\label{wIBCO-test}
\Omega =
\left.
\left(\dot{\varphi}_{\mathrm p}/\dot{t}_{\rm p}\right)\right|_{r_{\mathrm p} = 4M}
= \frac{1}{8 M} ,
\end{equation}
where we have made use of Eqs.\ (\ref{tEq}), (\ref{phiEq}) and (\ref{MBMS-test}).

The iZEZO and oZEZO asymptote the IBCO at $t\to\infty$ and $t\to -\infty$, respectively, and we note that both do so ``exponentially fast'' (in either coordinate time $t$ or proper time $\tau$): for $\delta r_{\rm p}:=r_{\rm p}-4M\ll M$, Eq.\ (\ref{radEq}) gives $\dot{\delta r}_{\rm p}\simeq \pm \kappa\delta r_{\rm p}$ [with $\kappa=(4\sqrt{2}M)^{-1}$], implying $\delta r_{\rm p}(\tau)\sim e^{\pm\kappa\tau}$. 

At the other end of the orbit, for $r_{\rm p}\to \infty$, the azimuthal angle $\varphi_{\rm p}$ approaches a constant limiting value: combining (\ref{phiEq}) and (\ref{radEq}) we find $d\varphi_{\rm p}/d r_{\rm p}\sim (M/r_{\rm p}^3)^{1/2}$, and hence 
$\varphi_{\rm p}\to  \varphi_{\infty}+O(M/r_{\rm p})^{1/2}$.
In that sense, the ZEZO is asymptotically ``radial'' at infinity.
Note, however, that the ``impact parameter'', defined as usual through 
\begin{equation}
b:=\lim_{r\to\infty} r \sin\left|\varphi_{\rm p}(r)-\varphi_\infty\right| ,
\end{equation}
is actually infinite for the ZEZO. This is unlike the case of unbound orbits with $E>\mu$, for which the impact parameter has a finite value, given by $b=L/\sqrt{E^2-\mu^2}$.

%%%%%%%%%%%%%%%%%%%%%%%%%%%%%%%%%%%%%%%%%%%%%%%%%%%%%%%%%%%%%%%%%%
%%%%%%%%%%%%%%%%%%%%%%%%%%%%%%%%%%%%%%%%%%%%%%%%%%%%%%%%%%%%%%%%%%
\section{Conservative self-force modification to the orbit}
\label{sec:MBMSsf}
%%%%%%%%%%%%%%%%%%%%%%%%%%%%%%%%%%%%%%%%%%%%%%%%%%%%%%%%%%%%%%%%%%
%%%%%%%%%%%%%%%%%%%%%%%%%%%%%%%%%%%%%%%%%%%%%%%%%%%%%%%%%%%%%%%%%%

%%%%%%%%%%%%%%%%%%%%%%%%%%%%%%%%%%%%%%%%%%%%%%%%%%%%%%%%%%
\subsection{Equation of motion}
%%%%%%%%%%%%%%%%%%%%%%%%%%%%%%%%%%%%%%%%%%%%%%%%%%%%%%%%%%%%%

When the first-order gravitational self-force is taken into account, the particle's equation of motion can be written in the form 
\begin{equation}\label{GSF-EOM}
\mu\, {\hat u}^{\beta} \nabla_\beta {\hat u}^\alpha = F^\alpha_{\rm self} ,
\end{equation}
where the covariant derivative $\nabla_\beta$ is the one compatible with the {\it background} (Schwarzschild)
metric $g_{\mu \nu}$,  and $F^\alpha_{\rm self}$ is the self-force. We let the self-accelerated (slightly nongeodesic) orbit in the background spacetime be represented by $x^\alpha=\hat x_{\rm p}^\alpha(\tau)$, with tangent four velocity ${\hat u}^{\alpha}:=d\hat x^\alpha/d\tau$ normalized with respect to the background metric: $g_{\alpha\beta} {\hat u}^\alpha {\hat u}^\beta=-1$. From symmetry, the orbit remains equatorial even under the effect of the self-force (in any gauge that respects the up-down symmetry of the setup), so we have $\hat\theta_{\rm p}\equiv \pi/2$ and $\hat u^{\theta}\equiv 0$. The other components of Eq.\ (\ref{GSF-EOM}) take the simple form 
\begin{align}
\label{EEq}
\dot{\hat E} &= -{F}^{\rm self}_{t} , \\
\label{LEq}
\dot{\hat L} &= {F}^{\rm self}_{\varphi} , \\
\label{rEq}
\mu\,\ddot{\hat r}_{\rm p} &=  {\cal F}^r(\hat r_{\rm p};\hat L)+ {F}_{\rm self}^r ,
\end{align}  
where we have defined ${\hat E}:=-{\hat u}_t$ and ${\hat L}:={\hat u}_\varphi$, 
and indices are lowered using the background metric $g_{\mu \nu}$. Recall ${\cal F}^r$ is the effective geodesic radial force, introduced in Eq.\ (\ref{geodacc}). Note ${\cal F}^r=O(\eta)$ while ${F}_{\rm self}^r=O(\eta^2)$, so $F_{\rm self}^r$ represents a small perturbation of the effective radial force. 
Note also that Eq.\ (\ref{radEq}) remains valid, subject to replacing all quantities with their hatted counterparts:
\begin{equation}\label{rdotEq}
\mu\dot{\hat r}_{\mathrm p} = \pm\left[\hat E^2 - V(\hat r_{\mathrm p};\hat L)\right]^{1/2} .
\end{equation}

The self-force can be written as a sum of conservative and dissipative components:
\begin{equation}
F^\alpha_{\rm self}=F^\alpha_{\rm cons}+F^\alpha_{\rm diss}.
\end{equation}
This split is unambiguously and uniquely defined for the first-order force, as follows. In self-force theory, the actual self-force can be expressed as a functional, $F^\alpha_{\rm self}(h_{\mu\nu}^{\rm ret})$, of the physical, retarded (first-order) metric perturbation. One can similarly construct a fictitious self-force $F^\alpha_{\rm self}(h_{\mu\nu}^{\rm adv})$ out of the {\it advanced} metric perturbation. Then 
\begin{align}
  \label{Fcons}
  F^{\alpha}_{\rm {cons}}&:=\frac{1}{2}\left[F^\alpha_{\rm self}(h_{\mu\nu}^{\rm ret})+F^\alpha_{\rm self}(h_{\mu\nu}^{\rm adv})\right] ,\\
  \label{Fdiss}
  F^{\alpha}_{\rm {diss}}&:=\frac{1}{2}\left[F^\alpha_{\rm self}(h_{\mu\nu}^{\rm ret})-F^\alpha_{\rm self}(h_{\mu\nu}^{\rm adv})\right]
  \end{align}
describe, respectively, the time-symmetric (``conservative'') and dissipative pieces of $F^\alpha_{\rm self}$.
Here we are interested in the effect of the conservative force alone, so in Eqs.\ (\ref{EEq})--(\ref{rEq}) we henceforth set $F^\alpha_{\rm diss}=0$, thus replacing the full force $F^\alpha_{\rm self}$ with $F^{\alpha}_{\rm {cons}}$. 

In the next subsection we solve Eqs.\ (\ref{EEq})--(\ref{rEq}) (with $F^\alpha_{\rm self}\to F^{\alpha}_{\rm {cons}}$) for the perturbed ZEZO. But let us first make two comments about the calculation of $F^{\alpha}_{\rm {cons}}$ in practice. First, since we work here in the first-order self-force approximation, with dissipation neglected, it should suffice to evaluate $F^{\alpha}_{\rm {cons}}$ along the {\it unperturbed}, geodesic ZEZO. This is based on the expectation (confirmed with our explicit calculation below) that the perturbed orbit remains forever ``close'' to the background geodesic ZEZO: $\hat x_{\rm p}^\alpha(\tau)-x_{\rm p}^\alpha(\tau)=O(\eta)$ for all $\tau$. In our numerical calculation we can therefore evaluate the self-force along the fixed ZEZO geodesic, and need not worry about the correction to the orbit due to the self-force.

Our second comment regards the extraction of $F^{\alpha}_{\rm {cons}}$ from the full self-force. From Eq.\ (\ref{Fcons}) it would seem that we require knowledge of both $h_{\mu\nu}^{\rm ret}$ and $h_{\mu\nu}^{\rm adv}$ (and their derivatives) along the orbit. However, there is a way to express $F^{\alpha}_{\rm {cons}}$ in terms of the retarded perturbation alone, taking advantage of the time-symmetry relation between the oZEZO and iZEZO. To see this, first observe that the two orbits are related via the transformation $(u^t,u^r,u^\varphi)\to (u^t,-u^r,u^\varphi)$, and further note the symmetry relation, valid at any given point along either orbit \cite{Mino:2003yg},
\begin{equation}
F^\alpha_{\rm adv}(u^t,u^r,u^\varphi)= q^{(\alpha)} F^\alpha_{\rm ret}(u^t,-u^r,u^\varphi), 
\end{equation}
with $q^t=-1=q^\varphi$ and $q^r=1$ (no summation over $\alpha$), and where $F^\alpha_{\rm ret/adv}\equiv F^\alpha_{\rm self}(h_{\mu\nu}^{\rm ret/adv})$. Now consider a point with four velocity $u^\mu$ along the iZEZO. The conservative self-force at that point is given by Eq.\ (\ref{Fcons}), which, using the above two symmetry relations, gives
\begin{multline}\label{Fconsformula}
 F_{\rm cons}^\alpha(u^\mu)\bigg\rvert_{\rm iZEZO} =
\\ 
 \frac{1}{2}\left(F^\alpha_{\rm ret}(u^t,u^r,u^\varphi)
+q^{(\alpha)} F^\alpha_{\rm ret}(u^t,-u^r,u^\varphi)\right)\bigg\rvert_{\rm iZEZO}=
\\
\frac{1}{2}\left(F^\alpha_{\rm ret}(u^\mu)\bigg\rvert_{\rm iZEZO}
+ q^{(\alpha)} F^\alpha_{\rm ret}(u^\mu)\bigg\rvert_{\rm oZEZO}\right).
\end{multline}
We can thus construct the conservative self-force along the iZEZO given the full (retarded) self-force along both iZEZO and oZEZO. This turns out to be computationally simpler than a calculation of both retarded and advanced perturbations for the iZEZO alone.

We finally note that our quantities ${\hat E}$ and ${\hat L}$ are {\em not} conserved along the ZEZO, even when dissipation is ignored. The conservative self-force components $F_{t}^{\rm cons}$ and $F_{\varphi}^{\rm cons}$ in Eqs.\ (\ref{EEq}) and (\ref{LEq}) are generally nonzero along the ZEZO; they only vanish on the asymptotic IBCO (as they do, from time-symmetry, along any circular orbit).

%%%%%%%%%%%%%%%%%%%%%%%%%%%%%%%%%%%%%%%%%%%%%%%%%%%%%%%%%%
\subsection{The perturbed ZEZO}\label{Sec:pertZEZO}
%%%%%%%%%%%%%%%%%%%%%%%%%%%%%%%%%%%%%%%%%%%%%%%%%%%%%%%%%%%%%

We define the ``perturbed ZEZO'' as a solution of Eqs.\ (\ref{EEq})--(\ref{rEq}) (with $F^\alpha_{\rm self}\to F^{\alpha}_{\rm {cons}}$), subject to 
\begin{align}
\label{pertZEZO1} \dot{\hat r}_{\rm p}(r\to\infty)= 0 , \\
\label{pertZEZO2} \dot{\hat r}_{\rm p}(r\to \hat R)= 0 , \\
\label{pertZEZO3} \ddot{\hat r}_{\rm p}(r\to \hat R)= 0 ,
\end{align}
for some constant radius $\hat R=4M+O(\eta)$. Our construction below shows that these three conditions pick out a unique solution that is a perturbation of the geodesic ZEZO. However, since $\hat r_{\rm p}$ is gauge dependent (just like the self-force itself), we need to be mindful of gauge-related ambiguities in the above definition. One way to remove such ambiguities is to reformulate the conditions (\ref{pertZEZO1})--(\ref{pertZEZO3}) in a geometrical language, alluding to invariant (asymptotic) Killing symmetries of the perturbed spacetime. Thus (referring to the iZEZO, for example), we can demand that at past timelike infinity ($i^-$) the perturbed sapcetime possesses 
a time-translation symmetry with a (normalized) generator $t^\alpha$ coinciding with the 
particle's four-velocity $\hat u^\alpha$; and that at future timelike infinity ($i^+$) the perturbed spacetime has an asymptotic helical symmetry, with $\hat u^\alpha$ lying tangent to a generator of it. In Appendix \ref{app:ADM} we will discuss an alternative definition of the perturbed ZEZO, in which the condition (\ref{pertZEZO1}) is replaced with a condition on the spacetime's ADM mass: instead of fixing the velocity at infinity, we fix the ADM mass at $M+\mu$ through $(\eta^2)$. This manifestly invariant way of fixing the initial conditions should prove convenient in future studies of hyperbolic-type orbits.

In practice, however, we need to translate such invariant conditions into a coordinate form such as in (\ref{pertZEZO1})--(\ref{pertZEZO3}), and to do so without ambiguity we must restrict the class of gauges in which these coordinate conditions apply. As we discuss in Sec.\ \ref{subsec:gauge} below, for our specific calculation it will suffice to require that the metric perturbation associated with the self-force is manifestly asymptotically flat, as well as helically symmetric at late time. The two physical self-force effects that we calculate in this work---the IBCO frequency shift and the shift in the critical value of the angular momentum at infinity---will be invariant within this class of gauges.

%restricting the class of gauges in which the metric perturbation and self-force are expressed. Specifically, we require that (1) the perturbed metric is manifestly asymptotically flat (this condition will be discussed in Sec.\ \ref{subsec:gauge} below, with a relevant counterexample shown); (2) At past timelike infinity the perturbed metric should admit a manifest timelike killing vector field $t^\alpha$ so that $\hat u^\alpha(r\to\infty)=t^\alpha $

%We remove such ambiguities by demanding that the gauge in which the self-force is expressed is manifestly ``asymptotically flat'' (a counterexample will be discussed in Sec.\ \ref{subsec:gauge} below), as well as helically symmetric near timelike infinity. Our perturbed-orbit solution $\hat x^\alpha_{\rm}(\tau)$ will still, of course, depend on the exact gauge of $F^{\alpha}_{\rm cons}$; however, the two physical self-force effects to be extracted from the solution---the IBCO frequency shift and the shift in the critical value of the angular momentum at infinity---will be invariant within this class of gauges.\footnote{In Appendix \ref{app:ADM} we will discuss an alternative definition of the perturbed ZEZO, in which the condition (\ref{pertZEZO1}) is replaced with a manifestly gauge-invariant condition on the system's ADM mass. The two definitions are equivalent under the above restrictions on the class of gauges. } 

We now look for a solution of Eqs.\ (\ref{EEq})--(\ref{rEq}) that is a perturbation of the geodesic ZEZO. We thus write 
\begin{align}\label{MBMS-GSF}
\hat{E} &= \mu + \delta E(r_{\rm p}), \\
\label{Lpert}
\hat{L} &= 4M\mu + \delta L(r_{\rm p}), \\
\label{Rpert}
\hat{R} &= 4M + \delta R ,
\end{align}
and consider the linearization of the equations of motion (\ref{EEq})--(\ref{rEq}) and normalization condition (\ref{rdotEq}) in the perturbations $\delta E(\propto \eta^2)$, $\delta L(\propto\eta^2)$ and $\delta R(\propto\eta)$. [We hereafter use $r_{\rm p}$ in lieu of $\tau$ as a parameter along the orbit, assuming $r_{\rm p}(\tau)$ is monotonic, on either the iZEZO or the oZEZO, even for the perturbed orbit.] Applying the three conditions (\ref{pertZEZO1})--(\ref{pertZEZO3}) then yields, respectively,
\begin{align}
\label{norm-inf}
\delta E(\infty)&= 0 , \\
\label{norm-IBCO}
\delta E(\hat R)  &=(8M)^{-1} \delta L(\hat R) ,\\
\label{ddot-IBCO}
\mu\,\delta R + \delta L(\hat R) &= -32M^2 F^r_{\rm cons}(\hat R) ,
\end{align}
where we have used
$(\partial V/\partial r)\big\vert_{r\to\infty}=0=(\partial V/\partial L)\big\vert_{r\to\infty}$, 
and also
$\partial V/\partial r=0$ for $(r,L)=(4M,4M\mu)$.
Within our linear approximation we may replace $\hat R\to 4M$ in the argument of all perturbative quantities in (\ref{norm-IBCO}) and (\ref{ddot-IBCO}).

Two more relations are obtained by integrating the self-force in Eqs. (\ref{EEq}) and (\ref{LEq}) along the geodesic iZEZO:
\begin{align}
\label{E4}
\delta E(\hat R) - \delta E(\infty)
&= -\int_{\infty}^{4M} \!\!\!\! {F}^{\rm cons}_{t} \,
\frac{d r_{\mathrm p}}{\dot{r}_{\mathrm p}}
=: {\Delta E}\,, \\
\label{E5}
\delta L(\hat R) - \delta L(\infty)
&= \int_{\infty}^{4M} \!\!\!\!{F}^{\rm cons}_{\varphi} \,
\frac{d r_{\mathrm p}}{\dot{r}_{\mathrm p}} 
=: {\Delta L}\,.
\end{align}
The five equations~\eqref{norm-inf}--\eqref{E5} form a closed algebraic system
for the five unknowns
$\delta E(\hat R)$, $\delta E(\infty)$, $\delta L(\hat R)$, $\delta L(\infty)$
and $\delta R$. Solving it, we find
\begin{eqnarray}
\label{deltaR}
\mu\, \delta R &=& -8 M{\Delta E} - 32 M^2 {F}^r_{\rm cons}(\hat R) , \\
\label{deltaLinf}
\delta L(\infty) &=& 8 M{\Delta E}-{\Delta L} , \\
\label{deltaER}
\delta E(\hat R) &=& {\Delta E} , \\
\label{deltaLR}   
\delta L(\hat R) &=& 8M{\Delta E} , 
\end{eqnarray}  
along with $\delta E(\infty)= 0$.
These expressions provide sufficient input for our calculation of invariant physical effects in the next two sections. 

But before we proceed to doing that, let us inspect the type of self-force input needed. It involves three bits of information: the value  ${F}^r_{\rm cons}(\hat R)$, and the two integrals ${\Delta E}$ and ${\Delta L}$. The quantity ${F}^r_{\rm cons}(\hat R)$ is the (constant) value of the $r$ component of the conservative self-force on the asymptotic IBCO. Within our first-order self-force approximation, this can equally well be evaluated on the geodesic IBCO at $r=4M$. The numerical value of ${F}^r_{\rm cons}$ on the IBCO can be obtained with great precision using standard (bound-orbit) self-force codes; we give this value below in Eq.\ (\ref{Fr}). 

The evaluation of the self-force integrals ${\Delta E}$ and ${\Delta L}$ is more involved and will be described in Sec.\ \ref{sec:Numeric-MBMS}. Here we comment on the expected convergence of these integrals. At $r_{\rm p}\gg M$ we expect the asymptotic form
%\footnote{This is based on assuming that, in the limit $r_{\rm p}\to\infty$, the conservative self-force is simply ($\mu\times$) the particle's acceleration, projected orthogonally to the four-velocity $u^\alpha$, as calculated with flat-space connections: $F^\alpha_{\rm cons}\simeq (\delta_{\beta}^{\alpha}+u^\alpha u_\beta)u^\gamma\nabla^{\rm flat}_{\gamma}u^\beta$. }
%$\dot{\Phi}\sim \dot{r}_{\rm p}/r_{\rm p}^2 \sim r_{\rm p}^{-5/2}$, 
%and $F_\varphi^{\rm cons}\sim r^2 \dot{a}^\varphi\sim r_{\rm p}^2 (\dot{r}_{\rm p}/r_{\rm p}^4)\sim %r_{\rm p}^{-5/2}$, where $\Phi \sim 1/ r_{\rm p}$ is the Newtonian gravitational potential at the particle's location, and $a^\varphi\sim ({\boldsymbol{\nabla}}\Phi)^\varphi\sim r_{\rm p}^{-3}$ is the $\varphi$ component of the Newtonian acceleration.} 
\begin{equation}\label{Fasympt}
F_t^{\rm  cons}\propto \dot{r}_{\rm p}/r_{\rm p}^{2},\quad\quad 
F_\varphi^{\rm  cons} \propto \dot{r}_{\rm p} / r_{\rm p}^{2}
\end{equation}
(see Appendix \ref{App:Newtonian}), where $\dot{r}_{\rm p}\simeq -(2M/r_{\rm p})^{1/2}$. %(this will be confirmed numerically in Sec.\ \ref{sec:Numeric-MBMS}). 
%\begin{equation}\label{Fasympt}
%F_t^{\rm  cons}\simeq -\frac{\sqrt{2}\,\eta^2}{(r_{\rm p}/M)^{5/2}} ,
%\quad\quad
%%F_\varphi^{\rm  cons}\simeq - \frac{4\sqrt{2}M\,\eta^2}{(r_{\rm p}/M)^{5/2}} ,
%F_\varphi^{\rm  cons}\simeq - \frac{(?)\times M\,\eta^2}{(r_{\rm p}/M)^{5/2}} ,
%\end{equation}
%which will be confirmed numerically in Sec.\ \ref{sec:Numeric-MBMS}. 
Hence, the integrands in Eqs.\ (\ref{E4}) and (\ref{E5}) fall off as $\sim r_{\rm p}^{-2}$, and both integrals converge well at infinity. Truncating the integration at some $r_{\rm max}\gg M$ should produce an error of $O(1/r_{\rm max})$, which could be reduced to $O(1/r_{\rm max}^2)$ using a Richardson-type extrapolation.  Near the IBCO, for $r_{\rm p}-4M\ll M$, we have \cite{CB} $F_t^{\rm cons}\sim \dot{r}_{\rm p}\tilde F_t(r_{\rm p})$, where $\tilde F_t$ is some smooth function of $r_{\rm p}$ with a generally nonzero limit $r\to 4M$, and similarly for $F_\varphi^{\rm cons}$.  The integrals therefore converge well also at their $r_{\rm p}\to 4M$ limit. Truncating at  $r_{\rm p}=4M+\epsilon$ should produce an error of $O(\epsilon)$.

%%%%%%%%%%%%%%%%%%%%%%%%%%%%%%%%%%%%%%%%%%%%%%%%%%%%%%%%%%%%%%%%
\section{Self-force correction to the IBCO frequency }
\label{sub2sec:IBCO-shift}
%%%%%%%%%%%%%%%%%%%%%%%%%%%%%%%%%%%%%%%%%%%%%%%%%%%%%%%%%%%%%%%%%
%For convenience, we set $M=1$ in what follows (this implies, in particular, $\mu=\eta$). 

The quantity $\delta R$ describes the shift in the coordinate radius of the IBCO due to the conservative piece of the self-force. It is by itself not a very useful measure of the self-force effect, because it is gauge dependent. A more useful measure is the associated shift in the IBCO frequency $\Omega$,
which {\it is} invariant, at least within a class of physically reasonable gauges (to be defined below).
The perturbed IBCO frequency is defined through $\hat\Omega:= (\hat u^\varphi / \hat u^t)\big\vert_{r_{\rm p}=\hat R}$, and we write it as
\begin{equation}
\hat\Omega:= \Omega + \delta\Omega ,
\end{equation}
where $\Omega=(8M)^{-1}$ is the geodesic IBCO frequency from Eq.\ (\ref{wIBCO-test}).
Our goal now is to derive an expression for the $O(\eta)$ self-force correction $\delta\Omega$.
 
Recalling $\hat u^\varphi=g^{\varphi\varphi}\hat L$ and $\hat u^t=-g^{tt}\hat E$ (where the background metric is evaluated on the perturbed orbit),  we have
%All the GSF quantities appearing in Eqs. ~\eqref{E1} -- \eqref{E5} 
%are inherently gauge-dependent and are of little use in cross-comparisons 
%with other frameworks to study binary dynamics.
%Instead, as suggested in Refs.~\cite{damour,CB}, 
%we obtain gauge-invariant information about GSF-perturbed MBMS orbits 
%by looking at the orbital frequency of the GSF-perturbed IBCO, 
%as described in what follows. 
\begin{equation}
\hat\Omega = 
\frac{1}{\hat R^2}\left(1-\frac{2M}{\hat R}\right)\frac{\hat L(\hat R)}{\hat E(\hat R)},
\end{equation}
which, upon substituting from Eqs.\ (\ref{MBMS-GSF})--(\ref{Rpert}), expanding in $\eta$, and dropping all terms beyond $O(\eta)$, gives 
\begin{equation}
\hat\Omega/\Omega =1+  \frac{1}{4M\mu}\left[\delta L(\hat R) -4M\delta E(\hat R) -\mu\, \delta R\right].
\end{equation}
Then substituting from Eqs.\ (\ref{deltaR}), (\ref{deltaER}) and (\ref{deltaLR}), we arrive at
\begin{equation}
\label{omegaSF}
\hat\Omega/\Omega=
1+ 3\eta \widetilde{{\Delta E}} +8\eta \widetilde{F}^r_{\rm ibco}  .
\end{equation}
Here we have made explicit the $\eta$ scaling of the self-force terms, by introducing the mass-rescaled dimensionless quantities 
\begin{equation}\label{tildeDeltaE}
\widetilde{\Delta E} := (M/\mu^2){\Delta E},
\quad\quad
\widetilde{F}^r_{\rm ibco} := \eta^{-2}{F}^r_{\rm cons}(\hat R) .
\end{equation} 
(For future use, we also introduce $\widetilde{\Delta L} :={\Delta L}/\mu^2$.)
The sum of the last two terms on the right-hand side in Eq.\ (\ref{omegaSF}) is the $O(\eta)$ relative frequency shift $\delta\Omega/\Omega$ of the IBCO. Notice it involves ${F}^r_{\rm ibco}$ and ${\Delta E}$, but not ${\Delta L}$. 

%%%%%%%%%%%%%%%%%%%%%%%%%%%%%%%%%%%%%%%%%%%%%%%%%%
\subsection{Conditions for gauge invariance}
%%%%%%%%%%%%%%%%%%%%%%%%%%%%%%%%%%%%%%%%%%%%%%%%%%%%

The frequency shift $\delta\Omega$ is what Ref.\ \cite{Barack:2018yvs} refers to as a ``quasi-invariant'' quantity (see Sec.\ 7.6 of that review for definition and a discussion): it is invariant within a class of ``physically reasonable'' gauges. We can identify the relevant gauge conditions by examining what effect a generic gauge transformation has on the form of Eq.\ (\ref{omegaSF}), as follows.

Consider a gauge transformation 
\begin{equation}\label{gauge transformation}
x^{\alpha}\to x^\alpha - \xi^\alpha
\end{equation}
with a generator $\xi^\alpha=O(\eta)$, and let $\delta_\xi X$ denote the change in a quantity $X$ under such a transformation. To evaluate  $\delta_\xi(\hat\Omega/\Omega)$, it is convenient to first write (\ref{omegaSF}) in the equivalent form 
\begin{equation}
\label{omegaSFalt}
\hat\Omega/\Omega=
1 - \frac{3}{8M}\delta R - 4\eta \widetilde{F}^r_{\rm ibco}  ,
\end{equation}
obtained using (\ref{deltaR}). The radial coordinate shift $\delta R$ transforms according to, simply, $\delta_\xi(\delta R) =-\xi^r$, where hereafter in this discussion $\xi^\alpha$ should be understood to be evaluated on the unperturbed IBCO. The transformation of
%we need $\delta_\xi(\widetilde{\Delta E})$ and $\delta_\xi \widetilde{F}^r_{\rm ibco}$. Starting with the former, we recall ${\Delta E}=\delta E(\hat R)-\delta E(\infty)=\hat E(\hat R)-\hat E(\infty)$ [using (\ref{MBMS-GSF})]. The term $\hat E(\hat R) =-\mu\hat u_t(\hat R)=\mu(1-2M/\hat R)\hat u^t(\hat R)$ transforms according to
%\begin{equation}
%\delta_\xi \hat E(\hat R) = 
%%-\frac{2M\mu\xi^r}{R^2\sqrt{1-3M/R}} - \mu(1-2M/R)\dot\xi^t
%%\nonumber\\
%%&=& 
%-\frac{\mu }{4M}\, \xi^r\Big\vert_{\rm ibco} - \frac{\mu}{2}\, \dot\xi^t\Big\vert_{\rm ibco} ,
%\end{equation}
%where we have used $\delta_\xi \hat R=-\xi^r$ and $\delta_\xi \hat u^t=-\dot\xi^t$ (evaluated on the IBCO), and, keeping to leading order in $\eta$, have evaluated the entire expression on the unperturbed IBCO, replacing $\hat R\to R=4M$ and recalling $u^t(R)=(1-3M/R)^{-1/2}=2$. Similarly,
%\begin{equation}
%\delta_\xi \hat E(\infty) = 
%-\mu\, \dot\xi^t(\infty),
%\end{equation}
%so, altogether,
%\begin{equation}
%\delta_\xi (\eta\widetilde{\Delta E}) = 
%-\frac{1}{4M}\, \xi^r  - \frac{1}{2}\, \dot\xi^t + \dot\xi^t(\infty).
%\end{equation}
$\delta_\xi \widetilde{F}^r_{\rm ibco}$ can be obtained from the standard formula for the gauge transformation of the self-force \cite{Barack:2001ph}, which, 
%\begin{equation}
%\delta_\xi F_{\rm self}^\alpha = -\mu\left[(g^{\alpha\lambda}+u^\alpha u^\lambda)\frac{D^2 \xi_\lambda}{d\tau^2} +R^\alpha_{\,\, \mu\lambda\nu}u^\mu \xi^\lambda u^\nu  \right],
%\end{equation}
%where $D/d\tau = u^\mu\nabla_\mu$ is a covariant derivative along the orbit, and $R^\alpha_{\,\, \mu\lambda\nu}$ is the Riemann tensor of the Schwarzschild background. 
applied to the geodesic IBCO, gives 
\begin{equation}\label{gaugeFr}
\delta_\xi (\eta \widetilde{F}^r_{\rm ibco})= 
\frac{3}{32M}\xi^r  - M\ddot\xi^r - \frac{1}{8}\dot\xi^t + M\dot\xi^\varphi .
\end{equation}
Combining the two results, we find 
\begin{equation}\label{gaugeOmega}
\delta_{\xi}\hat\Omega = \frac{1}{2}\left(\ddot\xi^r + \Omega\dot\xi^t - \dot\xi^\varphi\right) ,
\end{equation}
where, importantly, the two terms $\propto\xi^r$ got cancelled out, with all remaining terms being proportional to derivatives of $\xi^\alpha$ along the orbit. 

Equation (\ref{gaugeOmega}) makes it clear that the frequency is not a true invariant: it is sensitive to diffeomorphisms that induce radial acceleration $(\ddot\xi^r\ne 0)$, or are otherwise incompatible with the helical symmetry of the circular-orbit configuration $(\Omega\dot\xi^t \ne \dot\xi^\varphi)$. However, it is also clear that some restrictions are necessary on the class of allowable gauges, if we wish $\hat\Omega$ to make physical sense. For example, we wish $\hat\Omega$ to have a constant value along a circular orbit.   A natural requirement is for the metric perturbation $h_{\alpha\beta}$ to be manifestly helically symmetric, so that, in particular, $\dot h_{\alpha\beta}\equiv 0$ on the circular orbit. [For the iZEZO (oZEZO), this is replaced with a requirement that $h_{\alpha\beta}$ is helically symmetric in the vicinity of $i^+$ ($i^-$).] Can we say that $\hat\Omega$ is invariant under transformations $\xi^\alpha$ that preserve the helical symmetry of the metric perturbation? 
%Is $\delta\Omega$ invariant within the class of (asymptotically) helically symmetric perturbations?
%Gauge transformations with nonzero right-hand side in Eq.\ (\ref{gaugeOmega}) do not preserve $\delta\Omega$. This is hardly surprising: we do not expect transformations that do not respect the helical symmetry of the circular-orbit configuration to keep the frequency (as we have defined it) unaltered. In the case of a circular orbit, for our definition of frequency to even make physical sense, we must restrict to the class of gauges in which the perturbed metric $g_{\alpha\beta}+h_{\alpha\beta}$, and hence also the perturbation $h_{\alpha\beta}$ itself are manifestly helically symmetric: $\dot h_{\alpha\beta}=0$. For the iZEZO (oZEZO), where helical symmetry is only asymptotic, we require helical symmetry only in the vicinity of future (past) timelike infinity. 
%Is $\delta\Omega$ invariant within the class of (asymptotically) helically symmetric perturbations? In other words, is the requirement that the gauge perturbation $\delta_\xi h_{\alpha\beta}$ be helically symmetric {\em sufficient} to exclude all transformations $\xi^\alpha$ that give a nonzero $\delta_{\xi}(\delta\Omega/\Omega)$ in Eq.\ (\ref{gaugeOmega})?
It turns out that the answer is negative: it is not hard to find generators $\xi^\alpha$ that produce helically symmetric gauge perturbations while changing the value of $\delta\Omega$. All such generators have the form 
\begin{equation}\label{Xi_general}
\xi^\alpha = (\alpha_1 t+\alpha_2 \varphi)\delta^\alpha_t +(\alpha_3 t+\alpha_4 \varphi)\delta^\alpha_\varphi =: \Xi^\alpha ,
\end{equation}
where $\alpha_n$ are constants ($\propto\eta$). It can be checked that $\delta_{\Xi}h_{\alpha\beta}$ is helically symmetric, while, from Eq.\ (\ref{gaugeOmega}), we find a generally nonzero frequency correction
\begin{equation}\label{deltaXideltaOmega}
\delta_{\Xi}(\delta\Omega)=\Omega(\alpha_1+\Omega\alpha_2)-(\alpha_3+\Omega\alpha_4).
\end{equation}
(Here we have substituted the IBCO value $\dot t_{\rm p}=2$.)

Is there a physical reason to reject gauge transformations of the form (\ref{Xi_general})? The answer comes from examining the form of the metric perturbation generated by $\Xi^\alpha$, whose nonzero components work out to be
\begin{eqnarray}\label{deltaXh_tt}
\delta_\Xi h_{tt} &=&-2\alpha_1 (1-2M/r) \\
\label{deltaXh_tphi}
\delta_\Xi h_{t\varphi} =\delta_\xi h_{\varphi t}&=&  \alpha_3 r^2\sin^2\theta - \alpha_2 (1-2M/r) ,\\
\label{deltaXh_phiphi}
\delta_\Xi h_{\varphi\varphi} &=& 2 \alpha_4 r^2 \sin^2\theta .
\end{eqnarray}
Such a gauge perturbation is pathological at $r\to\infty$, where asymptotic flatness requires that $h_{tt}$, $h_{t\varphi}$ and $h_{\varphi\varphi}/r^2$ all vanish. In fact, the perturbation is pathological for any choice of $\alpha_n$, except $\alpha_n=0$ for all $n$. Thus, restricting to a class of manifestly asymptotically flat gauges excludes all $\Xi^\alpha$ transformations. And, since $\Xi^\alpha$ are the only transformations among helically symmetric perturbations that can change $\delta\Omega$, we find that imposing both helical symmetry and asymptotic flatness is {\em sufficient} for $\delta_\xi(\delta\Omega)=0$.

In other words, $\delta\Omega$ is invariant within the class of gauges in which the perturbed metric is both manifestly helically symmetric and manifestly asymptotically flat. For convenience, we hereafter take the point of view that this invariant value {\it defines} the frequency shift $\delta\Omega$ (instead of considering $\delta\Omega$ as a gauge-dependent quantity). If one chooses to work in a gauge that is not helically symmetric or asymptotically flat, one can still (in principle) calculate $\delta\Omega$, using a suitable gauge-adjusted version of Eq.\ (\ref{omegaSF})

%However, such transformations are excluded if one restricts attention to the class of gauges in which the metric perturbation is manifestly helically symmetric at asymptotically late time (for the iZEZO) or early time (for the oZEZO). The generators $\xi^\alpha$ of gauge transformations within this class satisfy $\dot{\xi^\alpha}=0=\ddot\xi^\alpha$ WRONG and they thus leave $\delta\Omega$ unchanged. If we apply Eq.\ (\ref{omegaSF}) in two different helically-symmetric gauges, we would generally fail to find an agreement on the individual values of $\tilde{{\cal E}}$ and $\tilde{F}^r_{\rm ibco}$, but their particular combination in that equation should return the same value, enabling comparison.  

%%%%%%%%%%%%%%%%%%%%%%%%%%%%%%%%%%%%%%%%%%%%%%%%%%%%%%%%
\subsection{Lorenz-gauge adjustment}
\label{subsec:gauge} 
%%%%%%%%%%%%%%%%%%%%%%%%%%%%%%%%%%%%%%%%%%%%%%%%%%%%%%%%%

Our numerical calculation of the metric perturbation and self-force in this work will be done in a Lorenz gauge. Subject to (retarded) boundary conditions, the Lorenz-gauge perturbation is determined uniquely, up to (1) mass and angular-momentum perturbations of the background Schwarzschild geometry, (2) a gauge displacement of the center-of-mass (CoM) location, and (3) certain monopolar and dipolar gauge modes that are linear in time $t$. The first type of ambiguity is resolved through conditions on the mass and angular momentum of the large black hole and of the entire spactime, as we discuss in Sec.\ \ref{subsec:low-ell}. The CoM ambiguity is discussed and resolved in Sec.\ \ref{CoM} via a condition on the mass-dipole content of the perturbation. Finally, the linear-in-$t$ modes are excluded using (in essence) a regularity condition at $i^\pm$, as we shall discuss at length in Sec.\ \ref{sec:Numeric-MBMS}.
With these specifications, the Lorenz-gauge perturbation and associated self-force are fully determined. 

% [specified through $O(\eta)$]. In practice, these can be conveniently imposed using the Abbott-Deser conserved integrals \cite{AbbDes} (see also \cite{Dolan}, which first introduced this method in the current context), applied on the (unperturbed) horizon and at spatial infinity ($i^0$). One typically requires that the large Schwarzschild black hole has a mass $M+O(\eta^2)$ and zero angular momentum, and that the full spacetime has a mass $M+E+O(\eta^2)$ and angular momentum $L+O(\eta^2)$.
%The latter modes can be excluded, in principle, from a regularity condition at $i^\pm$, as we shall discuss at length in Sec.\ \ref{sec:Numeric-MBMS}. The first type of perturbations, in turn, are fixed by conditions on the mass and angular momentum of the large black hole and of the entire spactime [specified through $O(\eta)$]. In practice, these may be conveniently imposed using the Abbott-Deser conserved integrals \cite{AbbDes} (see also \cite{Dolan}, which first introduced this method in the current context), applied on the (unperturbed) horizon and at spatial infinity ($i^0$). One typically requires that the large Schwarzschild black hole has a mass $M+O(\eta^2)$ and zero angular momentum, and that the full spacetime has a mass $M+E+O(\eta^2)$ and angular momentum $L+O(\eta^2)$. With these specifications, the Lorenz-gauge perturbation and associated self-force are fully determined. 

The thus-specified Lorenz-gauge perturbation is manifestly helically symmetric, but, as first noted in \cite{baracklousto}, it is not manifestly asymptotically flat.\footnote{There is a way, first suggested in \cite{Berndtson:2009hp}, to specify a manifestly asymptotically flat Lorenz-gauge perturbation. However, this comes at the expense of having to shift the black hole's mass away from $M$.} Specifically, one finds 
   \begin{align}\label{alpha}
   \lim_{r\rightarrow\infty}h_{tt}^{(L)}=\alpha,
   \end{align}
with a generally nonzero constant $\alpha$ whose value depends on the sourcing orbit, and where a script $(L)$ hereafter labels quantities expressed in the above specific Lorenz-gauge. Other components of $h_{\alpha\beta}^{(L)}$ do show the appropriate fall-off; the anomalous behavior only affects the monopolar piece of the $tt$ component. For a circular geodesic source, the monopole piece of the perturbation can be written down analytically, and $\alpha$ works out as 
$- 2\mu[R(R-3 M)]^{-1/2}$, where $R$ is the orbital radius. For the IBCO, with $R=4M$, we thus have $\alpha=-\eta$, namely 
  \begin{align} \label{htt_asymp}
   \lim_{r\rightarrow\infty}h_{tt}^{(L)}=-\eta \quad \text{(IBCO)}.
   \end{align}
 As we shall check numerically in Sec.\ \ref{sec:Numeric-MBMS}, this is also the value obtained for the iZEZO and for the oZEZO, as might be expected. 

Comparing (\ref{htt_asymp}) with (\ref{deltaXh_tt}), we see that the anomalous behavior can be attributed to a $\Xi^\alpha$-type gauge transformation from an asymptotically flat gauge, with $(\alpha_1,\alpha_2,\alpha_3,\alpha_4)=(\eta/2,0,0,0)$; that is
\begin{equation}\label{Xi}
\Xi^\alpha=\frac{1}{2}\eta\, t\delta^{\alpha}_t .
\end{equation}
According to (\ref{deltaXideltaOmega}), such a transformation modifies $\delta\Omega/\Omega$ by an amount $\delta_{\Xi}(\delta\Omega/\Omega)=\alpha_1=\eta/2$. The inverse transformation, $-\Xi^\alpha$, takes the Lorenz-gauge perturbation out of the Lorenz-gauge class and into the class of gauges that are both helically symmetric and asymptotically flat. Hence we have $(\delta\Omega/\Omega)^{(L)} = \delta\Omega/\Omega + \eta/2$, and thus 
\begin{equation}
\hat\Omega/\Omega = (\hat\Omega/\Omega)^{(L)} - \eta/2 .
\end{equation}
Equation (\ref{omegaSF}) can now be written in terms of Lorenz-gauge self-force quantities: 
\begin{equation}
\label{omegaSFLorenz}
\hat\Omega/\Omega=
1 -\frac{1}{2}\eta + 3\eta \widetilde{\Delta E}^{(L)} +8\eta \widetilde{F}^{r(L)}_{\rm ibco}  .
\end{equation}

In Sec.\ \ref{sec:Numeric-MBMS} we will use Eq.\ (\ref{omegaSFLorenz}) to calculate $\hat\Omega$ with Lorenz-gauge numerical self-force data as input; and in Sec.\ \ref{sec:1st} we will show that our calculated value agrees with that predicted by the first-law of black hole binary mechanics, as applied to the IBCO.

%%%%%%%%%%%%%%%%%%%%%%%%%%%%%%%%%%%%%%%%%%%%%%%%%%%%%%%%%%%%%%%%
\section{Self-force correction to the critical angular momentum}
\label{sub2sec:IBCO-AMshift}
%%%%%%%%%%%%%%%%%%%%%%%%%%%%%%%%%%%%%%%%%%%%%%%%%%%%%%%%%%%%%%%%%

We now turn to our second (quasi)invariant quantity: the fine-tuned value $\hat L$ of angular momentum needed for the iZEZO orbit to become asymptotically circular at late time (again, neglecting radiation). 
For definiteness, the quantity we shall consider is a certain Bondi-type {\em total} angular momentum of the spacetime in a center-of-mass (CoM) frame, which we define precisely in subsection \ref{ADMdef} below.
%As we shall discuss below, we are interested in the total ``mechanical"  angular momentum of the system
%(without the ill-defined contribution linked to the presence of time-symmetric radiation), 
%Let us denote this value by by ${\cal L}$.*  
Expressed as an expansion in $\eta$, it has the form 
%At leading order in $\eta$, i.e..\ neglecting the self-force, it is given simply by
\begin{equation}\label{LADM_leading}
\hat L=L +O(\eta^2)=  4M^2 \eta +O(\eta^2) .
\end{equation}
%where $L$ is the conserved geodesic angular momentum introduced in previous sections. 
There occurs no $O(\eta^0)$ term, since our black hole has neither intrinsic spin nor (for $\eta\to 0$) orbital angular momentum in the CoM frame. In that frame, the only contribution at $O(\eta)$ comes from the conserved geodesic orbital angular momentum $L$. 

We are interested in the $O(\eta^2)$ term of $\hat L$, associated with the effect of the conservative (time-symmetric) first-order self-force. We immediately encounter at least four complications.
First, there is the fundamental issue of choosing a definition of angular momentum that makes sense for the time-symmetric ZEZO spacetime even at $O(\eta^2)$, where time-symmetric radiative contributions render the usual ADM angular momentum ill defined. This problem will be discussed and addressed in Sec.\ \ref{ADMdef}.
% of these is that, starting at $O(\eta^2)$, the total, ADM angular momentum *(taking also into account the angular momentum of the radiation)* is actually ill-defined for the time-symmetric perturbed metric associated with our ``conservative'' ZEZO configuration. We will address and resolve this difficulty in Sec.\ \ref{ADMdef} below. 
The three other complications are more technical. First, the definition of angular momentum refers to a CoM frame. In the geodesic approximation (i.e.\ for $\eta\to 0$), the CoM trivially coincides with the centre of Schwarzschild coordinates, $r=0$. However, as we perturb the metric, it is no longer obvious where our ``center of coordinates'' lies with respect to the CoM;  this must be established for the particular gauge chosen, and a suitable transformation to a CoM frame must be performed if necessary.  A second complication is that, at $O(\eta^2)$, $\hat L$ contains a contribution from the recoil motion of the large black hole about the CoM, which must be accounted for. Finally, if we are to express our angular momentum in terms of Lorenz-gauge self-force quantities, we would need to carefully account for the gauge pathology at infinity discussed at the end of the previous section.  We will deal with these issues one by one in the following four subsections. 

%%%%%%%%%%%%%%%%%%%%%%%%%%%%%%%%%%%%%%%%%%%%%%%%%%%%%%%%%%%%%%%%%%%%%%%%%
\subsection{Definition of $\hat L$ as a Bondi-type angular momentum}
\label{ADMdef}
%%%%%%%%%%%%%%%%%%%%%%%%%%%%%%%%%%%%%%%%%%%%%%%%%%%%%%%%%%%%%%%%%%%%%%%%

In helically-symmetric spacetimes the conditions for asymptotic flatness are violated because these spacetimes
must involve, for an infinite time, an equal amount of incoming and outgoing radiation having a slow ($\sim 1/r$) spatial decay. 
In particular, this renders the ADM integrals at $i^0$ mathematically ill defined.  The perturbed ZEZO spacetime is not precisely helically symmetric, but it is so asymptotically at $i^+$ (iZEZO) or $i^-$ (oZEZO). As a result, the ``advanced'' iZEZO geometry is helically symmetric at $i^0$, as is the ``retarded'' oZEZO geometry. This means that, for both iZEZO and oZEZO, the time-symmetric spacetime (``half-retarded-plus-half-advanced'') fails to be asymptotically flat, just as in the case of an ``eternal'' circular orbit. This failure manifests itself first at $O(\eta^2)$ in the metric, in the form of quadratic combinations of first-order radiative terms that do not have a sufficiently rapid fall-off at spatial infinity \cite{Pound:2015wva}. In consequence, we cannot meaningfully speak of the ADM angular momentum of the time-symmetric iZEZO or oZEZO spacetimes. 

We seek a different definition of angular momentum, applicable to the iZEZO. We choose the following. Let $\hat L$ be defined as the {\it incoming Bondi angular momentum at infinite past advanced time}. By ``incoming'' we refer to the standard Bondi integral as calculated on a segment of past null infinity, and here we are evaluating this integral in the limit $v\to -\infty$, where $v$ is advanced time. We expect the radiation content of both retarded and advanced iZEZO spacetimes to be vanishingly small in this limit, and therefore our $\hat L$ to be mathematically well defined (finite) even for the time-symmetric iZEZO spacetime. Intuitively, this Bondi quantity, free of problematic radiative contributions, represents a purely ``mechanical'' angular momentum of the particle--black hole system. This angular momentum can be calculated in the framework of the post-Minkowskian theory of scattering particles, where, indeed, the notion of mechanical momentum has a precise formulation (to be reviewed below). This has an obvious advantage: our calculation of $\hat L$ for the iZEZO configuration will require no knowledge of the second-order metric perturbation [which would normally be needed for a direct evaluation of the Bondi integral at $O(\eta^2)$]. Instead, we will extract $\hat L$ from the orbital ``kinematics'' alone, given the (first-order) self-force along the orbit. Furthermore, our $\hat L$ naturally relates to the notion of angular momentum used in EOB and PN theories, and also in the first-law formulation; it is thus the relevant notion to consider for the purpose of comparison. And $\hat L$ has one more attractive property: it has the alternative interpretation of a total ADM angular momentum---not in the fictitious, time-symmetric spacetime, but in the physical iZEZO problem with retarded boundary conditions. Our choice to consider an angular momentum $\hat L$  as defined above is thus both practically useful and physically motivated. 

%\bl{As we shall see, besides being well-defined, the mechanical angular momentum $\cal L$ has a few more attractive properties. First, our calculation of $\cal L$ through $O(\eta^2)$, for the iZEZO configuration, will require no knowledge of the second-order metric perturbation (which would normally be needed for a direct evaluation of the Bondi integral at that order); we will extract $\cal L$ from the orbital kinematics alone, given the self-force along the orbit. Second, our $\cal L$ naturally relates to the notion of angular momentum used in EOB and PN theories, and (as our work demonstrates) also in the first-law formulation. It is thus the relevant notion to consider for the purpose of comparison. Finally, as will be discussed, $\cal L$ also has the interpretation of a total ADM angular momentum---not in the fictitious, time-symmetric spacetime, but in the physical iZEZO problem with retarded boundary conditions. Our definition of $\cal L$ is thus also physically motivated. }

In the rest of this subsection we briefly review the notion of mechanical angular momentum (in electromagnetism and post-Minkowskian gravity), relate it to our Bondi-type angular momentum $\hat L$ in the iZEZO case, and discuss the ADM reinterpretation.  

%%%%%%%%%%%%%%%%%%%%%%%%%%%%%%%%%%%%%%%%%%%%%%%%%%%%%%%%%%%%%%%%%%%%%%%%%%%%%%%
\subsubsection{Interpretation of $\hat L$ as a mechanical angular momentum}
The notion of mechanical momentum (and energy) for a system of particles interacting via a time-symmetric field exchange goes back to the classic work of Fokker \cite{Fokker1929} in (flat-space) electromagnetism. 
%The simple (linearly interacting) case of the electromagnetic time-symmetric interaction of charges has
%been treated in the classic work of Fokker \cite{Fokker1929}. 
In that work, Fokker computed the purely mechanical reduced action (``Fokker action'') describing the dynamics of a system of electric charges, after having ``integrated out" the electromagnetic field. 
The ``purely mechanical'', action-at-a-distance approach of Fokker was later pursued by Wheeler and Feynman \cite{Wheeler:1949hn}. 
The Fokker(-Wheeler-Feynman) action being manifestly Poincar\'e invariant leads to conservation laws both for the total four-momentum, 
$P^\mu$, and for the total tensorial angular momentum, $J^{\mu \nu}$, of the mechanical system. Explicit expressions for these
mechanical conserved quantities were derived by Schild and his collaborators \cite{Dettman:1954zz,Schild1963}. 
For a two-particle system of electric charges, these Fokker-Wheeler-Feynman {\it mechanical} conserved momentum and angular
momentum of the system are of the form 
\begin{eqnarray} \label{PJmech}
P^\mu_{\rm mech}(x_1, x_2) &=& m_1 u_1^\mu(\tau_1)+ m_2 u_2^\mu(\tau_2) + p^\mu_{\rm int}(x_1, x_2), \nonumber\\
J^{\mu \nu}_{\rm mech}(x_1, x_2) &=& 2 m_1x_1^{[\mu}(\tau_1)  u_1^{\nu]}(\tau_1) \nonumber\\
&+& 2 m_2x_1^{[\mu}(\tau_2)  u_2^{\nu]}(\tau_2)  + j^{\mu \nu}_{\rm int}(x_1, x_2),
\end{eqnarray}
where $x_1, x_2$ are arbitrary, spacelike-related points on the two worldlines, 
$\tau_1, \tau_2$ are the proper times corresponding to $x_1, x_2$, and the {\it interaction terms} 
$p^\mu_{\rm int}$ and $ j^{\mu \nu}_{\rm int}$ are  mildly nonlocal functionals of the two worldlines (involving only finite proper-time intervals). The crucial point for our present discussion is that the quasi-localized structure of the interaction contributions
$p^\mu_{\rm int}(x_1, x_2)$ and $ j^{\mu \nu}_{\rm int}$ imply the following properties:
(i) in a scattering situation, both $p^\mu_{\rm int}(x_1, x_2)$ and $ j^{\mu \nu}_{\rm int}$  vanish 
in the infinite past (for the incoming state) and in the infinite future (for the outgoing state),
and (ii) in a bound-state situation, \ie\    for an eternally (absorbing and) emitting time-symmetric system of two charges (\eg\  on circular orbits), 
both $p^\mu_{\rm int}(x_1, x_2)$ and $ j^{\mu \nu}_{\rm int}$ are {\it finite}, in spite of the presence of an infinite
amount of energy in the homogeneous radiation field 
$F_{\mu \nu}^{\rm rad}= \frac12 (F_{\mu \nu}^{\rm ret}- F_{\mu \nu}^{\rm adv})$. 
[The fact that $F_{\mu \nu}^{\rm rad}$ does not contribute to the mechanical conserved quantities can be seen from the
results of Ref.\ \cite{Wheeler:1949hn}, notably their Eq. (33).]

%as the integral $\int T^{\mu \nu}_{\rm WF} d\sigma_\nu$,
%where $\sigma$ is  any spacelike hypersurface containing the points 
%$x_1, x_2$, and where $T^{\mu \nu}_{\rm WF}$  is a certain modified electromagnetic stress-energy tensor,
%which differs in several ways from the usual Maxwell stress-energy tensor  $T^{\mu \nu}(F^{\rm sym})$
%constructed from the time-symmetric electromagnetic field $F_{\mu \nu}^{\rm sym}= \frac12 (F_{\mu \nu}^{\rm %ret}+ F_{\mu \nu}^{\rm adv})$. One can check that it is the integral of the difference 
%$T^{\mu \nu}(F^{\rm sym}) - T^{\mu \nu}_{\rm WF}$ that would account for the momentum,
%say $p^\mu_{\rm rad}(x_1, x_2)$, contained in the radiative part of the field
%(so that the total, ADM momentum of the particle-plus-field system
%is  $P^\mu_{\rm FWF}(x_1, x_2) + p^\mu_{\rm rad}(x_1, x_2)$). In a scattering situation 
%$p^\mu_{\rm rad}(x_1, x_2)$ is finite, while it would infinite when considering a (time-symmetric)
%bound state (emitting for an infinite time). It can also be checked that the above summarized results
%on the mechanical part of the momentum of the system extend to the corresponding mechanical
%part of the angular momentum of the system, with similar results. 

The case of relevance to us here, of a time-symmetric gravitationally interacting system of masses,
is much more involved (due to the nonlinear structure of Einstein's gravity) 
and cannot be treated in exact form as the electromagnetic case. As emphasized
in Ref.\ \cite{Damour:1995kt}, in a post-Minkowskian framework one can formally derive a gravitational analog of the electromagnetic Fokker action by perturbatively iterating a Fokker-type time-symmetric
Green function while integrating out the gravitational field. This leads to a (post-Minkowskian) 
expansion for the reduced gravitational action involving Feynman-like diagrams in which the
nonlinear vertices defined by the Einstein-Hilbert action are connected by time-symmetric propagators.
We are not aware of any explicit proof showing that there exist, at all post-Minkowskian orders,  gravitational analogs of the mechanical
conserved quantities $P^\mu_{\rm mech}$ and $J^{\mu \nu}_{\rm mech}$ having the same properties as in the
electromagnetic case. However, there are partial results confirming the probable
existence of such well-defined mechanical conserved quantities. For instance, at the first post-Minkowskian approximation
(first order in $G$), Ref.\ \cite{Friedman:2005rx} has explicitly constructed (following \cite{Dettman:1954zz}) 
$P^{\mu 1 \rm PM}_{\rm mech}$ and $J^{\mu \nu 1 \rm PM}_{\rm mech}$, and has shown, in particular, that they were finite for 
gravitationally interacting helically symmetric binary systems. They have also verified that the conserved 
mechanical energy and angular momentum satisfied the expected first law: $\delta E = \Omega \delta J$.
In addition, the Fokker-like time-symmetric, reduced gravitational action is explicitly known to the fourth post-Newtonian 
accuracy \cite{Damour:2014jta,Damour:2015isa,Bernard:2015njp,Damour:2016abl,Bernard:2016wrg,Foffa:2019rdf,Foffa:2019yfl}.
This 4PN action includes terms coming from the fifth post-Minkowskian approximation [$O(G^5)$]. At this high order there appear delicate
contributions  to the action (related to the emission of gravitational radiation) which are nonlocal-in-time. In spite of the highly nonlinear aspects of the gravitational two-body interaction
described by this action, it was again explicitly shown \cite{Damour:2014jta,Damour:2015isa,Bernard:2017ktp} that there existed conserved 
mechanical energy and angular momentum, $P^{\mu 4 \rm PN}_{\rm mech}$ and $J^{\mu \nu 4 \rm PN}_{\rm mech}$, 
having the same structure as in the electromagnetic case.
Namely: (i) in the scattering case, the interaction contributions
to the conserved energy, momentum and angular momentum vanish for infinite separations (which
is a direct confirmation that they do not include the usual contribution coming from the spatial integral of the 
energy density of the incoming or outgoing gravitational radiation); while, 
(ii) in the bound-state case they are all {\it finite}, despite the presence of
infinite radiative contributions in the corresponding ADM quantities. 
Note that the same results a fortiori apply to the EOB conserved quantities, which by construction are equal to their PN counterparts (considered in a CoM frame). In the EOB formalism,  both the second-post-Minkowskian Hamiltonian 
(second order in $G$ and all orders in $1/c$) \cite{Damour:2017zjx}, and the third-post-Minkowskian one [$O(G^3)$] 
\cite{Bern:2019nnu,Antonelli:2019ytb}, have been recently derived  and exhibit the same features.
Let us also note that the validity of the first law of binary dynamics has been also checked at
the fourth post-Newtonian approximation \cite{Blanchet:2017rcn}.

In our present problem, the iZEZO spacetime is not globally amenable to a post-Minkowskian treatment, because the gravitational interaction is very strong at late time. However, a post-Minkowskian description is perfectly suitable near $i^-$ and in the far past of past null infinity, where the interaction is vanishingly small. Since our Bondi-type definition of $\hat L$ involves only information from that far past of spacetime, we can evaluate this quantity within a post-Minkowskian framework. In fact, as we shall see in Sec.\ \ref{ADML}, a leading-order, i.e, Minkowskian, calculation  would do for our purpose. The important point for us is that, at least at that order, it is intuitively clear (in view of the asymptotic vanishing of the radiation field near 
$i^-$ and in the far past of past null infinity) that
the Fokker-Wheeler-Feynman-type mechanical angular momentum will coincide with the Bondi-type angular momentum $\hat L$ as we have defined it above. We leave a detailed technical check of this equality to future work.

%*The useful consequence of the above considerations for our present problem is that we can
%unambiguously identify the conserved energy and angular momentum relevant for our comparison.
%Namely, we can equate them to the corresponding {\it incoming Bondi} quantities, {\it i.e.}
%the limits in the infinite past of advanced time (i.e.\ at $i^-$) of Bondi quantities. We use here the fact
%that we are interested in a situation where the radiation recorded on past null infinity vanishes in the far
%advanced-time past. Considering the incoming Bondi quantities is then equivalent to capturing the %corresponding
%purely mechanical quantities, determined from the orbital kinematics of the iZEZO at $r_{\rm p}\to\infty$ (i.e.\ at $i^-$). The calculation we shall describe below involves only the local first-order self-force (and not the second-order metric perturbation), and yields, in particular, a perfectly well defined angular momentum through $O(\eta^2)$, say  ${\cal L}$. 

%%%%%%%%%%%%%%%%%%%%%%%%%%%%%%%%%%%%%%%%%%%%%%%%%%%%%%%%%%%%%%%%%%%%%%%%%%%
\subsubsection{Interpretation of $\hat L$ as an ADM angular momentum}
%%%%%%%%%%%%%%%%%%%%%%%%%%%%%%%%%%%%%%%%%%%%%%%%%%%%%%%%%%%%%%%%%%%%%%%%%%

We have defined $\hat L$ as a Bondi-type quantity in the time-symmetric iZEZO spacetime. But there is a more physically compelling interpretation of $\hat L$, as follows.
Consider the {\em physical} iZEZO problem, with the full self-force restored, and with retarded boundary conditions. Suppose that, in the physical problem, we set the particle to start off with the same fine-tuned initial conditions as in the time-symmetric problem, i.e., in particular, the same value of $\hat L$. Since in the physical problem there is no radiation coming in from past null infinity, that $\hat L$ would also be the (now well defined and finite) total ADM angular momentum of the physical iZEZO spacetime.  
% Specifically, consider the physical iZEZO orbit that starts with the same fine-tuned initial conditions as the ``conservative'' iZEZO, but is subsequently subject to the full self-force, including dissipation. The corresponding spacetime, of course, admits a well defined ADM angular momentum, through all orders in $\eta$: this is guaranteed, since the retarded field at $i^0$ is only sensitive to the initial conditions at $i^-$, when the particle is infinitely far from the black hole. 
Furthermore, the physical orbit remains ``close'' to the time-symmetric iZEZO until well into the whirl phase, on account of the facts that (i) during the infall from infinity, the specific parameters of the orbit (say, $\hat E/\eta,\hat L/\eta$) deviate by amounts of only $O(\eta)$ due to the dissipative effect, and (ii) the final whirl, before the particle scatters back to infinity or plunges into the black hole, takes an amount of time $\propto\log\eta$ \cite{gund}, during which the dissipative piece of the self-force changes the orbital parameters by an amount of only $O(\eta\log\eta)$. Thus, the physical (dissipating) iZEZO can be considered a perturbation of the conservative iZEZO up until and through the whirl; but, crucially, the former, unlike the latter, does admit a well defined ADM angular momentum. 

This all means that we can {\it identify} $\hat L$ (as defined in the time-symmetric spacetime) with the ADM angular momentum of the corresponding physical (dissipating) iZEZO system with the same initial condition. Such an identification is not only physically compelling, but will also be useful for us in practice: In the next two subsections will rely on it in defining a CoM frame, as part of our calculation of $\hat L$.

\subsection{Expression for $\hat L$ in a CoM-centered, asymptotically-flat gauge} 
\label{ADML}
%%%%%%%%%%%%%%%%%%%%%%%%%%%%%%%%%%%%%%%%%%%%%%%%%%%%%%%%%%%%%%%%%%%%%%%%%%%%%%%

Our goal now is to obtain an expression for the $O(\eta^2)$ piece of $\hat L$ in Eq.\ (\ref{LADM_leading}), in terms of calculable self-force quantities.\footnote{The analysis leading to the intermediate result (\ref{calLflat})  was already carried out by one of us in \cite{damour}, but for completeness we give it here again, in a slightly different form.} 
%[The following calculations agree with those already given in Ref.  \cite{damour}.]*
We focus on the iZEZO case, and (for our current purpose) make the above identification of the orbit with a physical one, such that the perturbed spacetime is asymptotically flat and admits well defined ADM integrals. We introduce the ``asymptotically Lorentzian'' system $(t,x,y,z)$ defined from the Schwarzschild coordinates through
\begin{equation}
x:=r\sin\theta\cos\varphi, \quad\ y:=r\sin\theta\sin\varphi, \quad z:=r\cos\theta ,
\end{equation}
and assume a gauge is chosen so that the perturbed metric is manifestly asymptotically flat in these coordinates. Then, one can unambiguously define the spacetime's CoM location, $x^i=R^i$ (see Sec.\ \ref{CoM} below), where hereafter $i$ runs over the 3 spatial coordinates. Importantly, $R^i$ can be obtained from an ADM-type integral at $i^0$, and it is thus determinable entirely from the initial conditions at $t\to-\infty$. 
For the Schwarzschild background, one trivially finds $R^i=0$. However, the value of $R^i$ in the perturbed spacetime depends on the gauge. For our calculation of $\hat L$ below, we assume that the gauge is further specified so that $R^i\equiv 0$ through $O(\eta)$ at all time. We refer to this as a ``CoM-centered'' gauge. (In such a gauge, the spacetime also has zero total linear momentum.)
%Let us choose coordinates $x^\alpha$, and a gauge, such that the perturbed metric is manifestly asymptotically flat, as well as CoM-centered, by which we mean $R^i\equiv 0$ through $O(\eta)$ at all time (this also guarantees that, in this system, the full spacetime has zero total linear momentum). Specifically, we take $x^\alpha$ to be Cartesian-like coordinates $(t,x,y,z)$ defined from the Schwarzschild coordinates in the usual way:
%\begin{equation}
%x:=r\sin\theta\cos\varphi, \quad\ y:=r\sin\theta\sin\varphi, \quad z:=r\cos\theta .
%\end{equation}
%This choice ensures the asymptotic flatness and CoM conditions are satisfied for $\eta\to 0$, and we further assume a gauge is chosen such that they are also satisfied at $O(\eta)$. 

We now let $x^\alpha=\hat x^{\alpha}_{\rm p}(\tau)$ represent the particle's iZEZO trajectory in the above asymptotically Lorentzian coordinates, as corrected by the conservative self-force associated with the asymptotically-flat, CoM-centered perturbation. We wish to map the iZEZO system, for $\hat r_{\rm p}\to\infty$, onto a problem of two point particles in flat space. To this end, we interpret the $(t,x,y,z)$ coordinates, in the limit $r\to\infty$, as Cartesian coordinates ($+$time) in flat space, and introduce the three-velocity $\hat v^i:=d \hat x_{\rm p}^i/dt$. The particle's contribution to $\hat L$ is then given by 
\begin{equation}\label{calLp}
{\hat L}_{\rm p} = \mu \left(\hat x_{\rm p} \hat v^y - \hat y_{\rm p} \hat v^x\right) 
= \mu \hat v_{\varphi},
\end{equation}
where, in obtaining the second equality, we have used $\hat v^x=(\hat x_{\rm p}/\hat r_{\rm p})\hat v^r-\hat y_{\rm p} \hat v^\varphi$ and 
$\hat v^y=(\hat y_{\rm p}/\hat r_{\rm p}) \hat v^r +\hat x_{\rm p} \hat v^\varphi$, followed by $\hat r_{\rm p}^2 \hat v^\varphi = \hat v_{\varphi}$. All quantities here are evaluated on the orbit in the limit $\hat r_{\rm p}\to\infty$. In an asymptotically-flat gauge, we have $\hat v_{\varphi}=(d\hat t_{\rm p}/d\tau)^{-1}\hat u_{\varphi}\to \hat u_{\varphi}$ in that limit. Therefore, recalling $\hat L =\mu \hat u_{\varphi}$ and Eq.\ (\ref{Lpert}), we arrive at
\begin{equation}
{\hat L}_{\rm p} = 4M\mu +\delta L_\infty.
\end{equation}
The $O(\eta^2)$ quantity $\delta L_\infty:=\delta L(\infty)$ is given in Eq.\ (\ref{deltaLinf}) in terms of the self-force integrals $\Delta E$ and $\Delta L$.

As mentioned already, the ADM angular momentum has a contribution from the motion of the black hole about the CoM, first appearing at $O(\eta^2)$. This contribution---call it ${\hat L}_{\rm bh}$---is easily obtained in the point-particle picture. For $r_{\rm p}\to\infty$, the black-hole's Cartesian coordinates in the above-defined CoM system  are $X=-\eta \hat x_{\rm p}$ and $Y=-\eta \hat y_{\rm p}$, with corresponding three-velocity components $V^x=-\eta \hat v^x$ and $V^y=-\eta \hat v^y$. Hence,
\begin{equation}
{\hat L}_{\rm bh} = M \left(X V^y - Y V^x\right) = M\eta^2 \hat u_{\varphi} = 4\mu^2,
\end{equation}
where we have omitted terms beyond the leading, $O(\eta^2)$ contribution.

The total ADM angular momentum is therefore ${\hat L}={\hat L}_{\rm p} + {\hat L}_{\rm bh}=4M\mu+4\mu^2+\delta L_\infty$, which we write in the form
\begin{equation}\label{calLflat}
\widetilde{\hat L}= 4\eta + 4\eta^2 + \eta^2\widetilde{\delta L}_{\infty} ,
\end{equation}
introducing the mass-rescaled dimensionless quantities 
\begin{equation}
\widetilde{\hat L}:={\hat L}/M^2,
\quad\quad
\widetilde{\delta L}_{\infty}:= {\delta L}_{\infty}/M^2.
\end{equation}
In Eq.\ (\ref{calLflat}) (which agrees with the expression derived in \cite{damour}), the first term is the ``background'' (geodesic) value, the second term is the contribution from the black hole's recoil motion, and the third term is due to the self-force acting on the particle.  

%%%%%%%%%%%%%%%%%%%%%%%%%%%%%%%%%%%%%%%%
\subsection{Center-of-Mass condition} 
\label{CoM}
%%%%%%%%%%%%%%%%%%%%%%%%%%%%%%%%%%%%%%%

Equation (\ref{calLflat}) is applicable in a CoM-centered gauge. We will now show how to choose our Lorenz gauge so that it is indeed CoM-centered. 

Our treatment is based on the Landau-Lifshitz formulation, as described, e.g., in Sec.\ 6.1 of \cite{poisson_will_2014}. For an asymptotically flat spacetime with metric $\hat g_{\alpha\beta}$ and ``gothic inverse metric''  $\mathfrak{g}^{\alpha\beta}=(-\hat g)^{1/2}\hat g^{\alpha\beta}$ (where $\hat g$ is the metric's determinant), the CoM position can be obtained via 
\begin{equation}\label{CoMformula}
R^i = \frac{1}{16\pi M}\oint_{i^0} \left(x^i \partial_j H^{tjtk}-H^{titk}\right)dS_k ,
\end{equation}
where
$H^{\alpha\beta\gamma\delta}:=\mathfrak{g}^{\alpha\gamma}\mathfrak{g}^{\beta\delta}-\mathfrak{g}^{\alpha\delta}\mathfrak{g}^{\gamma\beta}$,
and the integral is performed over a two-sphere with outward-pointing normal $dS_k$ in the limit $r\to\infty$. The expression is valid in asymptotically Lorentzian coordinates such as the ones defined above, with indices $i,j,k$ running over the three spatial Cartesian-like coordinates. Here we wish to apply Eq.\ (\ref{CoMformula}) with $\hat g_{\alpha\beta}=g_{\alpha\beta}+h_{\alpha\beta}$, where $g_{\alpha\beta}$ is the background Schwarzschild metric and $h_{\alpha\beta}$ a Lorenz-gauge metric perturbation. It is easy to show that the contribution to $R^i$ from $g_{\alpha\beta}$ vanishes, so we need only consider the piece of the integrand linear in $h_{\alpha\beta}$.

A few simplifications make this task manageable analytically. First, noting that the value of $R^i$ does not depend on how one chooses to approach $i^0$, we can choose to do so on an early-time hypersurface $t={\rm const}\ll -M$, on which the asymptotic perturbation from our iZEZO at $r\to\infty$ is expected to be dominated by a static, $t$-independent piece. Promoting this expectation to an assumption, it suffices to consider a static perturbation $h_{\alpha\beta}$. Second, we expect only a particular multipolar mode of the perturbation to contribute to $R^i$, i.e.\ the even-parity dipole mode $(\ell,m)=(1,\pm 1)$ (in a suitable tensor-harmonic decomposition such as the one to be introduced in Sec.\ \ref{sec:Numeric-MBMS} below); the contribution from other modes should vanish upon integration over the 2-sphere in Eq.\ (\ref{CoMformula}), at least in the limit $r\to\infty$. Third, the even-parity dipole mode is known to be a pure-gauge mode of the perturbation away from any sources \cite{Zerilli,lowmodes,Origauge}. 

These simplifications make it sufficient for us to consider vacuum perturbations of the form 
\begin{equation}
h_{\alpha\beta}=\nabla_{\alpha}\xi_{\beta}+\nabla_{\beta}\xi_{\alpha},
\end{equation}
where the generator $\xi^\alpha$ is subject to the Lorenz-gauge conditions 
\begin{equation}\label{Boxxi}
\nabla^{\alpha}\nabla_{\alpha}\xi_\beta =0 ,
\end{equation}
and assumes the static, even-party dipolar form 
\begin{eqnarray}\label{xi}
\xi_t &=& a(r)\sin\theta\cos\varphi, 
\nonumber\\
\xi_r &=& b(r)\sin\theta\cos\varphi, 
\nonumber\\
\xi_\theta &=& c(r)\cos\theta\cos\varphi, 
\nonumber\\
\xi_\varphi &=& -c(r)\sin\theta\sin\varphi.
\end{eqnarray}
[We have fixed here the azimuthal phase of $\xi_\alpha$ at a specific value, for convenience. The phase of the actual solution is determined by the initial orbital phase  $\varphi_{\rm p}(t\to-\infty)$, and our particular choice must correspond to {\it some} value of that phase; here, without loss of generality, we assume that value.] Equation (\ref{Boxxi}) then constitutes a coupled set of three second-order ordinary differential equations for $a(r)$, $b(r)$ and $c(r)$. The general solution is a linear combination of six independent ``basis'' solutions, which we give analytically in Appendix \ref{App:static_dipole}.\footnote{These solutions were previously derived, and five of them are given, in Ref.\ \cite{Origauge}.} We call these solutions 
$\big\{a_{(j)}^{\pm},b_{(j)}^{\pm},c_{(j)}^{\pm}\big\}$, and, correspondingly, $\xi^{\pm}_{\alpha(j)}$, where $j=1,2,3$. These are chosen so that the three solutions $\xi^{-}_{\alpha(j)}$ generate metric perturbations that are regular at the event horizon (by which we mean, perturbations whose components in a horizon-regular system, such as ingoing Eddington-Finkelstein coordinates, are smooth on the horizon); and the three solutions $\xi^{+}_{\alpha(j)}$ generate metric perturbations that are regular at infinity (by which we mean that their components in our asymptotically Lorentzian system fall off at least as $1/r^2$ for $r\to\infty$). All three of the solutions $\xi^{+}_{\alpha(j)}$ are irregular at the horizon, and the two solutions $\xi^{-}_{\alpha(1)}$ and $\xi^{-}_{\alpha(2)}$ are irregular at infinity. The solution $\xi^{-}_{\alpha(3)}$  is special, in that it generates a gauge perturbation that is globally regular, in the above sense. This solution, whose physical interpretation will be discussed momentarily, has the simple form
\begin{eqnarray}\label{xiCoM}
\xi_{t(3)}^- &=& 0, 
\nonumber\\
\xi_{r(3)}^- &=&  \sin\theta\cos\varphi, 
\nonumber\\
\xi_{\theta(3)}^- &=&  (r-M)\cos\theta\cos\varphi, 
\nonumber\\
\xi_{\varphi(3)}^- &=& - (r-M)\sin\theta\sin\varphi
\end{eqnarray}
(up to an arbitrary amplitude).

The actual even-dipole mode is determined by solving the inhomogeneous linearized field equations, sourced by the point particle. In practice, this amounts to matching the ``external'' solutions generated by $\xi^{+}_{\alpha(j)}$ to the ``internal'' solutions generated by $\xi^{-}_{\alpha(j)}$ on the surface $r=r_{\rm p}(t)$, using junction conditions determined from the form of the (distributional) source. This procedure guarantees that the actual solution satisfies both the junction conditions at the particle and the regularity conditions at infinity and on the horizon. However, the existence of the globally regular homogeneous solution $\xi^{-}_{\alpha(3)}$ means that no unique solution can be determined in this way: one can add the solution generated by $\xi^{-}_{\alpha(3)}$ with an arbitrary amplitude, without violating any of the junction or boundary conditions.

The physical significance of this arbitrariness will be discussed shortly, but for now let us return to our main thread, i.e.\ the evaluation of the CoM position $R^i$. For this, it turns out that we do not need to obtain the actual inhomogeneous dipole perturbation; all we need is the most general form of the perturbation near $i^0$ (at $t\ll -M$), which (in terms of the generator) reads
\begin{equation}
\xi_\alpha = C_1 \xi^{+}_{\alpha(1)}+C_2 \xi^{+}_{\alpha(2)}+C_3 \xi^{+}_{\alpha(3)}+C_4 \xi^{-}_{\alpha(3)},
\end{equation} 
with some constants $C_n$. We know the actual perturbation near $i^0$ is generated by a $\xi_\alpha$ of this form. As input for Eq.\ (\ref{CoMformula}), it will suffice to provide the $O(1/r^2)$ piece of this perturbation. At this order, the nonzero components work out to be (up to an arbitrary amplitude)
\begin{eqnarray}\label{dipole_asympt}
h_{tt}&=& -\frac{2C_4}{r^2}\sin\theta\cos\varphi ,
\nonumber\\
h_{rr}&=& \frac{2(C_3+C_4)}{r^2}\sin\theta\cos\varphi ,
\nonumber\\
h_{r\theta}&=& h_{\theta r} =\frac{C_3+2C_4}{r}\cos\theta\cos\varphi ,
\nonumber\\
h_{r\varphi}&=& h_{\varphi r}= -\frac{C_3+2C_4}{r}\sin\theta\sin\varphi ,
\nonumber\\
h_{\theta\theta}&=& -2C_4\sin\theta\cos\varphi, 
\nonumber\\
h_{\varphi\varphi}&=& -2C_4\sin^3\theta\cos\varphi.
\end{eqnarray} 
Evidently, the perturbations generated by $\xi^{+}_{\alpha(1)}$ and $\xi^{+}_{\alpha(2)}$ decay too fast at infinity to be able to produce a CoM shift.

It is straightforward to calculate the contribution to $R^i$ from the metric perturbation  (\ref{dipole_asympt}) via Eq.\ (\ref{CoMformula}). The result is very simple:
\begin{equation}
\{R^x,R^y,R^z\}=\{-C_4,0,0\} .
\end{equation}
Namely, the only gauge perturbation that shifts the CoM location is the one generated by $\xi^{-}_{\alpha(3)}$; and, with $\xi^{-}_{\alpha(3)}$ normalized as in Eq.\ (\ref{xiCoM}), it does so by an amount of $-1$ in the $x$ direction. It is now a good time to return to the question of the physical interpretation of $\xi^{-}_{\alpha(3)}$. That is made clear by examining the form of this generator at $r\to\infty$, in Lorenzian coordinates:
\begin{equation}
\left\{\xi^{t-}_{(3)},\xi^{x-}_{(3)},\xi^{y-}_{(3)},\xi^{z-}_{(3)}\right\}\sim \{0,1,0,0\}.
\end{equation}
That is, at large $r$, $\xi^{\alpha-}_{(3)}$ is a simple coordinate displacement $x\to x-1$ [recall our sign convention in Eq.\ (\ref{gauge transformation})]. Clearly, such a displacement shifts the CoM location by exactly $-1$ in the $x$ direction, consistent with the result of our calculation.
%\footnote{Ref.\ \cite{lowmodes} contains a similar analysis, for circular orbits.} 
The particular ($x$) direction of the CoM shift is, of course, inherited from our particular choice of phase in Eq.\ (\ref{xi}). The actual direction of the CoM shift will depend on the actual initial orbital phase $\varphi_{\rm p}(t\to-\infty)$. To determine this dependence we would need to construct the actual inhomogeneous solution, but there is no need for us to attempt this here. 

For our purpose, it suffices that we have established that the aforementioned arbitrariness in the Lorenz-gauge even-parity dipole solution corresponds precisely to the freedom of performing spatial gauge displacements away from the CoM system. This arbitrariness is removed with a condition on the location of the CoM. We can {\it choose} a CoM-centered Lorenz gauge, by restricting the support of the perturbation generated by $\xi^{-}_{\alpha(3)}$ to the region $r<r_{\rm p}(t)$ of spactime. This is, indeed, what we shall do in Sec.\ \ref{sec:Numeric-MBMS} when we construct our Lorenz-gauge perturbation, hence ensuring our gauge is CoM-centered as desired.\footnote{In Refs.\ \cite{lowmodes,Origauge}, where a Lorenz-gauge even-parity dipole perturbation was constructed for circular orbits, the support of the static mode $\xi^{-}_{\alpha(3)}$ was similarly restricted to $r<r_{\rm p}(t)$. But this was done there based on considerations of regularity at infinity (in \cite{Origauge}), or via the imposition of boundary conditions (in \cite{lowmodes}), rather than being interpreted as picking a CoM frame. We emphasize that the perturbation generated by $\xi^{-}_{\alpha(3)}$ is perfectly regular at infinity (and elsewhere)---cf.\ Eq.\ (\ref{dipole_asympt}).}

Finally, we address one natural question: would it not be simpler, for our purpose, to just ``gauge away'' the entire even-parity dipole perturbation? This would save us having to calculate it in a Lorenz gauge, but would guarantee just the same that we are in a CoM frame. The answer is that gauging away this mode in the vacuum regions $r<r_{\rm p}(t)$ and $r>r_{\rm p}(t)$ leaves a distribution (a delta function) in the metric on the surface $r=r_{\rm p}(t)$. The resulting ``singular gauge'', discussed in \cite{lowmodes}, is indeed (trivially) CoM-centered. However, its pathological nature makes the calculation of the corresponding self-force subtle. In the case of the iZEZO, the gauge pathology is exacerbated by the fact that the coefficient in front on the term $\propto\delta(r-r_{\rm p}(t))$ in the metric turns out to blow up in the limit $t\to -\infty$. It is not known to us how to calculate the self-force in such a gauge,  or, in particular, what contribution the singular-gauge dipole mode has to $\delta L_{\infty}$ in Eq.\ (\ref{calLflat}). Our numerical results in Sec.\ \ref{sec:Numeric-MBMS} show that this contribution, as calculated in the regular, CoM-centered Lorenz gauge, is nonzero.  

%%%%%%%%%%%%%%%%%%%%%%%%%%%%%%%%%%%%%%%%
\subsection{Lorenz-gauge adjustment} 
\label{Ladj}
%%%%%%%%%%%%%%%%%%%%%%%%%%%%%%%%%%%%%%%%

Having constructed a CoM-centered Lorenz gauge, it remains only to address the aforementioned gauge subtly at infinity. As discussed in Sec.\ \ref{subsec:gauge}, the anomalous behavior expressed in Eq.\ (\ref{alpha}) can be entirely accounted for in terms of a simple transformation $x^{\alpha}\to x^{\alpha(L)}=x^{\alpha}-\frac{1}{2}\eta t\delta_t^{\alpha}$ from a (non-Lorenz) gauge that {\it is} manifestly asymptotically flat.
We write
\begin{equation}
t^{(L)}=(1-\eta/2)t,
\end{equation}
where the non-labelled $t$ corresponds to the asymptotically flat gauge.
The only way in which such a transformation affects the discussion leading to our expression for $\hat L$ in Eq.\ (\ref{calLflat}) is through an $O(\eta)$ modification of the three-velocity components $\hat v^i$:
\begin{equation}
\hat v^i=\frac{dt^{(L)}}{dt}\hat v^{i(L)}= (1-\eta/2)\hat v^{i(L)}.
\end{equation}  
In terms of the Lorenz-gauge velocity $\hat v^{i(L)}$, the particle's angular momentum in Eq.\ (\ref{calLp}) becomes
\begin{equation}
{\hat L}_{\rm p}=\mu(1-\eta/2)\hat v^{(L)}_\varphi = \mu\hat v^{(L)}_\varphi -2\eta^2 M^2
\end{equation}
(using $\hat v^{(L)}_\varphi=4M$ at leading order). No correction enters ${\hat L}_{\rm bh}$ at the relevant order. 

We thus find that, when expressed in terms of Lorenz-gauge quantities, the total angular momentum $\hat L$ picks out a correction term equals to $-2\eta^2 M^2$. Equation (\ref{calLflat}) thus becomes
\begin{equation}\label{calLLorenz1}
\widetilde{\hat L}= 4\eta + 2\eta^2 + \eta^2\widetilde{\delta L}^{(L)}_{\infty} .
\end{equation}
Here, the term $2\eta^2$ is made up of a $+4\eta^2$ contribution from the black hole recoil motion, and a $-2\eta^2$ contribution from the Lorenz gauge correction. Finally, substituting from Eq.\ (\ref{deltaLinf}), we obtain
\begin{equation}\label{calLLorenz}
\widetilde{\hat L}= 4\eta + 2\eta^2 + \eta^2\Big(8\widetilde{\Delta E}^{(L)} -\widetilde{\Delta L}^{(L)}  \Big).
\end{equation}

In Sec.\ \ref{sec:Numeric-MBMS} we will use Eq.\ (\ref{calLLorenz}) to calculate $\hat L$ with Lorenz-gauge numerical self-force data as input; and in Sec.\ \ref{sec:1st} we will show that our calculated value agrees with that predicted by the first-law of black hole binary mechanics, as applied to the IBCO.

%%%%%%%%%%%%%%%%%%%%%%%%%%%%%%%%%%%%%%%%%%%%%%%%%%%%%%%%%%%%%%%%%%%%
%%%%%%%%%%%%%%%%%%%%%%%%%%%%%%%%%%%%%%%%%%%%%%%%%%%%%%%%%%%%%%%%%%%
\section{Numerical method}
\label{sec:Numeric-MBMS}
%%%%%%%%%%%%%%%%%%%%%%%%%%%%%%%%%%%%%%%%%%%%%%%%%%%%%%%%%%%%%%%%%%% 
%%%%%%%%%%%%%%%%%%%%%%%%%%%%%%%%%%%%%%%%%%%%%%%%%%%%%%%%%%%%%%%%%%%%

We remind that the calculation of the self-force corrections to $\Omega$ and $L$, via Eqs.\ (\ref{omegaSFLorenz}) and (\ref{calLLorenz}) respectively, requires three bits of self-force input: The value $\widetilde{F}^{r(L)}_{\rm ibco}$ and the two integrals $\widetilde{\Delta E}^{(L)}$ and $\widetilde{\Delta L}^{(L)}$. The former is relatively easy to obtain, as it requires  only the evaluation of the self-force on a circular geodesic orbit, for which methods and codes have been in existence for over a decade.  Lorenz-gauge calculations for circular orbits have been performed in the time domain \cite{baracklousto,bsago1} as well as in the frequency domain \cite{Akcay2,Akcay}. As part of the calculation in \cite{Akcay}, three of us (LB, TD and NS), with S.\ Akcay, have computed the self-force component $F^r$ as a function of the circular-orbit radius $R$ in the range $3M<R\leq 6M$, and, in particular, obtained  $F^r$ for the IBCO, $R=4M$. This value is not given in \cite{Akcay} (or elsewhere in print), but let us cite it here:
\begin{equation}\label{Fr}
\widetilde{F}^{r(L)}_{\rm ibco}=-0.003088(1),
\end{equation}
where the parenthetical figure indicates the estimated error in the last displayed decimal [i.e, $-0.003088(1)=-0.003088\pm 0.000001$]. We have confirmed this value using a new implementation (described below) of the time-domain method of \cite{bsago1}, which gives the less accurate---but reassuringly consistent---value of $\widetilde{F}^{r(L)}_{\rm ibco}=-0.00309(3)$.
Incidentally, $F^r(R)$ appears to change its sign near $R=4M$ (at around $4.1M$), making it harder to compute $\widetilde{F}^{r(L)}_{\rm ibco}$ with a good fractional accuracy. Fortunately, however, the relative contribution of the $\widetilde{F}^{r(L)}_{\rm ibco}$ term in Eq.\ (\ref{omegaSFLorenz}) turns out to be very small, since, as we shall see, the integral $\widetilde{\Delta E}^{(L)}$ is more than a hundred times large than $\widetilde{F}^{r(L)}_{\rm ibco}$. As a result, it is sufficient to obtain $\widetilde{F}^{r(L)}_{\rm ibco}$ with only a modest accuracy, and the value given in (\ref{Fr}) above will do for our purpose.
 
In the rest of this section we will describe our calculation of $\widetilde{\Delta E}^{(L)}$ and $\widetilde{\Delta L}^{(L)}$, using a specially adapted new implementation of the time-domain method of \cite{baracklousto,bsago1,bsago2}. Section \ref{review} reviews this method on general, and in Sec.\ \ref{details} we describe the details of our implementation of it in the ZEZO case. The computation of the monopole and dipole modes of the metric perturbation is particularly challenging in this case, and required much new development, to be described in Sec.\ \ref{subsec:low-ell}. 

%%%%%%%%%%%%%%%%%%%%%%%%%%%%%%%%%%%%%%%%%%%%%%%%%%%%%%%%%%%%%%%%%%%%%%%%%%%%%%
\subsection{Self-force via time-domain integration of the Lorenz-gauge perturbation equations}
\label{review}
%%%%%%%%%%%%%%%%%%%%%%%%%%%%%%%%%%%%%%%%%%%%%%%%%%%%%%%%%%%%%%%%%%%%%%%%%%%%%

We start with a brief review of the formalism and numerical implementation as they were developed in Refs.\ \cite{baracklousto,bsago1,bsago2}, referring readers to these papers for details. 

Einstein's equations, linearly perturbed about a Schwarzschild geometry, take a relatively simple form under the Lorenz gauge conditions $\nabla^\alpha\bar h_{\alpha\beta}=0$, where $\bar h_{\alpha\beta}$ is the trace-reversed metric perturbation. The angular dependence of the perturbation can be separated by writing $\bar h_{\alpha\beta}$ as a sum over multipole harmonics, each having the form $\sim \sum_{i=1}^{10}\bar h^{(i)\ell m}(t,r) Y_{\alpha\beta}^{(i)\ell m}(\theta,\phi)$, where $Y_{\alpha\beta}^{(i)\ell m}$ is a basis of tensor harmonics. For each $(\ell,m)$, one thus obtains a set of 10 coupled wave-like differential equations for the time-radial variables $h^{(i)\ell m}(t,r)$. The set decouples into two subsets: seven equations for the even-parity piece of the perturbation ($i=1,\ldots,7$ in the notation of \cite{baracklousto}) and three for the odd parity ($i=8,9,10$). In the self-force problem, one has a delta-function source on the right-hand side of the linearized Einstein equation, which, upon multipole decomposition, translates to a source $\propto \delta[r-r_{\rm p}(t)]$ in the field equations for $h^{(i)\ell m}$. For an equatorial source, modes with even (odd) values of $\ell+m$ are of pure even (odd) parity. 

In the implementation of \cite{baracklousto,bsago1,bsago2}, the equations for each $\ell,m$ are solved numerically in the time domain, using a finite-difference scheme with characteristic coordinates on a uniform mesh in $1+1$-dimensions. The trajectory of the particle, assumed given, splits the mesh into two disjoint parts. At each time step, suitable jump conditions (which are derived analytically, in advance, from the form of the source) are used to integrate the numerical field across the particle. Since the evolution is characteristic, no boundary conditions are needed. However, one has to specify characteristic initial data. The standard choice is to set all field variables $\bar h^{(i)\ell m}$ to zero on the initial characteristic rays. This results in a burst of ``junk'' radiation sourced by the particle initially, but such radiation decays over time (typically as $t^{-2\ell -3}$), and one later simply discards the early, junk-dominated part of the data.  

As the evolution proceeds, one records the value of the fields $\bar h^{(i)\ell m}$ and their ($r$ and $t$) derivatives on the particle, and from these the physical self-force is constructed using the procedure of mode-sum regularization~\cite{Barack:1999wf,Barack:2001bw,Barack:2001gx}. In this procedure, one first constructs the ``bare'' force associated with each $\ell,m$ as a certain linear combination of $\bar h^{(i)\ell m}$ and its first derivatives. Each vectorial components of the bare force is then decomposed into a basis of standard (scalar) spherical harmonics, each of which couples between several of the original tensorial-harmonic modes, and the outcome is summed over $m$ for a given $\ell$ (where $\ell$ now labels the scalar harmonic). The resulting quantity, evaluated on the particle, is the ``$\ell$-mode bare force'', denoted $F^{\alpha\ell}_{\pm}$, where the two signs correspond to an evaluation from $r\to r_{\rm p}^{\pm}$, which, in general, yields two different values. The total, physical self-force at each point along the orbit is then given by the mode-sum formula 
\begin{equation}\label{mode-sum}
F^{\alpha}= 
\sum_{\ell=0}^{\infty}
\left[ F^{\alpha \ell}_{\pm} - 
\left( \ell + \frac{1}{2} \right) A^{\alpha}_{\pm} - B^{\alpha} 
\right]\,,
\end{equation}
where $A^{\alpha}_\pm$ and $B^{\alpha}$ are the ``regularisation parameters'', first derived for Schwarzschild in \cite{Barack:2001gx}. For a fixed geodesic orbit (i.e., fixed $E,L$), $A^{\alpha}_\pm$ and $B^{\alpha}$ are simple, analytically given functions of $r_{\rm p}$ and $\dot{r}_{\rm p}$. The particular form of these functions in the ZEZO case ($E=\mu$ with $L=4\mu M$) can be directly read off the expressions given in \cite{Barack:2001gx}. In Eq.\ (\ref{mode-sum}), $F^{\alpha \ell}_{+} - 
(\ell+1/2) A^{\alpha}_{+}=F^{\alpha \ell}_{-} - (\ell+1/2) A^{\alpha}_{-}$, so the full summand is insensitive to the direction in which the limit $r\to r_{\rm p}$ is taken. At large $\ell$, the summand usually falls off as $\ell^{-2}$, and the mode sum converges as $\ell^{-1}$. 

In principle, the above scheme can be applied with little change for any kind of sourcing orbit (modulo complications with the monopole and dipole modes, discussed below), and in this work we apply it for the ZEZO.  As discussed, it will suffice, for our purpose, to consider a fixed, geodesic ZEZO orbit as a source of the perturbation, and there is no need to account for the orbit's self-acceleration. We do, however, need to calculate the {\it conservative} piece of the self-force (specifically, the components $F_t^{\rm cons}$ and $F_{\varphi}^{\rm cons}$), and, as also discussed, this requires the evaluation of the self-force along both iZEZO and oZEZO. This, in turn, required two separate numerical evolutions, once with the iZEZO as a source and again with the oZEZO as a source. The conservative components are then constructed post-process at each point along the orbit using Eq.\ (\ref{Fconsformula}).

We have developed two independent implementations of this approach. One represents an evolution of the original code by two of us (LB and NS) \cite{bsago1,bsago2}, and the other is an entirely new code developed by one of us (MC) for the purpose of the present calculation. While both codes use a similar method, the ability to cross-check our results provided much reassurance and has proven valuable. All numerical results to be presented in this paper have been confirmed using both codes.

%%%%%%%%%%%%%%%%%%%%%%%%%%%%%%%%%%%%%%%%%%%%%%%%%%%%%%%%%
\subsection{Implementation details}
\label{details}
%%%%%%%%%%%%%%%%%%%%%%%%%%%%%%%%%%%%%%%%%%%%%%%%%%%%%%%

%%%%%%%%%%%%%%%%%%%%%%%%%%%%%%%%%%
\subsubsection{Junk radiation}
%%%%%%%%%%%%%%%%%%%%%%%%%%%%%%%%%%%

\begin{figure}[htb]
  \begin{center}
    %\includegraphics[width=\columnwidth]{./graphics/junk.pdf} 
    %% logn-r data [Nori: 13/03/19]
        \includegraphics[width=\columnwidth]{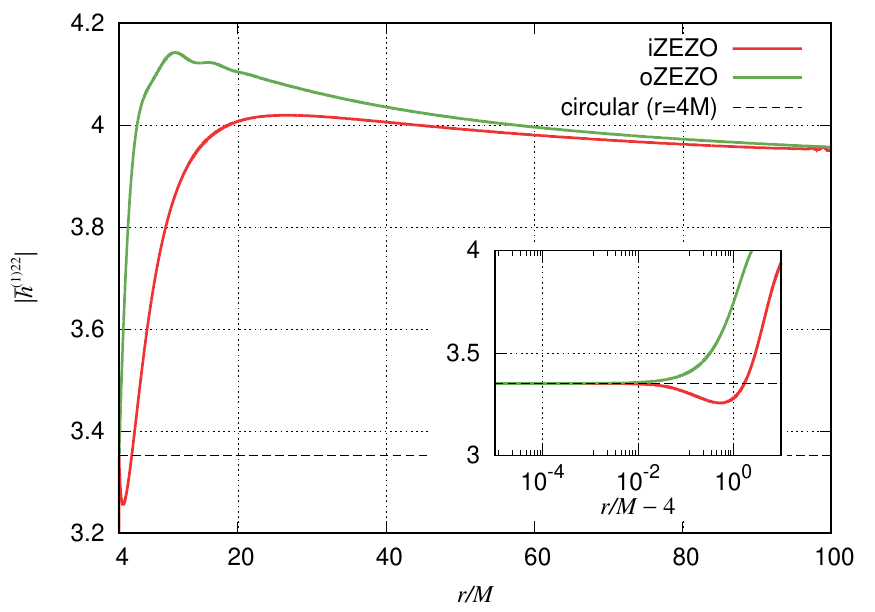} 
        %% short-r data  [Nori: 31/08/18]
    \caption{ Treatment of junk radiation, illustrated here for the mode $(i,\ell,m)=(1,2,2)$ of the perturbation (other modes exhibit a similar behavior). In the oZEZO case (green) we release the particle at $r_{\rm init}=(4+10^{-11})M$, letting junk radiation dissipate away while the particle is still in a tight circular whirl around the black hole; we then discard the $r_{\rm p}< (4+10^{-4})M$ portion of the data, which is dominated by junk. In the iZEZO case (red), the particle is released from $r=133M$, giving usable junk-free data for $r\lesssim 90M$. The thick horizontal dashed line marks the value of the perturbation mode on a strictly circular geodesic at $r=4M$ (the IBCO); reassuringly, the perturbations along both iZEZO and oZEZO asymptotically approach this value, as desired.  }
    \label{fig:junk}
  \end{center}
\end{figure}

In previous implementations, for bound (periodic) orbits, it was shown that initial junk radiation usually subsides to negligible levels after one or two periods of orbital revolution. Owing to the periodicity of the setup, one can then simply read off the relevant self-force data during the (say) third revolution period. Not so for the ZEZO, which is not periodic. Here, we deal with junk radiation in the following manner. In the case of the oZEZO, we simply start our orbit very close to $r_{\rm p}=4M$, so that it initially performs a good number of near-circular whirl orbits, letting all junk radiation dissipate away before the particle leaves the IBCO. Starting at $r_{\rm init}=(4+\epsilon)M$, the number of subsequent whirl orbits is $\propto\ln(1/\epsilon)$, and in practice we have found that taking $\epsilon=10^{-11}$  suffices for ensuring junk-free data in the range $r_{\rm p}\gtrsim (4+10^{-4})M$.
 Figure \ref{fig:junk} illustrates how this works for a particular mode of the perturbation. 

The iZEZO case is potentially more problematic. Here, initial junk contaminates an important part of the data at large $r_{\rm p}$, and there appears to be no way of mitigating this, except, possibly, via direct filtering or by improving the initial data. However, we have found that even a (hypothetical) complete elimination of the junk would only have a marginal effect on the accuracy of our calculation, for the following reason. Since, for the iZEZO, the infall time from $r=r_{\rm max}$ scales as $r_{\rm max}^{3/2}$, the run time of our 1+1-dimensions evolution code scales as $r_{\rm max}^3$. This puts a stringent constraint on how far out we can start our iZEZO orbit. In practice, given the computational resources committed within this project, we have found it computationally prohibitive to set $r_{\rm max}$ far above $\sim100M$. Taking $r_{\rm max}= 133M$ appeared to leave us with a clean, junk-free stretch of data in the range $r_{\rm p}\lesssim 90M$, as also illustrated in Figure \ref{fig:junk}. Truncating the integrals in Eqs.\ (\ref{E4}) and (\ref{E5}) at $r=r_{\rm max}$ produces a relative error of $O(1/r_{\rm max})$ in $\Delta E$ and $\Delta L$ (recall our discussion in the last paragraph of Sec.\ \ref{sec:MBMSsf}), which is not much larger for $r_{\rm max}=90M$ than it is for $r_{\rm max}=133M$. Thus, even a complete elimination of the junk would only mean reducing a (say) $1\%$ relative error to, perhaps, $0.7\%$. 
%Even if we improve the convergence of the integrals via a Richardson extrapolation, so that the truncation error becomes $\propto r_{\max}^{-2}$, still the improvement gained from a total elimination of initial junk would reduce the error by a factor of only 2 or so.  

We have therefore opted, for simplicity, to set $r_{\rm max}$ as far out as we practically could, and simply discard the junk-contaminated initial stretch of data. To measure the magnitude of residual junk, we have compared data from runs with varying values of $r_{\rm max}$. We have thus selected a usable stretch of data where the magnitude of junk was deemed smaller than that of other sources of numerical error. In practice, we have put the cuttoff at $r_{\rm max}=90M$.

%through numerical experiments we have found that the problem is sufficiently mundane in our case (i.e., starting at rest at infinity) that we need not resort to such measures here. Instead, we simply start the iZEZO orbit at sufficiently large $r_{\rm p}$ that the stretch of clean, junk-free data extends as far out as desired. In practice, we have found that starting the iZEZO orbit at $r_{\rm p}=150M$ leaves us with clean data for $r_{\rm p}\lesssim 100M$, which appeared sufficient for our purpose. 

%We speculate that the problem of initial junk is not as severe for the iZEZO as it probably is for higher-energy scattering problem, thanks to a combination of two mitigating factors. First, the amplitude of junk is probably relatively small to begin with, owing to the fact that the particle starts at rest at infinity, with a vanishing coordinate acceleration: we have $\dot{r}_{\rm p}\sim r_{\rm p}^{-1/2}$ and $\dot{\varphi}_{\rm p}\sim r_{\rm p}^{-2}$, with $\ddot{r}_{\rm p}\sim r_{\rm p}^{-2}$ and $\ddot{\varphi}_{\rm p}\sim r_{\rm p}^{-7/2} $ for large $r_{\rm p}$. Second, precisely because the particle is very slow at large $r_{\rm p}$, it has an ample time to ``shed'' the junk off itself; contrast this with the case of a highly relativistic particle, which falls in almost at the same speed as the junk radiation that accompanies it. Also to our aid is that fact that we are able to analytically predict the form of the self-force at large $r_{\rm p}$, reducing the range of $r_{\rm p}$ for which clean numerical data is necessary.  

To recap: we have run our oZEZO evolution with the orbit starting at $r_{\rm p}=(4+10^{-11})M$ and ending at $100M$; and we have run our iZEZO evolution with the orbit starting at $r_{\rm p}=133M$ and ending at $r_{\rm p}=(4+10^{-5})M$. This produced clean, sufficiently junk-free self-force data over the radial interval $(4+10^{-4})M \leq r_{\mathrm p} \leq 90M$. As we describe in the next section, an extrapolation for the self-force on $r_{\mathrm p} > 90M$ was obtained by fitting to an analytical power-law model. Similarly, the small whirl contribution from $4M< r_{\mathrm p} \leq (4+10^{-4})M$ was estimated using a simple extrapolation. The uncertainty from the large-$r$ fitting procedure ended up dominating the overall error budget in our calculation of $\hat\Omega$ and ${\hat L}$.

%%%%%%%%%%%%%%%%%%%%%%%%%%%%%%%%%%%%%%%%%%%%%%%%%%%%%%%%%%%
\subsubsection{Large-$\ell$ contribution to the mode sum}
%%%%%%%%%%%%%%%%%%%%%%%%%%%%%%%%%%%%%%%%%%%%%%%%%%%%%%%%%%%

Another unavoidable truncation involved in our calculation is that of the mode sum in Eq.\ (\ref{mode-sum}). The computation burden increases sharply with $\ell$, both because there are $2\ell +1$ $m$-modes to compute for each $\ell$, and because the resolution requirements fast increase with $\ell$.  Limited by computational resources, in this work we were able to calculate the first 16 (scalar-harmonic) modes, truncating the mode sum at $\ell_{\rm max}=15$. (This is comparable with $\ell_{\rm max}$ values taken in previous time-domain work for periodic orbits \cite{bsago1,bsago2}.) Because of mode coupling, obtaining the first 16 scalar-harmonic mode contributions required data for the first 18 tensor-harmonic $\ell$-modes. This, in turn, required a total of 648 separate numerical evolutions of individual $\ell,m$ modes: $2\ell+1$ evolutions for each $\ell$ and for each of the two orbits (i/oZEZO). 
%In the mode-sum scheme that we adopt here, 
%$F^{\mu}_{\mathrm {cons}}(r_{\mathrm p})$ is 
%constructed from an infinite sum of suitably regularised spherical harmonics 
%modes.  

A straight truncation of the mode sum at $\ell=\ell_{\rm max}$ would produce a very large relative error, of $O(\ell_{\rm max}^{-1})$. Instead, we follow the method of Ref.\ \cite{bsago2}, in which an approximation is obtained for the truncated modes by fitting the summand in Eq.\ (\ref{mode-sum}) to an expression of the form $a_{0}/(\ell+1/2)^2+a_{1}/(\ell+1/2)^4$, where $a_{0}$ and $a_{1}$ are fitting parameters (see \cite{bsago2} for details, including a motivation for this form). This extrapolation procedure effectively brings the truncation error of the mode sum down to $O(\ell_{\rm max}^{-5})$, which, for $\ell_{\mathrm {max}} = 15$, translates to $\sim O(10^{-6})$. The error from the large-$\ell$ tail fitting procedure was estimated from the covariance matrix of the fitting parameters, and found to be subdominant in our calculation (as compared to the error from the integral truncation, discussed above).

%The above tail-fitting procedure leaves one with an error on the GSF of order 
%$\sim \sum_{\ell = \ell_{\mathrm {max}}+1}^{\infty}{\ell^{-6}} 
%= O(\ell_{\mathrm {max}}^{-5})$: when $\ell_{\mathrm {max}} = 15$, 
%this translates into an absolute error $\sim O(10^{-6})$, 
%which we can afford to tolerate in this work.

%%%%%%%%%%%%%%%%%%%%%%%%%%%%%%%%%%%%%%%%%%%
\subsubsection{Numerical convergence}
%%%%%%%%%%%%%%%%%%%%%%%%%%%%%%%%%%%%%%%%%%%

As mentioned, we have used the fourth-order-convergent finite-difference scheme developed in Ref.\ \cite{bsago2}, as detailed in Sec.\ III.B of that paper. This means that our field variables $\bar h^{\ell m(i)}$ are designed to converge with a finite-resolution residual that scales as $O(\Delta^4)$, where $\Delta\times\Delta$ are the coordinate dimensions of a single grid cell [in null coordinates $v=t+r_*$ and $u=t-r_*$, where $r_*=r+2M\ln[r/(2M)-1]$]. To achieve this global convergence property,  our finite-difference formula has a {\em local} error of $O(\Delta^6)$ in vacuum points away from the particle, and $O(\Delta^5)$ on the particle and its immediate vicinity (see \cite{bsago2} for details). The latter is achieved with the help of suitable jump conditions for $\bar h^{\ell m(i)}$ and its first four derivatives, which Appendix E of \cite{bsago2} gives analytically for generic geodesic orbits.

By running our ZEZO codes several times with varying resolution ($\Delta=\{0.32,0.16,0.08,0.04,0.02\}M$), we have convinced ourselves that (1) our two codes each has the intended fourth-order global convergence, and that (2) with the highest resolution in the set, the error from the finite-difference approximation is sub-dominant in the total error budget (the total error being dominated by integral truncation). 

%%%%%%%%%%%%%%%%%%%%%%%%%%%%%%%%%%%%%%%%%%%%%%%%%%%%%%%%%%%%%%%%%%%%%%%%%%%%%%%
\subsection{Monopole and dipole modes}
\label{subsec:low-ell}
%%%%%%%%%%%%%%%%%%%%%%%%%%%%%%%%%%%%%%%%%%%%%%%%%%%%%%%%%%%%%%%%%%%%%%%%%%%%%%%

For all modes $\ell\geq 2$ we find a stable numerical evolution with our fourth-order-convergent finite-difference scheme. Moreover, in the case of the iZEZO, the numerical solutions all appear to  approach at late time the same solution one obtains for an evolution sourced by a strictly circular orbit of radius $r=4M$---as expected.  

Unfortunately, the modes $\ell=0,1$ do not behave in this manner, and have to be tackled separately. The problem with the odd-parity dipole mode $[(\ell,m)=(1,0)]$ is a minor one: the mode does evolve stably, and the iZEZO evolution does reproduce the circular-orbit solution at late time, but in the oZEZO case the numerical solution appears to contain a gauge mode that is irregular at the event horizon. Our simple solution to this problem is described further below. 

\begin{figure}[htb]
  \begin{center}
    \includegraphics[width=\columnwidth]{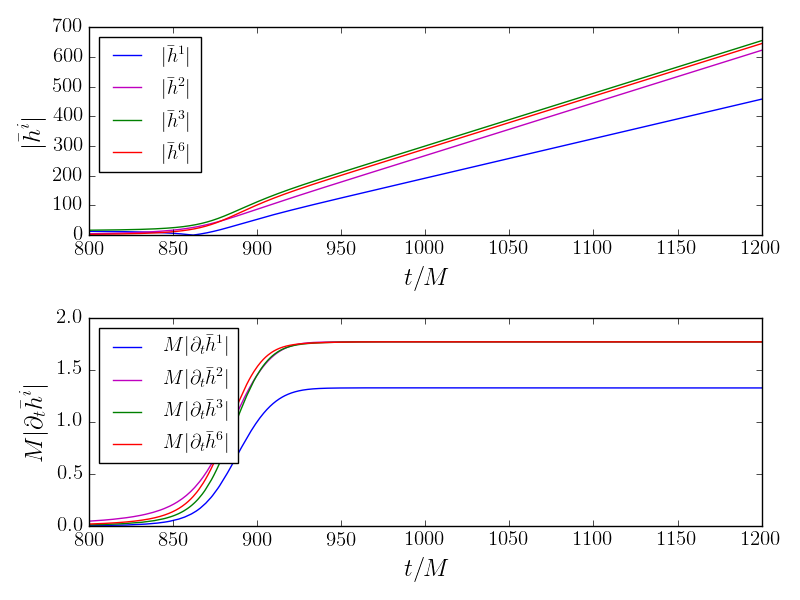}
    \caption{Raw numerical data for the monopole ($\ell=0$) mode along the iZEZO orbit, as it settles into a circular whirl (at around $t\sim 920M$).
    %in a time window where the orbit spans the range $22 M \lesssim r_p\lesssim (4 +10^{-12})M$. 
    The upper and lower panels show our numerical variables $\bar h^{(i)00}$ (refer to the first paragraph of Sec.\ \ref{review}] and their first time derivatives on the particle, respectively. During the whirl, we expect the metric perturbation to assume a constant value on the orbit (in any reasonable gauge); we see instead the characteristic behaviour of a linear-in-time gauge mode, evidently present in the data.}.
    \label{fig:mono}
  \end{center}
\end{figure}

The problem with the monopole $[(\ell,m)=(0,0)]$ and even-party dipole $[(\ell,m)=(1,\pm 1)]$ is more acute: the numerical solutions are found to develop a linear growth in $t$ during the circular whirl (at any fixed $r$, including on the orbit), which is clearly unphysical. This behavior, illustrated in Fig.\ \ref{fig:mono}, is similar to that observed in previous time-domain implementations for circular and other bound orbits, and also in vacuum. It was thoroughly analysed in Ref.\ \cite{Dolan}, where it was attributed to certain (analytically identifiable) homogeneous gauge modes that satisfy both the Lorenz gauge conditions and regularity conditions at infinity and on the horizon. They thus represent a true ambiguity in the Lorenz-gauge solution, unless additional conditions are imposed (such as regularity at $i^\pm$, or, when appropriate, helical symmetry). In Refs.\ \cite{bsago1,bsago2}, this problem was circumvented simply by computing these two troublesome modes in the frequency domain, where a suitable periodicity condition can be explicitly imposed, to the effect of disallowing any linear-in-$t$ behavior. Ref.\ \cite{Dolan} sought to resolve the issue in a time-domain framework, making considerable progress via a combination of gauge-damping techniques and direct post-process filtering. However, the method of \cite{Dolan} is customized specifically to circular orbits. 
Others have been working towards more systematic solutions to the problem \cite{JT_capra21}, but these ideas are yet to fully mature. Here we will present our own remedy, customized specifically to the ZEZO problem, but making crucial use of the analytical solutions obtained in \cite{Dolan}.

In what follows we discuss each of the three problematic modes in turn. We start with the most straightforward case, that of the odd-parity dipole mode.

%%%%%%%%%%%%%%%%%%%%%%%%%%%%%%%%%%%%%%%%%%%
\subsubsection{The mode $(\ell,m)=(1,0)$}
\label{sub:dipodd}
%%%%%%%%%%%%%%%%%%%%%%%%%%%%%%%%%%%%%%%%%%%%

This mode encapsulates any angular-momentum perturbation to the background Schwarzschild  geometry. It is uniquely fixed by the combination of (i) the Lorenz-gauge condition, (ii) regularity at infinity and on the horizon, and (iii) conditions on the (ADM) angular-momentum of the large black hole and of the entire spacetime. The latter can be conveniently imposed using the Abbott-Deser formalism of conserved integrals \cite{AbbDes}, applied on the (unperturbed) horizon and at $i^0$ (see \cite{Dolan}, where this method was introduced in the current context). Specifically, we demand that the black hole has zero angular momentum [through $O(\eta)$], and that the full spacetime has angular momentum $L$ [through $O(\eta)$]. As far as we know, this mode does not admit any linear-in-$t$-type solutions. 

In the case of {\em circular} (geodesic) orbits, the unique $(\ell,m)=(1,0)$ solution satisfying the above conditions can be written down analytically. Specialized to the IBCO ($R=4M$), is reads
\begin{align}
h_{t\varphi} &= 
 - \mu\sin^2\theta \times 
\begin{cases}
 \frac{1}{8}{r^2}/{M^2}\,, \quad  & r < 4M \,, \nonumber \\
 8M / r\,, \quad & r > 4M \,, 
\end{cases} \\
\label{L01:barh8}
h_{r\varphi}
&= 
-\frac{2\mu M^2 \sin^2\theta}{r^2-2Mr}\, ,
\end{align}
with all other components equal to zero. Despite appearance, this solution is physically regular at the event horizon, in the sense that its components are regular (smooth) there in any horizon-regular frame.\footnote{In some previous related work \cite{baracklousto, lowmodes,Dolan}, a different Lorenz-gauge solution was adopted, legacy of Zerilli's work \cite{Zerilli}. That solution, which differs from ours by a gauge transformation, is, however, physically irregular at the horizon: introducing advanced Eddington-Finkelstein coordinates $(v=t+r_*,\tilde r=r,\tilde\theta=\theta,\tilde\varphi=\varphi)$, one finds for that solution $h_{\tilde r\tilde\varphi}\propto (r-2M)^{-1}$ near the horizon. It is easily checked that, in contrast, our solution (\ref{L01:barh8}) is perfectly smooth in these coordinates. }

We expect our numerical iZEZO $(\ell,m)=(1,0)$ perturbation to approach the solution (\ref{L01:barh8}) at late time, after the orbit has settled into near-circular motion. We find empirically, and reassuringly, that this is indeed the case. Thus, we find, the odd-parity dipole mode with an iZEZO source is amenable to time-domain evolution (using our particular scheme), without any problem. However,  in the oZEZO case, we find (starting, as usual, with zero initial data) that the solution does not spontaneously settle into (\ref{L01:barh8}) during the initial whirl, but instead it settles into a different solution that is not horizon-regular. A cure to this problem immediately suggests itself: simply use (\ref{L01:barh8}) as initial conditions for the oZEZO evolution. Implementing this cure, we indeed find that the evolution is correctly ``guided'' towards the desired, horizon-regular solution.

%%%%%%%%%%%%%%%%%%%%%%%%%%%%%%%%%%%%%%%%%%%%
\subsubsection{The mode $(\ell,m)=(0,0)$}
\label{sub:monopole}
%%%%%%%%%%%%%%%%%%%%%%%%%%%%%%%%%%%%%%%%%%%

This mode encapsulates any mass perturbation to the Schwarzschild background geometry. We again impose the Lorenz-gauge conditions and regularity at infinity and on the horizon, and supplement these with conditions on the ADM mass of the central black hole and of the entire spacetime: specifically, we require that the Abbott-Deser mass integral is $M$ when evaluated on the horizon and $M+E$ when evaluated at infinity. As already noted, these conditions alone specify the perturbation only up to certain linear-in-$t$ homogeneous gauge modes that are everywhere regular (except at $i^\pm$). These are eliminated, and a unique monopole solution is finally fixed, with a boundedness condition at $i^\pm$. In the case of a circular (geodesic) orbit, this static solution---call is $M_{\alpha\beta}^{\rm circ}(r;R)$---can be written down analytically as a function of the orbital radius $R$; the expressions, which are rather lengthy, can be found in Sec.\ III.D of \cite{baracklousto}.

We have found that, in the iZEZO evolution, the monopole perturbation does not settle to the static solution $M_{\alpha\beta}^{\rm circ}(r;R)$ at late time as desired, but rather it grows linearly in $t$; see Fig.\ \ref{fig:mono}. Similarity, the oZEZO evolution with zero initial data  shows a linear-in-$t$ growth during the initial whirl, when stationarity is expected. However, in the oZEZO case, starting with the solution $M_{\alpha\beta}^{\rm circ}(r;R)$ itself as an initial condition seems to provide a sufficient remedy: the solution appears to be stationary all through the initial whirl, with no sign of the problematic linear mode manifesting itself in the data. 

We cannot apply a similar remedy in the iZEZO case, where the physical initial conditions are not known. Instead, we resolve the issue at a post-processing level, taking advantage of the analytical insight given in \cite{Dolan} about the form of the problematic linear mode. There, a Lorenz-gauge homogeneous monopole solution was analytically derived, having all the properties of the linear mode that appears to contaminate the data:  it is linear in $t$ but has a constant trace; it is a pure gauge mode and has a zero Abbott-Deser mass; and it is physically regular on the horizon. This solution reads (setting $\mu=1=M$ for brevity) 
\footnote{We correct here a typo in $h_{tr}^{\rm lin}$ in Eq. (128) of Ref.\ \cite{Dolan}.}
%%%%%%%%%%%%%%%%%%%%%%%%%%%%%%%%%%%%%%%%%%%%%%%%%%%%%%%%%%%%%%%%%
%\bl{[Sis: typo in $M_{tr}^{\rm lin}$ is fixed.]}
\begin{align}\label{monopole:lin}
M_{tt}^{\rm lin} &=A\frac{-r^4+4(t-t_0)+r^2+4r+8\ln(rf)}{r^4},
\nonumber \\
M_{tr}^{\rm lin}&=A\frac{3(t-t_0)-3+6\ln(2f)}{3r^2f}=M_{rt}^{\rm lin}
,\nonumber \\
M_{rr}^{\rm lin}&=A\frac{4(t-t_0)(2r-3)+5r^2-12r+8(2r-3)\ln(rf)}{r^4 f^2},
\nonumber \\
M_{\theta\theta}^{\rm lin}&=-A\frac{4(t-t_0)+r^2+4r+8\ln(rf)}{r}=\frac{M_{\varphi\varphi}^{\rm lin}}{\sin^2\theta},
\end{align}
where $f:=1-2M/r$, $A$ and $t_0$ are arbitrary parameters, and all other components vanish. The idea is to identify the mode $M_{\alpha\beta}^{\rm lin}$ in our iZEZO evolution data, with $A$ and $t_0$ accurately fitted for, and then simply subtract it off. 

To identify $M_{\alpha\beta}^{\rm lin}$ in the data, we choose a late-time $t=$const slice of the numerical solution, such that the entire slice is contained in the future light-cone of a worldline point where the orbit can be said to be essentially circular [say, a point with $r_{\rm p}=(4+10^{-3})M$]. We wish to demonstrate that, on such a slice, the data is consistent with, simply, $M_{\alpha\beta}^{\rm circ}(r;4M)+M_{\alpha\beta}^{\rm lin}(t,r;A,t_0)$, for some $A$ and $t_0$. We found it convenient to do this by first looking at the $t$ derivative of the numerical solution near the horizon: since $M_{\alpha\beta}^{\rm circ}$ is $t$-independent, and recalling (\ref{monopole:lin}), we expect to find
\begin{equation}
\partial_t\left\{M_{tt},fM_{tr},f^2M_{rr},M_{\theta\theta}\right\}\sim \frac{1}{4}A\{1,1,1,-8\} 
\end{equation}
for some $A$.
This we indeed find, and from this asymptotic form we extract the amplitude parameter $A$. (In practice, we extract $A$ independently from each of the 4 independent metric components, and then average.) With $A$ now known, we next determine the time shift parameter $t_0$ by fitting the entire solution $M_{\alpha\beta}^{\rm circ}(r;4M)+M_{\alpha\beta}^{\rm lin}(t,r;A,t_0)$ to the numerical data on the chosen late-time slice. Finally, we clean the data by subtracting the fitted solution $M_{\alpha\beta}^{\rm lin}$. 

\begin{figure}[htb]
 	\centering
 		\includegraphics[width=\columnwidth]{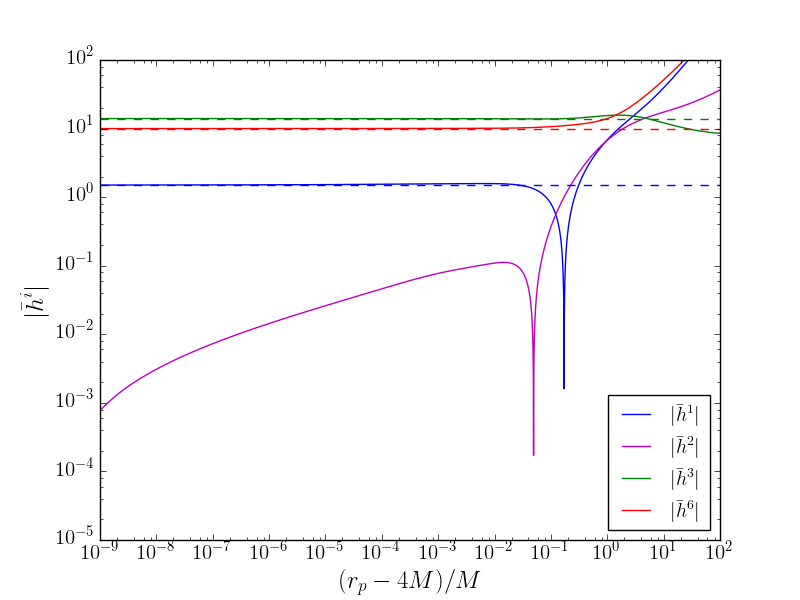}
 		 \caption[ Numerical filtering for the monopole solution along the iZEZO.]{\label{fig:h00R} Numerical filtering of the monopole solution along the iZEZO. The plot shows our numerical metric variables, evaluated on the orbit, after filtering out the linear-in-\textit{t} gauge mode present in the raw data. Horizontal dashed lines mark the (constant) values of the metric variables on the IBCO. Reassuringly, these values are approached as the orbit settles into a circular whirl at the IBCO. (Note the variable $h^{(2)}$ is zero for the IBCO; the residual value of the numerical $h^{(2)}$ solution can serve as an error estimate.)} 
 \end{figure}

As a check, we have verified that the resulting filtered solution is perfectly stationary, and that, moreover, it is consistent with the analytical solution $M_{\alpha\beta}^{\rm circ}(r;4M)$ over the entire whirl phase (and not only on our selected time slice). This is shown in Fig.\ \ref{fig:h00R}.

Our filtered monopole perturbation is fed into the mode-sum formula for the self-force. Since, by construction, our monopole perturbation coincides with the standard circular-orbit Lorenz-gauge solution at late (iZEZO) or early (oZEZO) times, we expect it to exhibit the anomalous feature described in Eq.\ (\ref{htt_asymp}), i.e.\ $h_{tt}\to -\eta$ for $r\to\infty$; and, since this feature is attributed to a static piece of the solution, we expect to see this novanishing limit at all times (not only during the whirl). We have indeed verified this against our data. In our calculation of $\hat\Omega$ and $\hat L$ we shall therefore have to apply the gauge adjustment described in Secs.\ \ref{subsec:gauge} and \ref{Ladj}.

%%Ffidown

%%%%%%%%%%%%%%%%%%%%%%%%%%%%%%%%%%%%%%%%%%%%%%%%%%%%%%%%%%%
\subsubsection{The modes $(\ell,m)=(1,\pm 1)$}
\label{sub:dipeven}
%%%%%%%%%%%%%%%%%%%%%%%%%%%%%%%%%%%%%%%%%%%%%%%%%%%%%%%%

As noted in Sec.\ \ref{CoM}, the even-parity dipole mode of the perturbation is pure gauge in vacuum, and (except on the particle) can be locally derived from a gauge generator $\xi_{\alpha}$ via $h_{\alpha\beta}=\nabla_{\alpha}\xi_{\beta}+\nabla_{\beta}\xi_{\alpha}$. The Lorenz-gauge and regularity conditions do not on their own specify a solution: they leave a freedom of gauge-shifting the CoM location and adding linear-in-$t$ modes. In the case of (geodesic) circular orbits, a unique Lorenz-gauge, regular, stationary and CoM-centered solution was constructed semi-analytically in Ref.\ \cite{lowmodes}, to be referred to here as $D_{\alpha\beta}^{\rm circ}(R)$. (This is a CoM-centered solution on account of the fact that the only CoM-shifting mode, $\xi^-_{\alpha(3)}$, has no support at $i^0$ within this solution---recall our discussion in Sec.\ \ref{CoM}.) We wish our numerical solution to coincide with $D_{\alpha\beta}^{\rm circ}(4M)$ at late time (for the iZEZO) or early time (for the oZEZO). However, in both cases we find the behavior to be dominated by linear-in-$t$ growth. In the oZEZO case we have tried to remedy this as we have done for the monopole, by using the correct circular-orbit solution, $D_{\alpha\beta}^{\rm circ}(4M)$, as initial data. However, for reasons that remain unclear to us, this does not seem to work in the dipole case: a linear growth becomes quickly manifest even with the correct initial conditions. 

In the dipole case, therefore, we have resorted to post-precess filtering for both the iZEZO and the oZEZO. Again, we make use of an explicit linear-in-$t$ solution derived analytically in Ref.\ \cite{Dolan}, which exhibits all the right characteristics: It is a pure gauge homogeneous perturbation that is globally regular and grows linearly in $t$, but whose trace remains stationary (in fact, zero), consistent with the empirical behavior of the numerical solution. The solution derived in \cite{Dolan}, which we call here $D_{\alpha\beta}^{\rm lin}(t,r;A,t_0)$, is generated by the gauge vector 
\begin{equation}\label{Dlin}
\xi^\pm_{\alpha} = A \left(\nabla_\alpha\Phi^{\pm}-2rf\delta^t_{\alpha}{\cal Y}_{\pm}\right) , 
\end{equation}
where ${\cal Y}_{\pm}:=\sin\theta e^{\pm i\varphi}$, and 
\begin{equation}
\Phi^\pm = \Big[(t-t_0)(r-M)+2M\left[2M+(r-M)\ln f \right]\Big] {\cal Y}_{\pm}.
\end{equation}
Here the signs correspond to $m=\pm 1$, and $A$ and $t_0$ are again arbitrary parameters. 

\begin{figure}[htb]
  \begin{center}
    \includegraphics[width=\columnwidth]{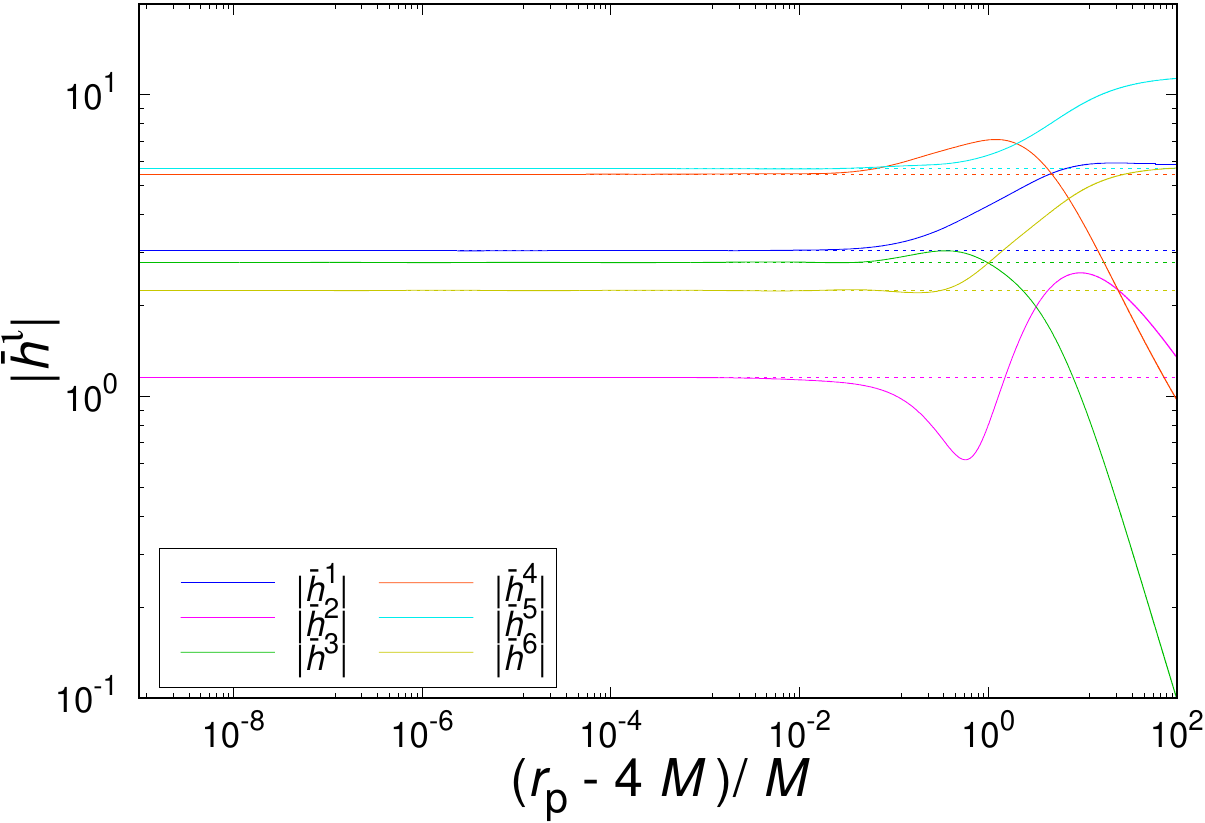}
    \caption[Numerical filtering of the $\ell=1,m=1$ mode for an inbound MBMS orbit.]{Numerical filtering of the even-parity dipole mode for the iZEZO. The plot shows the dipole field along the orbit, after subtraction of a suitable gauge mode with a generator of the form \eqref{Dlin}. Horizontal dashed lines mark the (constant) absolute values of the metric functions along the IBCO. Reassuringly, our filtered dipole solution approaches these values as the iZEZO settles into a circular whirl at the IBCO.}
    \label{fig:dipoleReg_iZEZO}
  \end{center}
\end{figure}

The filtering procedure proceeds as in the monopole case. For the iZEZO we select a suitable late-time $t$=const slice on which to fit for the parameters $A$ and $t_0$ against the numerical data. This time we also apply our filter in the oZEZO case, and for this we fit for $A$ and $t_0$ on a suitable early-time $t$=const slice, after the initial junk has subsided but well before the particle emerges from the whirl (ensuring the entire extent of the slice is contained within the future light cone of a whirl point on the worldline). In the oZEZO case we start from correct initial conditions, given by $D_{\alpha\beta}^{\rm circ}(4M)$. The fitted linear modes $D_{\alpha\beta}^{\rm lin}$ are then subtracted from the data, and we check that the filtered solution is stationary and consistent with $D_{\alpha\beta}^{\rm circ}(4M)$ during the whirl. Fig.\ \ref{fig:dipoleReg_iZEZO} shows the results of the filtering procedure for the iZEZO; similar results are obtained for the oZEZO.

%\begin{figure}
%  \begin{center}
%    \includegraphics[width=\columnwidth]{./graphics/h11Reg.png}
%    \caption[Numerical filtering of the $\ell=1,m=1$ mode for the oZEZO.]{\marta need a %similar plot for the oZEZO}
%    \label{fig:dipoleReg_oZEZO}
%  \end{center}
%\end{figure}

Our filtered dipole perturbation is fed into the mode-sum formula for the self-force. Since, by construction, it coincides with the standard circular-orbit Lorenz-gauge dipole at $i^\pm$, it sets our Lorenz gauge to be a CoM-centered one. It will therefore be appropriate to use Eq.\ (\ref{calLLorenz}), which assumes a CoM-centered gauge, in our calculation of $\hat L$.

%%%%%%%%%%%%%%%%%%%%%%%%%%%%%%%%%%%
%%%%%%%%%%%%%%%%%%%%%%%%%%%%%%%%%%
\section{Results}
\label{results}
%%%%%%%%%%%%%%%%%%%%%%%%%%%%%%%%%%
%%%%%%%%%%%%%%%%%%%%%%%%%%%%%%%%%%

\begin{figure}[htb]
	\begin{center}
        \includegraphics[width=\columnwidth]{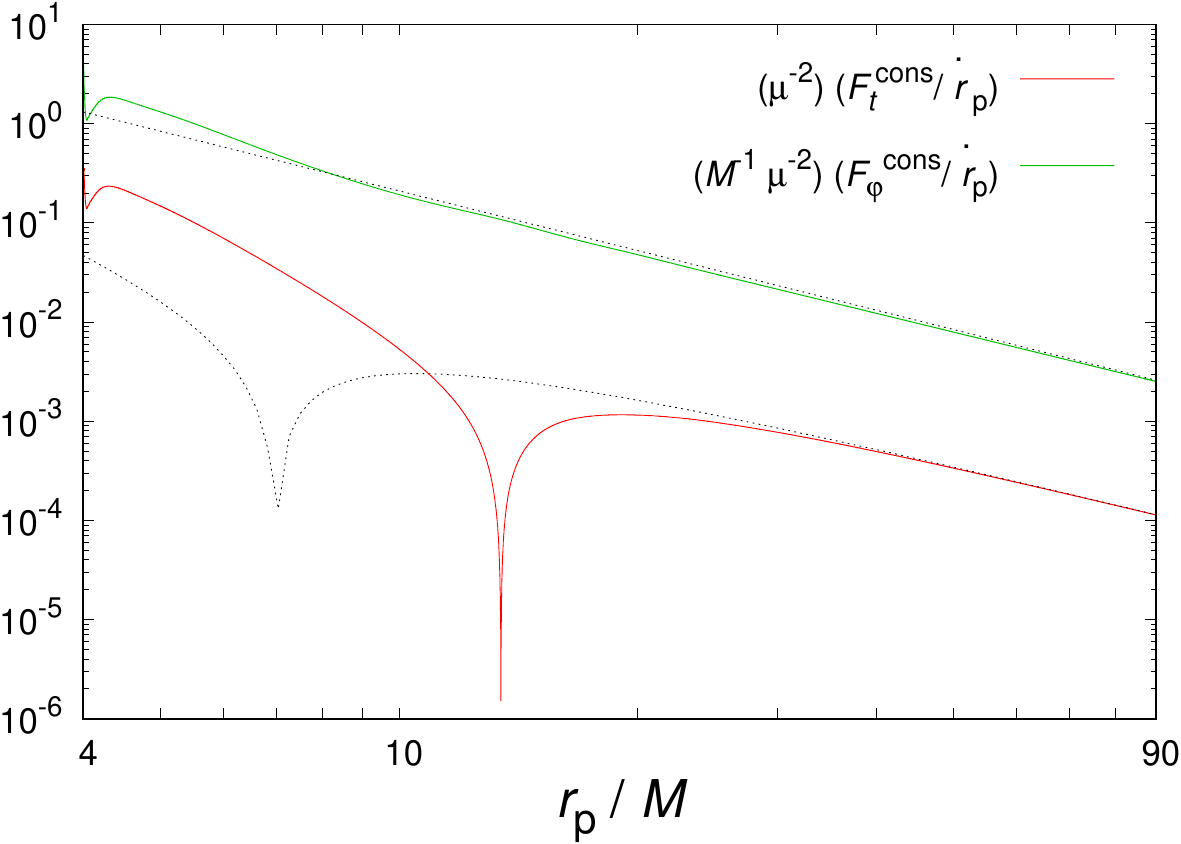}
\caption{Numerical results for the relevant self-force components, $F_{t}^{\mathrm {cons}}$ and $F_{\varphi}^{\mathrm {cons}}$. We present here, on a log-log scale,
$F_{t}^{\mathrm {cons}} / {\dot{r}_{\rm p}}$ and 
$F_{\varphi}^{\mathrm {cons}} / {\dot{r}_{\rm p}}$---the quantities that form the integrands in Eq.\ (\ref{E4}) and (\ref{E5})---as functions along the iZEZO orbit. The dashed curves are analytical fits to the asymptotic models (\ref{FtasymptFit}) and (\ref{FphiasymptFit}) at large $r$. 
The integrand $F_{\varphi}^{\mathrm {cons}} / {\dot{r}_{\rm p}}$ is negative throughout the domain, while $F_{t}^{\mathrm {cons}} / {\dot{r}_{\rm p}}$ flips its sign from positive to negative at $r_{\rm p}\sim 13.44M$.
%In the case of the $t$ component, the dominant, $\propto 1/r_{\rm p}^{2}$ fall-off represents a Newtonian contribution to the conservative self-force; see the discussion in Appendix \ref{App:Newtonian}.}
}
        \label{fig:Fcons}
	\end{center}
\end{figure}

Figure \ref{fig:Fcons} displays our numerical results for the self-force components $F_{t}^{\mathrm {cons}}$ and $F_{\varphi}^{\mathrm {cons}}$, as functions along the iZEZO orbit from $r_{\rm p}=r_{\rm max}=90M$ down to $r_{\rm p}=r_{\rm min}=(4+10^{-4})M$.  In the plot, the self-force components are shown divided by $\dot{r}_{\rm p}(<0)$, so as to form the integrands in Eq.\ (\ref{E4}) and (\ref{E5}); the quantities $\Delta E$ and $-\Delta L$ are then just the integrals with respect to $r_{\rm p}$, taken from $r_{\rm p}=4M$ to $r_{\rm p}=\infty$. We write each integral as a sum of three contributions, in the form  
\begin{equation}\label{split}
\Delta E=\Delta E_{\rm whirl}+\Delta E_{\rm num}+\Delta E_{\rm tail}
\end{equation}
(and similarly for $\Delta L$), corresponding to $\int_{4M}^{r_{\rm min}}$,  $\int_{r_{\rm min}}^{r_{\rm max}}$ and $\int_{r_{\rm max}}^{\infty}$, respectively.

The main contributions, $\Delta E_{\rm num}$ and $\Delta L_{\rm num}$, are obtained via numerical integration of our data. Note that our raw data is not uniformly sampled in $r_{\rm p}$, as the sampling intervals of the self-force along the orbit are inherited from the characteristic evolution grid. 
%
%To prepare the data for integration, we first interpolate it with a polynomial %fit. Then we integrate the interpolated data using Simpson's 3-point rule \mart%a. We estimate the error from this procedure by varying over the order of the p%polynomial fit and over the integration method. 
%
To prepare the data for integration, we first interpolate it with a cubic spline using \textit{Maple}'s \textsf{Spline}. Then we integrate the interpolated data using \textsf{evalf/Int} with appropriate controls to achieve sufficient integration precision. Each self-force data point comes with an error bar, estimated from a variation of numerical resolution ($\Delta$) and mode-sum cut-off ($\ell_{\rm max}$). The errors are combined in quadrature to estimate the total integration error. We obtain 
\begin{align}\label{DeltaEL_num}
\Delta E_{\mathrm {num}} &= 0.370111(2) \mu^2/M,
\nonumber\\
\Delta L_{\mathrm {num}} &= 5.86015(4) \mu^2.
\end{align}

The contributions $\Delta E_{\rm tail}$ and $\Delta L_{\rm tail}$ are obtained by fitting a large-$r_{\rm p}$ segment of the self-force data against analytic models of the form
\begin{align}\label{FtasymptFit}
\frac{F_t^{\rm  cons}}{\dot{r}_{\rm p}}&= 
-\frac{\mu^2}{r_{\rm p}^2}
\left(1+\alpha_t/r_{\rm p}+\cdots \right),
\\ \label{FphiasymptFit}
\frac{F_\varphi^{\rm  cons}}{\dot{r}_{\rm p}}&= 
\frac{ M\mu^2}{r_{\rm p}^2}\left(\alpha_\varphi+\beta_\varphi/r_{\rm p}+\cdots \right).
\end{align}
Here, the leading term of $F_t^{\rm  cons}$ represents a Newtonian-order contribution, whose form and coefficient can both be predicted using a simple asymptotic analysis---see Appendix \ref{App:Newtonian}. The form of the leading term of $F_\varphi^{\rm  cons}$ is strongly suggested from the numerical results (but we were not able to analytically calculate its coefficient). 
The error from the fitting procedure is estimated from the variation of the results under a change of the numerical data segment used for the fit, and of the number of terms included in the power-law fit models. Best-fit values for the leading coefficients in (\ref{FtasymptFit}) and (\ref{FphiasymptFit}) are 
$\alpha_t \simeq -7.0(7)$ and $\alpha_\varphi \simeq 21(1)$, giving
\begin{align}\label{DeltaEL_tail}
\Delta E_{\rm tail} &= -0.01068(4) \mu^2/M,
\nonumber\\
\Delta L_{\rm tail} &=  0.23(1) \mu^2. 
\end{align}

The final contributions to consider are $\Delta E_{\rm whirl}$ and $\Delta L_{\rm whirl}$. We do not have an analytical model of the behavior near the whirl, but we expect $F^{\rm cons}_t/\dot{r}$ and $F^{\rm cons}_\varphi/\dot{r}$ to be smooth functions of $r_{\rm p}$, approaching nonzero values for $r_{\rm p}\to 4M$ (these are IBCO values, which, unfortunately, we do not possess).  Thus, a rough estimate of these contributions is given by $\Delta  E_{\rm whirl}\cong \epsilon\times(F_{t}/\dot{r}_{\rm p})\big\vert_ {4M} $ and $\Delta L_{\rm whirl}\cong -\epsilon\times(F_{\varphi}/\dot{r}_{\rm p})\big\vert_ {4M}$, where  $\epsilon =10^{-4}$ is the radial extent of the whirl integration, and the IBCO values are estimated by extrapolating our numerical data to $r=4M$. We thus estimate
\begin{align}\label{DeltaEL_whirl}
\Delta E_{\rm whirl} &\simeq 0.00002(2)\mu^2/M,  
\nonumber\\
\Delta L_{\rm whirl} &\simeq -0.0001(1) \mu^2,
\end{align}
where the error bars conservatively bound the uncertainty from this procedure.

Finally, collecting our results (\ref{DeltaEL_num}), (\ref{DeltaEL_tail}) and (\ref{DeltaEL_whirl}), we obtain
\begin{align}\label{DeltaEL_values}
\Delta E^{(L)} &= 0.3594(1) \mu^2/M,
\nonumber\\
\Delta L^{(L)} &= 6.09(1) \mu^2 ,
\end{align}
where total errors where taken as simple sums of the three individual errors, conservatively. The superscripts $(L)$ remind us that these are Lorenz-gauge values.
We note that our fractional error in $\Delta L$ is an order of magnitude larger than that in $\Delta E$. This traces back to the fact that the leading-order term of $F^{\rm cons}_t$ at large $r$ (which is Newtonian) is known to us, whereas the leading-order term of $F^{\rm cons}_\varphi$ (which is post-Newtonian) is not. 
%where, recall, the tilded quantities are mass-rescaled and a-dimensionalized versions of $\Delta E$ and $\Delta L$. 

We now have at hand all the necessary input to obtain the IBCO frequency $\hat\Omega$ via Eq.\ (\ref{omegaSFLorenz}), and the  angular momentum $\hat L$ via Eq.\ (\ref{calLLorenz}). Substituting the numerical values from Eqs.\ (\ref{Fr}) and (\ref{DeltaEL_values}), we arrive at our final results as they are stated in Eqs.\ (\ref{Omega_final}) and (\ref{L_final}).

%
% %%%%%%%%%%%%%%%%%%%%%%%%%%%%%%%%%%%%%%%%%%%%%%%%%%%%%%%%%%%%%%%%%%%%%%%%%%%
% %%%%%%%%%%%%%%%%%%%%%%%%%%%%%%%%%%%%%%%%%%%%%%%%%%%%%%%%%%%%%%%%%%%%%%%%%%
\section{Comparison with (first-law-aided) EOB predictions}
\label{sec:EOB}
% %%%%%%%%%%%%%%%%%%%%%%%%%%%%%%%%%%%%%%%%%%%%%%%%%%%%%%%%%%%%%%%%%%%%%%%%%%
% %%%%%%%%%%%%%%%%%%%%%%%%%%%%%%%%%%%%%%%%%%%%%%%%%%%%%%%%%%%%%%%%%%%%%%%%%%%%
%
%

Ref.\ \cite{damour} derived from EOB theory the following simple theoretical predictions for the self-force-corrected angular momentum and frequency of the IBCO: \begin{eqnarray}\label{EOB1}
{\hat L} &=& 
4M \mu \left[1- 2 \, a\left(\frac14\right) \eta + O(\eta^2)\right],
\nonumber\\
{\hat \Omega} &=& (8 (M+\mu))^{-1} \left[1+ \frac12  a'\left(\frac14\right) \eta + O(\eta^2)\right]  \nonumber\\
&=& (8 M)^{-1} \left[1+ \left( \frac12  a'\left(\frac14\right) -1 \right) \eta + O(\eta^2)\right].
\end{eqnarray}
Here, the function $a(u)$ [with derivative $a'(u) := da(u)/du$]
 is the self-force correction to the main EOB radial potential $A(u;\nu)$, which is a $\nu$-deformed
avatar of the usual, $1-2 u$, Schwarzschild potential. Namely, $A(u;\nu) = 1- 2u + \nu a(u) + O(\nu^2)$, where $u= (M+\mu)/r_{\rm EOB}$, while $\nu:= \mu M/(M+\mu)^2 =\eta/(1+\eta)^2$ denotes the symmetric mass ratio. The argument $\frac14$ entering Eqs. \eqref{EOB1} is the value of $u$ at the unperturbed  IBCO.

As recalled in the introduction, at the time of Ref.\  \cite{damour}, the numerical values of $a(\frac14)$ and  $a'(\frac14)$ could only be approximately estimated by using PN theory, together with early results from self-force theory  and numerical relativity. [However, as we have indicated above, they do nicely agree with our accurate numerical results.] The later discovery of the  first law of binary black hole mechanics~\cite{LeTiec:2011ab}, and of its EOB reformulation \cite{Barausse:2011dq} provided an accurate way of numerically 
computing the function $a(u)$ in terms of the self-force contribution  to Detweiler's redshift. Ref.\ \cite{Akcay} computed a sample of accurate values of $a(u)$ over the interval $ 2/300 \leq u \leq 99/300$. The specific value $u= \frac14$ was not included in the study of Ref.\ \cite{Akcay}, but that work provided an accurate, global representation of the variation of the function $a(u)$ by means of several analytic models. One of the best analytical representations of $a(u)$ worked out in Ref.\ \cite{Akcay} is a 16-parameter model labelled as ``model 14" in Table II there. Using this analytical fit to $a(u)$, one gets
\begin{eqnarray} \label{EOB2}
a_{\rm model \, 14}\left(\frac14 \right)&=&0.15233714391(3) , \nonumber\\
a'_{\rm model \, 14}\left(\frac14 \right)&=&3.107206061(3) \,.
\end{eqnarray}
Here the error bar on $a_{\rm model \, 14}(\frac14)$ was estimated by comparing $a_{\rm model \, 14}(u)$
to the numerical values listed in Table IX of Ref.\ \cite{Akcay} for the neighboring values $u=74/300$, $u=76/300$.
The error bar on $a'_{\rm model \, 14}(\frac14)$ was estimated as the error that would result from 
the numerical errors listed in the last column of Table IX in Ref.\ \cite{Akcay} (treated as independent Gaussian errors) had one used a five-point stencil to estimate  $a'_{\rm model \, 14}(\frac14)$ from the four neighboring data points.

Inserting the numerical values in Eqs.\ \eqref{EOB2} in Eqs.\ \eqref{EOB1} yields
\begin{eqnarray}\label{EOB3}
{\hat L}&=&4M \mu \left[1-  0.30467428782(6) \eta + O(\eta^2)\right], 
\nonumber\\
{\hat \Omega}&=& (8 M)^{-1} \left[1+ 0.553603030(2) \eta + O(\eta^2)\right].
\end{eqnarray}
These are the  theoretical predictions from EOB theory, as computed through the crucial use of the first law. They agree with our result (\ref{Omega_final}), obtained via a direct integration of the self-force from infinity along the iZEZO. The agreement is well within the (larger) numerical error bars of our direct integration.

%
% %%%%%%%%%%%%%%%%%%%%%%%%%%%%%%%%%%%%%%%%%%%%%%%%%%%%%%%%%%%%%%%%%%%%%%%%%%%
% %%%%%%%%%%%%%%%%%%%%%%%%%%%%%%%%%%%%%%%%%%%%%%%%%%%%%%%%%%%%%%%%%%%%%%%%%%
 \section{IBCO frequency and angular momentum directly 
 from the first law of binary mechanics} \label{sec:1st}
% %%%%%%%%%%%%%%%%%%%%%%%%%%%%%%%%%%%%%%%%%%%%%%%%%%%%%%%%%%%%%%%%%%%%%%%%%%
% %%%%%%%%%%%%%%%%%%%%%%%%%%%%%%%%%%%%%%%%%%%%%%%%%%%%%%%%%%%%%%%%%%%%%%%%%%%%
%
%

In this section we provide an alternative, complementary derivation of $\hat\Omega$ and $\hat L$, by starting directly from the expressions [valid through $O(\eta^2)$] for the ``binding energy'' and angular momentum of a circular-orbit binary of black holes, as derived in Ref.\ \cite{LeTiec:2011dp} from the first law of binary black hole mechanics (hereafter ``the first law''). These expressions only require the values of the local (Detweiler's) redshift variable $\hat z(x):=1/\hat u^t(x)$, and its derivative $d\hat z/dx=:z'(x)$, through $O(\eta)$, on the circular orbit. The dimensionless variable $x$ (replacing $\Omega$ as 
a convenient gauge-invariant parametrization of circular orbits) is defined as\begin{equation}\label{x}
x:=\left[(M+\mu)\Omega\right]^{2/3} .
\end{equation}
[In the $\eta \to 0$ limit, the variable $x$ becomes equal to the EOB variable $u = (M+\mu)/r_{\rm EOB}$ used in the previous Section.] The energy and angular momentum were referred to in Ref.\ \cite{LeTiec:2011dp} as ``ADM'', though, as we saw above, we think that they should rather be viewed as the gravitational analogs of the Fokker-Wheeler-Feynman conserved energy and angular momentum, as appropriate to a conservative, time-symmetric dynamics. 

Ref.\ \cite{LeTiec:2011dp}'s expression for the total ADM mass of the circular-orbit binary spacetime reads [after suitable notational adjustments, and modulo a $O(M \eta^3)$ error term]
\begin{eqnarray}\label{EADM}
{\cal M}&=&M+\left(\frac{1-2x}{\sqrt{1-3x}}\right)\mu
\nonumber\\
&&+\left[\frac{x(1-6x)}{6(1-3x)^{3/2}}+\frac{1}{2}\delta z(x)-\frac{x}{3}\delta z'(x)\right]\frac{\mu^2}{M} ,
\end{eqnarray}
where 
\begin{eqnarray}
 \delta z(x) :=  \frac1\eta \left[\hat z(x)- \sqrt{1-3x}\right]
\end{eqnarray}
is the self-force piece of $\hat z$ at a fixed $x$. 
%\red{TD: There was an ambiguity between $\delta z$ being $O[\eta]$ or $O[1]$.}
The IBCO is identified via the requirement of ``zero binding energy'', i.e.\ ${\cal M}=M+\mu$---as in Eq.\ (\ref{ZEZOADM}) of Appendix \ref{app:ADM}. Imposing this in Eq.\ (\ref{EADM}) gives $x=1/4+\delta x$, with 
\begin{equation}\label{delta x}
\delta x = \frac{\eta}{24} \left[2 - 6\delta z(1/4)+\delta z'(1/4)\right].
\end{equation}
This is the self-force shift in the IBCO's inverse-radius $x$ away from the geodesic value of $1/4$. It was derived within
EOB theory in Ref.\ \cite{damour} with the result
\begin{equation}\label{deltaxEOB}
\delta x = \frac{\eta}{12} a'(1/4),
\end{equation}
showing, in passing, the link
\begin{equation}\label{a'vsz}
a'(1/4) = 1 - 3\delta z(1/4)+ \frac12 \delta z'(1/4),
\end{equation}
which is indeed a simple consequence of the general link between $\delta z(x)$ and $a(x)$ given in Eq. (2.14) of \cite{Barausse:2011dq}.

The values $\delta z(1/4)$ and $\delta z'(1/4)$ are [like $a(1/4)$ and $a'(1/4)$] gauge invariant (within a class of manifestly helically symmetric and asymptotically flat gauges), and can be obtained numerically with great accuracy using standard frequency-domain circular-orbit self-force codes. These values may be extracted from the Lorenz-gauge numerical results presented in \cite{Akcay}, but we quote here more recent, highly accurate values made available to us by M. van de Meent \cite{priv_comm_Maarten}, which were produced using the semi-analytical, radiation-gauge method of Ref.\ \cite{PhysRevD.94.044034}:
%
%% \red{TD: these values are WRONG !! It seems you forgot a packet of digits; as%% shown by my model14 results !!
%%I added my guesses in red for $\delta z$. For its derivative, I find a quite d%%ifferent result from $a'$ and your corrected $\delta z$ namely  9.04245784852.%% \red{?????} Is there a big problem in the derivative of model 14 or a big err%%or in your values ??? In addition, I think one should cite SOME TYPE OF ERROR %%BARS ON THESE VALUES.}
%
%\bl{[Sis: fixed due to MdvM (2019/08/14)]}
\begin{align}\label{zvalues}
\delta z(1/4) &= 0.804674287863142(6),
\nonumber\\
\delta z'(1/4) &= 9.0424578439(1).
\end{align}
%displaying only significant digits.
%\red{TD??: (We have checked that these values agree with the ones extracted fro%m \cite{Akcay}, to within the error bars of the latter.)}
Substituting these values in Eq.\ (\ref{delta x}) gives 
%\bl{[Sis: fixed.]}
\begin{equation}
\delta x = 0.258933838197(4).
\end{equation}

For the sake of comparison with our Eq.\ (\ref{Omega_final}), we need to express the IBCO shift in terms of $M \Omega$ rather than $x =\left[(M+\mu)\Omega\right]^{2/3}$. This leads, through $O(\eta)$, to
\begin{equation}
\hat\Omega = (8M)^{-1}\left[1+\frac{\eta}{4}\left[-2 - 6\delta z(1/4)+\delta z'(1/4)\right]\right].
\end{equation}
This is the direct first-law ``prediction'' for the IBCO frequency, as corrected by the first-order self-force. Using the link \eqref{a'vsz} it is seen to be totally equivalent to the EOB-derived expression \eqref{EOB1}. The numerical values in (\ref{zvalues}) then give
%\bl{[Sis: fixed.]}
\begin{equation}
\hat\Omega = 
(8M)^{-1}\left[1+ 0.55360302918(2) \eta\right].
\end{equation}
This agrees with our direct self-force result (\ref{Omega_final}), to within the (large) error bar of the latter.

Ref.\ \cite{LeTiec:2011dp} also gives an expression for the ADM angular momentum. Using our notation, it reads 
\begin{equation}
{\hat L} = \frac{M\mu}{x\sqrt{1-3x}}+\left[\frac{4-15x}{6\sqrt{x}(1-3x)^{3/2}}-\frac{1}{3\sqrt{x}}\delta z'(x)\right]\mu^2.
\end{equation}
On the IBCO, at $x=1/4+\delta x$, this evaluates to 
\begin{equation} \label{Lvsz}
{\hat L}
%&=& 4\eta+16\eta\delta x +\frac{2}{3}\eta^2 \left[1-\delta z'(1/4)\right]
%\nonumber\\
= 4M\mu\left[1+\frac{1}{2}\eta \left(1-2\delta z(1/4)\right)\right],
\end{equation}
where we have substituted for $\delta x$ from Eq.\ (\ref{delta x}). 
This is the direct first-law ``prediction'' for the angular momentum, as corrected by the first-order self-force. 
Comparing with \eqref{EOB1},
we get the link [which can also directly follows from Eq. (2.14) of \cite{Barausse:2011dq}]
\begin{equation}\label{avsz}
a(1/4) = \frac12 \delta z(1/4) - \frac14.
\end{equation}
Inserting the numerical values in \eqref{zvalues} into \eqref{Lvsz} gives 
\begin{equation}
{\hat L} = 4M\mu\left[1 -0.304674287863142(6) \eta\right],
\end{equation}
consistent with our direct self-force result \eqref{L_final}.
%The agreement, once more, is well within the numerical error bar. 

Let us finally note that the newly available redshift values \eqref{zvalues} translate, when using the links \eqref{a'vsz} and \eqref{avsz}, into the EOB values %\bl{[Sis: fixed.]}
\begin{align} \label{EOB4}
a({1}/{4}) &=  0.152337143931571(3), \nonumber\\
a'({1}/{4}) &= 3.10720605836(5) \,,
\end{align}
which agree, within the error bars, with the values \eqref{EOB2} deduced above from the accurate analytical fits of Ref.\ \cite{Akcay}.

%
% %
% %Our result is consistent with the recent results reported by Zimmerman~\etal, 
% %which also imply the validity of the first law holds for 
% %the circular-orbit binaries with a moderate mass ratio, 
% %making use of the full numerical-relativistic simulation
% %~\cite{Zimmerman:2016ajr}. 
%
%
%
%
%
% %%%%%%%%%%%%%%%%%%%%%%%%%%%%%%%%%%%%%%%%%%%%%%%%%%%%%%%%%%%%%%%%%%
% %%%%%%%%%%%%%%%%%%%%%%%%%%%%%%%%%%%%%%%%%%%%%%%%%%%%%%%%%%%%%%%%%%%
\section{Summary and discussion}\label{sec:conclusion}

We presented here a first direct calculation of two new physical quantities associated with the gravitational self-force in Schwarzschild spacetime. Ignoring dissipation and focusing on the conservative effect of the self-force, we numerically computed the $O(\eta)$ shift in the values of the critical angular momentum and the frequency of the asymptotic circular orbit (IBCO) for a finely-tuned zoom-whirl-type orbit that starts from rest at infinity. Our final results are stated in Eqs.\ (\ref{Omega_final}) and (\ref{L_final}). Our numerical error is of order $\sim 0.1\%$ for the frequency shift and $\sim 1\%$ for the angular-momentum shift, dominated by error from the truncation of the relevant self-force integral at large radius. 

An attractive feature of the marginally-bound ZEZO configuration considered here is that it admits well-defined notions of global angular momentum and binding energy, which involve the first-order self-force alone (with no reference to the second-order metric perturbation), as discussed in Sec.\ \ref{ADMdef}. This allows our results to be directly and unambiguously compared with corresponding results obtained in the framework of other approaches to the two-body problem, specifically EOB and the 1st-law of black hole binaries. We find an impressive agreement with the predictions of Ref.\ \cite{damour} using an early EOB model, and our results are in full agreement (within our error bars) with the later predictions of a much more accurate EOB model \cite{Akcay}, which was calibrated using self-force data along circular orbits and assuming the validity of the first law. A {\em direct} comparison with first-law predictions for the IBCO also shows a full agreement to within our error bars. This is significant, since no previous direct comparison has been made that deep inside the gravitational potential well: previous consistency tests were restricted to the exterior of the innermost circular stable orbit (ISCO, at $r=6M$), except the recent second-order self-force calculation of \cite{Pound:2019lzj}, which, however, quotes results only down to $r=5M$. The agreement illustrated here, at $r=4M$, reaffirms the now-well-established expectation that the first law provides (at least) a very good approximate description of the conservative dynamics even in the near-horizon region.

We caution, however, that our results here only test the accuracy of the first-law prediction to within our $\sim 1\%$ error bar. Interestingly, the recent direct calculation in Ref.\ \cite{Pound:2019lzj} of the circular-orbit binding energy using second-order perturbation theory reports a (numerically significant) deviation from the first-law predictions in the strong field: the apparent difference is at a level of $1\%$ around the ISCO, and $\sim 3\%$ at $r=5M$.  Ref.\ \cite{Pound:2019lzj} remains agnostic about the possible origin of this difference, noting that their set-up was quite different from the one considered in the 1st-law context: Ref.\ \cite{Pound:2019lzj}'s analysis was based on a fully systematic and fully GR-consistent two-timescale treatment of the perturbation equations for an adiabatically inspiralling object, including dissipation (or, in the case of orbits below the ISCO, a fine-tuned orbit on an adiabatic quasi-circular {\em outspiral}); the first-law, on the other hand, is a postulated variational formula that ignores dissipation. Ref.\ \cite{Pound:2019lzj} suggests that discerning the cause of the apparent discrepancy would require a better understanding of how the 1st-law formula might be generalized to account for radiation. Our results here, unfortunately, cannot shed new light on this matter, partly because our numerical error happens to be at the same, $\sim 1\%$ level of the reported discrepancy, partly because Ref.\ \cite{Pound:2019lzj} does not provide a result for $r=4M$, and partly because our treatment, too, ignores radiation. The issue provides motivation for work to improve the accuracy of our calculation.

At a more fundamental level, we have proposed here a precise {\em definition} of the notions of energy and angular momentum that feature in the first-law formula, valid for circular orbits below the ISCO. In this, we have taken advantage of the observation that such orbits are approached asymptotically by zoom-whirl-type orbits coming from infinity. We have thus argued that the 1st-law notions should be correctly interpreted as Fokker-Wheeler-Feynman-type quantities in a post-Minkowskian context, and as incoming-Bondi quantities in the context of perturbation theory. We have also suggested an effective interpretation in terms of ADM quantities in the full perturbed spacetime. It may be possible to extend these interpretations to circular orbits above the ISCO through an analytical-extension argument.  
 
Returning to the issue of numerical accuracy, let us discuss how it might be improved in future work.  Our error bars are predominantly from truncation of the self-force integrals $\Delta E$ and $\Delta L$ at $r_{\rm max}=90M$. As mentioned, the $\propto r_{\rm max}^3$ scaling of actual runtime is highly penalizing, so there is only a limited scope for a brute-force push to higher values of $r_{\rm min}$ using our existing numerical method. It is probably more productive, instead, to focus on obtaining an improved analytical formula for the behavior of the relevant self-force components at large $r$. In appendix \ref{App:Newtonian} we have taken a first step in that direction, deriving the leading-order, Newtonian term of the $t$ component, which already enabled us to reduce the truncation error (for $\Delta E$) by about an order of magnitude. To obtain a similar formula for the $\varphi$ components, and higher-order terms for both components, would require a systematic post-Newtonian or post-Minkowskian calculation, which we have left for future work. 

Our numerical method also encountered difficulties at the whirl end of the integration, in the form of bad convergence properties below around $r=4.0001M$. We have not been able to fully understand the cause for this failure, and so opted to simply truncate our numerical integration at that radius, replacing it with a rough extrapolation to the IBCO. It may be that a more sophisticated numerical method could be used to integrate further into the whirl. However, here too it may prove more productive to instead devise an analytical approximation for the self-force during the whirl, based on an expansion in the small parameter $r-4M$. Such an analysis could be modelled, for example, upon the method of Sec.\ V.B.2 of Ref.\ \cite{bsago2}, in which the perturbation equations themselves are expanded in a small parameter representing deviation from circularity. This calculation, too, we leave for future work. 

A step-function improvement in accuracy could also be achieved through a change of strategy for the numerical method. In the past few years there has been progress in the development of time-domain methods based on the Teukolsky formalism, with the idea of computing the self-force from a radiation-gauge metric perturbation constructed from numerical, time-domain solutions of the spin-$\pm 2$ Teukolsky equation \cite{GSFradgauge,Keidl:2006wk,PMB,Barack:2017oir,Bini:2019xwn}. This offers improved computational efficiency (since one has to solve a single scalar-like equation instead of 10 coupled equations in the Lorenz-gauge method), and also entirely circumvents the complications involved in computing the Lorenz-gauge monopole and dipole modes \cite{Merlin:2016boc,vandeMeent:2017fqk}. The implementation of this method in 1+1-dimensions appears to be numerically efficient even in the Kerr case, where mode-coupling has to be accounted for \cite{PacoThesis,priv_comm_Conor}. The method offers a promising alternative route to self-force calculations for unbound orbits, including a ZEZO configuration.  

Our ZEZO analysis provides but a first example of how interesting physics can be extracted from self-force calculations along unbound orbits. In future work one could consider the more general, one-parameter family of fine-tuned Schwarzschild orbits that start at infinity with some nonzero velocity and at $t\to\infty$ asymptote to an unstable circular orbit at radius $3M<r<4M$. Parametrizing such orbits by their initial $\gamma$ factor or energy, one could then calculate the conservative self-force-induced shift in the critical values of the angular momentum and asymptotic orbital frequency, just as in the ZEZO case. Such orbits are interesting because they probe the extremely strong gravitational field right down to the light ring. They will provide new, more challenging tests for the first-low formula, and set new benchmarks for EOB calibration (independent of the first law). A numerical code for tackling this kind of orbits could be developed from our existing codes in a straightforward manner. The only foreseeable issue is that of initial junk radiation at large $r$, which could be harder to deal with at large initial velocities and may require the development of suitable mitigation techniques (as the one employed in \cite{zimm}). We note, however, that the runtime scaling with $r_{\rm max}$ becomes slightly more favourable at nonzero initial velocity, scaling as $\propto r_{\rm max}^2$ (instead of  $\propto r_{\rm max}^3$ in the special case of the ZEZO).

Another interesting unbound configuration to consider is that of the two-parameter family of hyperbolic-type scatter orbits (this was first proposed by one of us in Ref.\ \cite{damour}). Here, one can compute the self-force correction to the scatter angle (as a function of, say, energy and impact parameter), providing an entirely new diagnostic of the post-geodesic dynamics in the strong field. Scatter orbits, too, can probe the black-hole geometry right down to the light ring. A unique advantage of scatter-angle calculations is that they can be performed {\em with or without} dissipation, thus providing a handle on both conservative and dissipative aspects of the dynamics. This also raises an interesting prospect for comparison with results from scatter-orbit simulations in full Numerical relativity \cite{Damour:2014afa,Ossokine:2015yla}. Finally, there has been much recent progress in quantum-field-theory ``amplitude'' calculations for the
gravitational scatter problem   (see \cite{Bern:2019nnu} and references therein). Self-force calculations of scatter angles can provide much-needed benchmarking for this program.

%%%%%%%%%%%%%%%%%%%%%%%%%%%%%%%%%%%%%%%%%%%%%%%%%%%%%%%%%%%%%%%%%%%
%%%%%%%%%%%%%%%%%%%%%%%%%%%%%%%%%%%%%%%%%%%%%%%%%%%%%%%%%%%%%%%%%%%%
\acknowledgments
We thank Eric Poisson, Adam Pound and Maarten van de Meent for useful discussions. We are also grateful to Maarten for providing us accurate redshift numerical data for the  analysis in Sec.~\ref{sec:1st}.  
Part of the research leading to his work received funding from the European Research Council under the European Union's Seventh Framework Programme (FP7/2007-2013)/ERC grant agreement No.\ 304978. 
LB and SI acknowledge additional support from STFC through Grant No.~ST/R00045X/1.
SI also acknowledges financial support of Ministry of Education, MEC, during his stay at IIP-Natal-Brazil, and he is grateful to Riccardo Sturani for his continuous encouragement.
MC acknowledges funding from the European Union's Horizon 2020 research and innovation programme, under the Marie Skłodowska-Curie grant agreement No.~751492.
NS thanks to JSPS Grant-in-Aid for Young Scientists (B), No.~25800154 and 
JSPS Grant-in-Aid for Scientific Research (C), No.~16K05356. 
%

%%%%%%%%%%%%%%%%%%%%%%%%%%%%%%%%%%
%%%%%%%%%%%%%%%%%%%%%%%%%%%%%%%%%%%
\appendix
%%%%%%%%%%%%%%%%%%%%%%%%%%%%%%%%
%%%%%%%%%%%%%%%%%%%%%%%%%%%%%%%

%%%%%%%%%%%%%%%%%%%%%%%%%%%%%%%%%%%%%%%%%%%%%%%%%%%%%%%%%%%%%%%%%%%%%%
\section{Gauge-invariant characterization of the ZEZO 
in terms of total ADM mass}
\label{app:ADM} 
%%%%%%%%%%%%%%%%%%%%%%%%%%%%%%%%%%%%%%%%%%%%%%%%%%%%%%%%%%%%%%%%%%%%%%%

In Sec.\ \ref{Sec:pertZEZO} we have defined the perturbed iZEZO via the coordinate condition $\dot{\hat r}_{\rm p}(t\to-\infty)= 0$ (in addition to a circularity condition at $t\to\infty$). This condition makes sense in a broad class of physically reasonable gauges, but it is, after all, gauge dependent. The purpose of this appendix is to comment that this condition can be replaced with a truly gauge-invariant condition on the total ADM mass of spacetime (or, in EOB or PN applications, the Fokker-Wheeler-Feynman-like mechanical mass), $\cal M$. In the ZEZO case, the two ways of specifying the orbit are equivalent (again, with suitable restrictions on the gauge), and equally convenient. However, the mass condition should do much better in avoiding ambiguity when dealing with hyperbolic-type orbits that start with a non-zero velocity at infinity. 

To speak of the  mass of the ZEZO spacetime, we must first address the problem, discussed in Sec.\ \ref{ADMdef}, that the ADM mass integral is mathematically ill defined for the time-symmetric ZEZO geometry.  Focusing on the iZEZO case, we resolve this as we did in Sec.\ \ref{ADMdef} for the angular momentum, by defining ${\cal M}$ either as the incoming Bondi mass (at $v\to\infty$) of a time-symmetric iZEZO, or as the ADM mass of the physical problem, with the full self-force and retarded boundary conditions, but with the same initial conditions as the for the time-symmetric iZEZO setup. In the latter case, we have a well defined notion of $\cal M$, calculable from the metric at $i^0$. The value of that $\cal M$ depends only on the initial conditions, near $i^-$,
when $r_{\rm p}\to\infty$, and can thus be derived using special-relativistic kinematics of point particles (as we did for $\hat L$ in Sec.\ \ref{sub2sec:IBCO-shift}). 

 In Ref.\ \cite{CB}, this method was applied to obtain an expression for $\cal M$, through $O(\eta^2)$, in terms of the quantity $\hat E(\infty):= \mu \hat u_t(r_{\rm p}\to\infty)$, for a particle falling from infinity with arbitrary initial conditions, and assuming $\hat E$ is given in a CoM-centered gauge.  It reads
\begin{equation}\label{def-ADM}
{\cal M}
= 
M +  {\hat {E}}(\infty) + 
\frac{1}{2M} \left({\hat {E}}^2(\infty) -\mu^2\right) +O(\eta^3),
\end{equation}
in which the first and second terms on the right are the black hole's and particle's ``rest masses'', respectively, and the third, $O(\eta^2)$ term accounts for both objects' initial ``kinetic energies'' in the CoM frame.  In the iZEZO case, the condition $\dot{\hat r}_{\rm p}(t\to-\infty)= 0$ (``no kinetic energy at infinity'') implies [recalling (\ref{MBMS-GSF}) with (\ref{norm-inf})] $\hat {E}(\infty)=\mu+O(\eta^3)$, so the ADM mass is, simply,
\begin{equation}\label{ZEZOADM}
{\cal M}
= 
M +  \mu +  O(\eta^3) ,
\end{equation}
as one expects intuitively. 

We can now reverse the point of view, and consider (\ref{ZEZOADM}) to be (part of) the {\it definition} of the ZEZO, in place of the condition $\dot{\hat r}_{\rm p}(t\to-\infty)= 0$ (the latter now being a consequence, valid within a class of gauges). This alternative specification of the ZEZO conditions is advantageous in that it is gauge invariant. (Note, however, that the particular form of the relation between $\cal M$ and the initial velocity, or $\hat E$, still, of course, depends on the gauge.)
In other words, we are now parametrizing the initial conditions in terms of the invariant quantity $\cal M$ (in addition to, say, $\hat L$), instead of the gauge-dependent velocity. The pair $\{{\cal M},{\hat L}\}$, we propose, provides a natural and convenient, gauge-invariant parametrization of unbound configurations of either the zoom-whirl or the scattering types.

%%%%%%%%%%%%%%%%%%%%%%%%%%%%%%%%%%%%%%%%%%%%%%%%%%%%%%%%%%%%%%%%%%%%%%
\section{Asymptotic behavior of the self-force at large $r$}
\label{App:Newtonian}
%%%%%%%%%%%%%%%%%%%%%%%%%%%%%%%%%%%%%%%%%%%%%%%%%%%%%%%%%%%%%%%%%%%%%%%

In this appendix we obtain an analytical prediction for the large-$r$ asymptotic behavior of the Lorenz-gauge self-force along the ZEZO orbit. The results provide a test of the numerical data, and are also used (in Sec.\ \ref{results}) for improving our estimation of the large-$r$ tail contribution to the self-force integrals that feature in our calculation [the quantities $\Delta E_{\rm tail}$ and $\Delta L_{\rm tail}$ introduced in Eq.\ (\ref{split})].

The idea behind our analysis is simple, and based on the assumption that the leading-order term of the conservative self-force at $r_{\rm p}\gg M$ comes entirely from expressing the usual $\propto r^{-2}$ Newtonian gravitational force in a suitable coordinate system (consistent with our Lorenz-gauge choice), and then identifying any resulting $O(\eta^2)$ terms as ``self-force''. The coordinate adjustment has two components: first, a transformation from the usual ``separation'' radial coordinate used in Newton's gravitation law to the CoM-centered radial coordinate employed in our Lorenz-gauge calculation; and, second, a gauge correction accounting for the non-asymptotic-flatness of the Lorenz gauge (discussed in Sec.\ \ref{subsec:gauge}). As we shall see, this predicts a ``Newtonian'', $\propto r^{-2}$ term of the Lorenz-gauge self-force, which we find to be in agreement with our numerical results. 

Following this strategy, we consider, for $r_{\rm p}\gg M$, a mapping of the true iZEZO orbit in Schwarzschild spacetime into an (accelerated) trajectory in flat space. The mapping is defined by identifying the Schwarzschild coordinates $x^{\alpha}_{\rm p}$ along the orbit with the usual polar coordinates (and time $t$) on flat spacetime, centered at the large black hole. The mapped trajectory experiences a Newtonian gravitational force with a 4-force counterpart
\begin{equation}\label{F_newt}
F^\alpha_{\rm Newt}= \mu(\delta_{\beta}^{\alpha}+u^\alpha u_\beta)a^\beta
=-\mu(\delta_{\beta}^{\alpha}+u^\alpha u_\beta)\Gamma^\beta_{\gamma\delta}u^{\gamma}u^{\delta}.
\end{equation}
Here, the spatial projection of $a^\beta:=d^2x^\beta/d\tau^2$ is the ``Newtonian'' gravitational acceleration in flat space, and $\Gamma^\beta_{\gamma\delta}$ are the Schwarzschild connections evaluated on the particle. Focusing first on the $r$ and $t$ components (the $\varphi$ components will be considered later), Eq.\ (\ref{F_newt}) gives, at leading order in $1/{r_{\rm p}}$,
\begin{equation}\label{FtrNewt}
F^r_{\rm Newt} \simeq -\frac{\mu M}{r_{\rm p}^2},\quad\quad
\frac{F_t^{\rm Newt}}{\dot{r}_{\rm p}} \simeq +\frac{\mu M}{r_{\rm p}^2},
\end{equation}
where $\dot{r}_{\rm p}\simeq -(2M/r_{\rm p})^{1/2}$, and the expressions are 
applicable to both oZEZO ($\dot{r}_{\rm p}>0$) and iZEZO ($\dot{r}_{\rm p}<0$). As expected, $F^r_{\rm Newt}$ has the standard form of the Newtonian force acting between two point masses.

In the expressions (\ref{FtrNewt}), the coordinate $r_{\rm p}$ represents the separation between the two masses; it is different from the Lorenz-gauge radial coordinate (also denoted by $r_{\rm p}$ in the bulk of this work), which is CoM-centered by construction---recall the discussion in Sec.\ \ref{CoM}. Let us, in this appendix only, denote the Lorenz-gauge radial coordinate along the orbit by $r_{\rm com}$, to distinguish it from the separation $r_{\rm p}$. At leading order, the two radii are related via $r_{\rm com}=(1-\eta)r_{\rm p}$. In terms of the CoM radial coordinate, the Newtonian force components thus become
\begin{eqnarray}\label{FtrNewtCoM}
F^r_{\rm Newt} &\simeq& -\frac{\mu M}{r_{\rm com}^2} + \frac{2\mu^2}{r_{\rm com}^2} ,
\nonumber\\
\frac{F_t^{\rm Newt}}{\dot{r}_{\rm com}} &\simeq& +\frac{\mu M}{r_{\rm com}^2}
- \frac{2\mu^2}{r_{\rm com}^2} ,
\end{eqnarray}
omitting terms of $o(\eta^2)$ and of $o({r}^{-2}_{\rm com})$. The $O(\mu^2)$ terms in Eqs.\ (\ref{FtrNewtCoM}) are interpreted as (conservative) self-force.
 
To obtain $F^\alpha_{\rm Newt}$ in the Lorenz-gauge, we must also account for the gauge pathology in the monopole, discussed in Sec.\ \ref{subsec:gauge}. We have seen that the Lorenz-gauge perturbation not asymptotically flat, but as simple monopole gauge transformation takes it to a (non-Lorenz) gauge that is manifestly asymptotically flat. The generator $\Xi^\alpha$ of the inverse gauge transformation (from the asymptotically-flat gauge to the Lorenz gauge) was given in Eq.\ (\ref{Xi}). It generates a gauge perturbation 
\begin{equation}
\delta_\Xi h_{\alpha\beta}= -\eta(1-2M/r)\delta_\alpha^t\delta_\beta^t
\end{equation}
[Eqs.\ (\ref{deltaXh_tt})--(\ref{deltaXh_phiphi}) with $(\alpha_1,\alpha_2,\alpha_3,\alpha_4)=(\eta/2,0,0,0)$]. It is straightforward to calculate the contribution to the self-force from this gauge transformation, either starting with $\delta_\Xi h_{\alpha\beta}$ and using Eq.\ (16) of \cite{gauge}, or starting with $\Xi^\alpha$ itself and using Eq.\ (6) of \cite{gauge}. Either way, the gauge correction (flat$\to$Lorenz) to the Newtonian self-force  works out as 
\begin{equation}\label{FtrNewtgauge}
\delta_\Xi F^r_{\rm Newt} = -\frac{\mu^2}{r_{\rm com}^2},\quad\quad
\frac{\delta_\Xi F_t^{\rm Newt}}{\dot{r}_{\rm com}} = +\frac{\mu^2}{r_{\rm com}^2},
\end{equation}
at leading order in $r^{-1}_{\rm com}$.
The (``Newtonian'' term of the) Lorenz-gauge self-force is the sum of the asymptotically-flat-gauge self-force from Eq.\ (\ref{FtrNewtCoM}) and the gauge correction (\ref{FtrNewtgauge}):
\begin{eqnarray}\label{FtrNewtLor}
\left(F^r_{\rm cons}\right)_{\rm Lor} &\simeq&  + \frac{\mu^2}{r_{\rm com}^2} ,
\nonumber\\
\frac{\left(F_t^{\rm cons}\right)_{\rm Lor}}{\dot{r}_{\rm com}} &\simeq& 
- \frac{\mu^2}{r_{\rm com}^2} .
\end{eqnarray}

The leading-order behavior expressed in (\ref{FtrNewtLor}) is found to be consistent with that of our numerical data, for both $r$ and $t$ components (the agreement is illustrated for the $t$ component in Fig.\ \ref{fig:Fcons}). As an additional check, we have confirmed that the leading, $r_{\rm p}^{-2}$ fall off of our numerical results comes entirely from the (tensor-harmonic) monopole and dipole modes of the metric perturbation (without those contributions, the numerical self-force is found to fall off as $r_{\rm p}^{-3}$ instead). This confirms our assumption that the $r_{\rm p}^{-2}$ term of the self-force is entirely due to a transformation to a CoM gauge (dipole mode) and a $\Xi$ transformation (monopole mode). 

Unfortunately, the leading-order fall-off of the self-force component $F_\varphi^{\rm cons}$, also needed in our analysis, cannot be determined using the above method. Given the $r$ and $t$ components of the Newtonian force (and recalling the $\theta$ component is zero for our orbit), we can attempt to obtain the $\varphi$ component directly from the orthogonality condition $u^\alpha F^{\rm Newt}_{\alpha}=0$, giving
\begin{equation}
\frac{F_\varphi^{\rm Newt}}{\dot{r}_{\rm com}}
=-r^2_{\rm com}\left(F^r_{\rm Newt}+
\frac{F_t^{\rm Newt}}{\dot{r}_{\rm com}} \right).
\end{equation}
From Eqs.\ (\ref{FtrNewtgauge}) and (\ref{FtrNewtLor}) we see, however, that the right-hand side here vanishes---at both $O(\mu)$ and $O(\mu^2)$---when inserting the leading-order Newtonian force. Hence, we can expect the leading-order term of $F_\varphi^{\rm const}$ to be post-Newtonian rather than Newtonian. We have not attempted here the post-Newtonian analysis required to extract that leading-order term. All we can say based on our Newtonian-order analysis (and assuming that the first post-Newtonian terms of $F^r$ and $F_t/\dot{r}_{\rm com}$ fall off at least as $r^{-3}_{\rm com}$), is that $F_\varphi^{\rm Newt}/\dot{r}_{\rm com}$ should fall off at least as $1/r_{\rm com}$. In fact, we numerically find a $r^{-2}_{\rm com}$ fall off. See Fig.\ \ref{fig:Fcons} and Eq.\ (\ref{FphiasymptFit}).

%%%%%%%%%%%%%%%%%%%%%%%%%%%%%%%%%%%%%%%%%%%%%%%%%%%%%%%%%%%%%%%%%%%%%%%%%%%%%
\section{General solution for the static piece of the even-parity dipole mode}
\label{App:static_dipole} 
%%%%%%%%%%%%%%%%%%%%%%%%%%%%%%%%%%%%%%%%%%%%%%%%%%%%%%%%%%%%%%%%%%%%%%%%%%%%%%

We give here explicitly the general solution of Eq.\ (\ref{Boxxi}) for the static even-parity dipole mode, i.e.\ the six-parameter family of homogeneous solutions $\xi_{\alpha(j)}^{\pm}$ ($j=1,2,3$). Our calculation of the CoM shift in Sec.\ \ref{CoM} involves only the four solutions $\xi_{\alpha(j)}^{+}$ and $\xi_{\alpha(3)}^{-}$, but for completeness we nonetheless give here all six. Five of the solutions (all but $\xi_{\alpha(3)}^{+}$) where given previously by Ori in \cite{Origauge}.

According to Eq.\ (\ref{xi}), each of the solutions $\xi_{\alpha(j)}^{\pm}$ is determined by three functions: $a^{\pm}_{(j)}(r)$, $b^{\pm}_{(j)}(r)$ and $c^{\pm}_{(j)}(r)$. These are given by the following expressions (where we have set $M=1$ for convenience; the missing factors of $M$ can be easily retrieved using dimensional analysis).
\begin{eqnarray}
a^-_{(1)}&=& r-2 ,
\nonumber\\
b^-_{(1)}&=& 0 ,
\nonumber\\
c^-_{(1)}&=& 0 ,
\end{eqnarray}
\begin{eqnarray}
a^-_{(2)}&=& 0 ,
\nonumber\\
b^-_{(2)}&=& 12r^2+6r-8/r+8\ln r ,
\nonumber\\
c^-_{(2)}&=& -6r^3+3r^2-8r-12+8(r-1)\ln r ,
\end{eqnarray}
\begin{eqnarray}
a^-_{(3)}&=& 0 ,
\nonumber\\
b^-_{(3)}&=& 1 ,
\nonumber\\
c^-_{(3)}&=& r-1,
\end{eqnarray}
\begin{eqnarray}
a^+_{(1)}&=& 2(r-1)/r+rf\ln f ,
\nonumber\\
b^+_{(1)}&=& 0 ,
\nonumber\\
c^+_{(1)}&=& 0,
\end{eqnarray}
\begin{eqnarray}
a^+_{(2)}&=& 0 ,
\nonumber\\
b^+_{(2)}&=& \frac{2(r-1)}{r^2 f}+\ln f ,
\nonumber\\
c^+_{(2)}&=& 2+(r-1)\ln f,
\end{eqnarray}
\begin{eqnarray}
a^+_{(3)}&=& 0 ,
\nonumber\\
b^+_{(3)}&=& \frac{6(r-1)(2r+1)-(6r^2-9r-4)r\ln r}{rf}
\nonumber\\
&& +\left[2f+3r(2r+1)+4\ln(r/4)\right]\ln(rf)+\Lambda(r) ,
\nonumber\\
c^+_{(3)}&=& -3r(2r+1)-5-\frac{1}{2}(6r^2-3r+4)r\ln f
\nonumber\\
&& +4\left[(r-1)\ln(r/4)-2\right]\ln(rf)
 + (r-1)\Lambda(r) . \nonumber\\
\end{eqnarray}
Here $f:=1-2M/r$, and 
\begin{equation}
\Lambda(r):= 8\,{\rm Li}_2(1-r/2)+4\pi^2/3+4(\ln 2)^2 ,
\end{equation}
where
\begin{equation}
{\rm Li}_n(z)=\sum_{k=1}^\infty z^k/k^n
\end{equation}
is the polylogarithm function.

%%%%%%%%%%%%%%%%%%%%%%%%%%%%%%%%%%%%%%%%%%%%%%%%%%%%%%%%%%%%%%%%%%%
%%%%%%%%%%%%%%%%%%%%%%%%%%%%%%%%%%%%%%%%%%%%%%%%%%%%%%%%%%%%%%%%%%%
\bibliographystyle{unsrt}
\bibliography{newbiblio}
%%%%%%%%%%%%%%%%%%%%%%%%%%%%%%%%%%%%%%%%%%%%%%%%%%%%%%%%%%%%%%%%%%
%%%%%%%%%%%%%%%%%%%%%%%%%%%%%%%%%%%%%%%%%%%%%%%%%%%%%%%%%%%%%%%%%%%

\end{document}